\documentstyle[12pt]{book}

\def\deg{^\circ}
\def\arcmin{^\prime}
\def\arcsec{^{\prime\prime}}

\def\@biblabel#1{\relax}
\def\@cite#1#2{#1\if@tempswa , #2\fi}

\def\@citex[#1]#2{\if@filesw\immediate\write\@auxout{\string\citation{#2}}\fi
\def\@citea{}\@cite{\@for\@citeb:=#2\do
{\@citea\def\@citea{,\penalty\@m\ }\@ifundefined
{b@\@citeb}{\@warning
{Citation `\@citeb' on page \thepage \space undefined}}%
{\csname b@\@citeb\endcsname}}}{#1}}
     \def\@evenhead{\parbox{\textwidth}{{\itshape\thepage\hfil\itshape\leftmark}\vspace{-3mm} {\rule{\textwidth}{0.1mm}}}}
     \def\@oddhead{\parbox{\textwidth}{{\itshape\rightmark\hfil\thepage}\vspace{-3mm} {\rule{\textwidth}{0.1mm}}}}

\newcommand{\et}{{\em et al} }
\newcommand{\dg}{\nobreak^\circ}
\newcommand{\degg}{\ifmmode^\circ \else$^\circ $\fi }
\newcommand{\lta}{{\small\raisebox{-0.6ex}{$\,\stackrel
{\raisebox{-.2ex}{$\textstyle <$}}{\sim}\,$}}}

\newcommand{\gta}{{\small\raisebox{-0.6ex}{$\,\stackrel
{\raisebox{-.2ex}{$\textstyle >$}}{\sim}\,$}}}

\newcommand{\simlt}{\mbox{$\stackrel{<}{_{\sim}}$} }
\newcommand{\simgt}{\mbox{$\stackrel{>}{_{\sim}}$} }

\newcommand{\kms}{\,\hbox{km}\,\hbox{s}^{-1}}

\newcommand{\ie}{{\it i.e.\ }}

\input psfig
\psfull

\makeindex
\begin{document}
\begin{titlepage}
\begin{center}
\noindent
{\Huge\bf Application of novel}
 
\noindent
{\Huge\bf analysis}

\noindent
{\Huge\bf techniques to}
 
\noindent
{\Huge\bf Cosmic Microwave Background}

\noindent
{\Huge\bf astronomy}
 
 
 
\vspace{1in}
\noindent
{\large Aled Wynne Jones}
 
\vspace{0.5in}
\noindent
{\footnotesize\it Mullard Radio Astronomy Observatory}
 
\noindent
and
 
\noindent
{\footnotesize\it King's College, Cambridge.}
 
\end{center}
 
%
%
\begin{center}
\noindent
A dissertation submitted for the degree of Doctor
 
\noindent
of Philosophy in the University of Cambridge.
\end{center}
\end{titlepage}

\frontmatter

\noindent
{\Huge{\bf Preface}}

\vspace{1cm}

\noindent
This dissertation is the result of work undertaken at the Mullard Radio 
Astronomy Observatory, Cambridge between October 1994 and September 1997. 
The work described here is my own, unless specifically stated otherwise. 
To the best of my knowledge it has not, nor has any similar dissertation,
been submitted for a degree, diploma or other qualification at this, or 
any other university. This dissertation does not exceed 60 000 words in 
length. 

\vspace{2in}
\hspace{3in} Aled Wynne Jones


\rule{0cm}{3in}
 
\begin{center}
\noindent
To Isabel and Ffion
\end{center}


\rule{0cm}{2.5in}
{\center
\noindent
{\em I am not sure how the universe was formed. But it knew
how to do it, and that is the important thing.}
 
\hspace{2.7in} Anon. (child)
}
 
\vspace{2in}
{\center
\noindent
{\em It is enough just to hold a stone in your hand. The universe would have
been equally incomprehensible if it had only consisted of that one stone the
size of an orange. The question would be just as impenetrable: where did
this stone come from?}
 
\hspace{2.7in} Jostein Gaarder (in `Sophie's World')
}

\noindent
{\Huge\bf Acknowledgements}
\label{ack}

\vspace{0.5in}

Firstly I would like to thank the two people who have introduced me 
to the immense field of microwave background anisotropies, Anthony 
Lasenby and Stephen Hancock. Without them I would not have begun to 
uncover the beauty at the beginning of time. I would also like to 
thank Joss Bland-Hawthorn whose supervision and enthusiasm 
during my time in Australia 
has made me more inquisitive in my field. The many 
collaborations involved in this project have introduced me to many 
people without whom this thesis would not have been written; 
Graca Rocha and Mike Hobson at MRAO, Carlos Gutierrez, 
Bob Watson, Roger Hoyland and Rafael Rebolo at Tenerife, and 
Giovanna Giardino, Rod Davies and Simon Melhuish at Jodrell Bank. 

I want to say a special thank you to the two women in my life that have kept 
me going for the last few years. Thanks to Ffion, my sister, whose insanity
has kept me sane and thanks to Isabel whose support and encouragement I 
could not have done without and whose love has made it all worth while. 

Martina Wiedner, Marcel Clemens and Dave St. Jacques deserve a special 
mention for making my time in the department a little more bearable. 
Anna Moore for putting up with me for three months in Australia.
Cynthia Robinson, Martin Gunthorpe, Nicholas Harrison, Liam Cox and 
Dafydd Owen for putting up with me for the first years of my research. 

I could not finish thanking people without mentioning Pam Hicks
and David Titterington (special thanks for all the colour overhead 
transparencies) who have kept the department running smoothly.

I am also very grateful to PPARC for awarding me a research studentship. 
May they know better next time. 

\vspace{1in}

Yn olaf diolch yn fawr i fy rhieni sydd wedi rhoi i fyny efo fi am yr
ugain mlynedd diwethaf.

\tableofcontents

\mainmatter

\chapter{Introduction}
\label{mainintro}

The Big Bang is the name given to the theory that describes how
the Universe came into existence about 15 billion years ago at infinite
density and temperature and then expanded to its present form. 
The very early Universe was opaque due to the constant interchange
of energy between matter and radiation. About 300,000 years after the
Big Bang, the Universe cooled to a temperature of $\sim 3000\deg$C
because of its expansion. At this stage the
matter does not have sufficient energy to remain ionised. 
The electrons combine with the protons to form atoms and the cross
section for Compton scattering with photons is dramatically reduced.
The radiation from this point in time has been travelling towards us 
for 15 billion years and has now cooled to a blackbody temperature of
2.7 degrees Kelvin. At this temperature the Planck spectrum has its
peak at microwave frequencies ($\sim 1-1000$~GHz) and its study
forms a branch of astronomy called Cosmic Microwave Background
astronomy (hereafter CMB astronomy). In 1965 Arno Penzias
and Robert Wilson (Penzias \& Wilson 1965) 
were the first to detect this radiation. It is seen from
all directions in the sky and is very uniform. This uniformity creates a 
problem. If the universe is so smooth then how did anything form? There must
be some bumps in the early universe that could grow to 
create the structures we see today. 

In 1992 the NASA Cosmic Microwave Background Explorer (COBE) satellite 
was the first experiment to detect the bumps. These initial measurements were in
the form of a statistical detection and no physical features could be
identified. Today, experiments 
all around the world are finding these bumps that eventually grew into galaxies
and clusters of galaxies. 
The required sensitivities called for new techniques in astronomy. 
The main principle behind all of these experiments is that, instead of 
measuring the actual brightness, they measure the difference in brightness 
between different regions of the sky. The experiments at Tenerife produced the 
first detection of the real, individual CMB fluctuations.

There are many different theories of how the universe began its life 
and how it evolved into the structures seen today. Each of these theories make
slightly different predictions of how the universe looked at the very 
early stages which up until now have been impossible to prove or disprove. 
Knowing the structure of the CMB, 
within a few years it should be possible for astronomers to tell us where
the universe came from, how it developed and where it will end up. 

\subsection{Outline of thesis content}
\label{outline}

This thesis covers four main topics: the detection of anisotropies of
the CMB with
data taken by the Tenerife differencing experiments and Jodrell
Bank interferometers, the analysis of this data to produce actual sky
maps of the fluctuations, the potential of future CMB experiments 
and subsequent analysis of the sky
maps. 

Chapter 2 introduces the processes involved in the formation of
the fluctuations and the resultant radiation 
we expect to see in our sky maps. This
includes both the CMB and other emission processes that operate at the
frequencies covered by CMB experiments. These other processes are
the Sunyaev--Zel'dovich effect, point source emissions and various
Galactic foregrounds (namely dust, bremsstrahlung and synchrotron
emission). 

Chapters 3 and 4 present the experiments discussed in the thesis and the data
obtained from them. Preliminary analysis is done on the data to put
constraints on Galactic emissions and point source contributions.
The data discussed covers the 5~GHz to 33~GHz frequency
range which is at the lower end of the useful CMB spectral
window. Lower frequency surveys are used to put constraints on the
spectral index of some of the Galactic foregrounds (the frequency
range is not large enough to put any useful constraints on dust
emission which is expected to dominate at higher frequencies). The Tenerife 
experiments are used to put constraints on the level of the 
CMB anisotropy.

Chapter 5 introduces the concepts of the Maximum
Entropy Method (MEM), the Wiener filter, CLEAN and Singular Value Decomposition. 
These four methods offer
different alternatives to find the best underlying cosmological signal
within the noisy data. The usual approach of a positive--only Maximum
entropy is enlarged to cover both positive and negative fluctuations,
as well as multifrequency and multicomponent observations. 
Simulations performed (Chapter 6) have shown that MEM is
the best method of the four tested when attempting a reconstruction
of the CMB signal. 

Chapter 7 presents the final sky maps of the CMB produced with the
Maximum Entropy algorithm,
as well as maps of the various contaminants that the experiments are
also sensitive to. 
The maps from a range of different experiments 
can be used to put constraints on various cosmological
parameters such as the density parameter, $\Omega_\circ$, Hubble's
constant, $H_\circ$, and the spectral index of the large scale CMB
power spectrum, $n$.

Chapter 8 presents subsequent analysis 
performed on the sky maps. These include examining the topology 
using genus as well as looking at the power
spectrum and correlation functions. The methods discussed are
first applied to simulations to test their usefulness at distinguishing
between the origins of the fluctuations and then applied to the
reconstructed CMB sky maps.

New constraints on the power spectrum and some of the cosmological
parameters will be given in the final chapter. Here, the data
and analysis described will be brought together and the future of
CMB experiments discussed.

\noindent
{\Large \bf Rhagarweiniad}
 
\vspace{0.5in}
O ble rydym ni i gyd yn dod? Pa brosesau wnaeth ddigwydd i greu popeth a 
welwn o'n cwmpas?
I ateb y cwestiynau hyn byddai'n fanteisiol gallu teithio yn \^ol mewn amser.
Yn \^ol deddfau ffiseg mae hyn yn amhosibl, ond maent yn caniat\'au rhywbeth sy'n ail orau i hynny. 
Gallwn edrych yn \^ol mewn amser. Mae golau'n teithio ar gyflymder penodol,
felly os edrychwn yn \^ol ddigon pell gallwn weld yn \^ol i ddechreuad y
bydysawd. Er enghrifft, pan edrychwn ar yr haul mae mor bell fel ein bod yn
edrych arno fel ag yr oedd wyth munud ynghynt.

Heddiw mae seryddwyr yn gallu gweld cyn belled yn \^ol mewn amser ag sy'n bosibl.
Cymysegedd o f\'as ac ynni ymbelydrol yw'r bydysawd. Dros dymheredd arbennig 
($\sim 4000^\circ$C), oherwydd ynni uchel yr ymbelydriad, mae'r rhyngweithiad
rhwng m\'as ac ynni yn gwneud y bydysawd yn dywyll. Yn nechrau'r bydysawd
roedd popeth wedi ei wasgu'n belen fechan poeth a ddechreuodd ehangu ac oeri 
wedyn. Felly, os, edrychwn yn \^ol mewn amser dylem weld y pwynt pryd y 
daeth y bydysawd yn glir. 
Mae'r ymbelydriad o'r pwynt hwn mewn amser wedi bod yn teithio tuag atom am 
15 biliwn o flynyddoedd ac mae wedi oeri i dymheredd o $-270^\circ$C erbyn hyn. Mae'r tymheredd hwn yn cyfateb i ymbelydriad microdon. Arno Penzias a Robert
Wilson oedd y rhai cyntaf i ddarganfod yr ymbelydriad hwn yn 1965. Mae'n 
ymddangos o bob cyferiad yn yr awyr ac mae'n unffurf iawn. Mae'r unffurfedd
hwn yn achosi problem. Os yw'r bydysawd mor wastad sut y gwnaeth unrhyw beth 
ffurfio? Mae'n rhaid fod yna rai gwrymiau yn y bydysawd cynnar i greu'r adeileddau
a welwn heddiw.

Lloern {\em Cosmic Microwave Background Explorer (COBE)} NASA yn 1992 oedd 
yr arbrawf cyntaf i ddarganfod y gwrymiau. Ni llwyddod i dynnu lluniau o'r
gwrymiau mewn gwirionedd oherwydd fod cymaint o s\'wn yn gwneud hynny'n amhosibl.
Daeth ar draws s\'wn na'r arfer a'r unig eglurhad y gellid ei roi am hynny
oedd presenoldeb gwrymiau yn y bydysawd. Roedd hyn yn ffodus i seryddwyr neu byddent
wedi gorfod newid eu holl ddamcaniaethau. Erbyn heddiw mae mapiau o'r bydysawd
cynnar hyd yn oed yn cael eu cynhyrchu. Led led y byd mae arbrofion yn 
darganfod gwrymiau a dyfodd yn y diwedd yn alaethau a chlystyrau o alaethau.

Roedd y gwaith hwn mor sensitif fel bod rhaid dyfeisio technegau newydd
mewn seryddiaeth. Datblygwyd dulliau newydd o dynnu lluniau o'r awyr gyda
thelegopau newydd. Yr egwyddor sylfaenol y tu \^ol i'r holl arbrofion hyn oedd,
yn hytrach na'u bod yn mesur y disgleirdeb gwirioneddol, eu bod yn mesur
y gwahaniaeth mewn disgleirdeb rhwng gwahanol ranbarthau o'r 
awyr. Cynhyrchodd yr arbrofion yn Tenerife y mapiau cyntaf o'r awyr. Mae 
Telesgop Anisotropy Caergrawnt hefyd yn cynhyrchu mapiau o ranbarthau llai 
o'r awyr, gan weld gwrymiau llai na'r rhai a welwyd o Tenerife.

Bydd yr holl arbrofion hyn yn rhoi prawf ar ddamcaniaethau'r seryddwyr. Ceir 
cannoedd o wahanol ddamcaniaethau yngl\^yn \^a'r modd y dechreuodd y
bydysawd a sut y datblygodd i'r adeileddau a welir heddiw. Mae pob un o'r damcaniaethau
hyn yn cynnig syniadau ychydig yn wahanol yngl\^yn \^a'r modd yr edrychai'r
bydysawd yn gynnar yn ei hanes a hyd yma bu'n amhosibl eu profi neu eu gwrthbrofi.

O fewn ychydig flynyddoedd dylai fod yn bosibl i seryddwyr fedru dweud wrthym o 
ble y daeth y bydysawd, sut y datblygodd ac ym mhle y bydd yn gorffen. Mae'n 
gyfnod cyffrous i seryddwyr gyda holl gyfrinachau'r bydysawd yn disgwyl i
gael eu darganfod.

\chapter[The Universe and its evolution]{The Universe and its
evolution: the origin of the CMB and foreground emissions}
\label{chap2}

In this chapter I summarise the various processes that go into forming
the power spectrum of fluctuations in the Cosmic Microwave
Background. This includes primordial effects as well as foreground
effects like the Sunyaev-Zel'dovich effect. 
I also describe the various foregrounds that have significant
emissions at the frequencies of interest to Microwave Background
experiments. 

The theory of the Big Bang stems from Edwin Hubble's observations that every
galaxy is moving away from every other galaxy (providing they are not
in gravitational orbit about each other). Hubble's law tells us
that the velocity of recession away from a point in the Universe is
proportional to the distance to that point. Today that constant of
proportionality is called Hubble's constant, $H_\circ$. If we
extrapolate this law back in time, there comes a point where everything
in the Universe is very close together. To get everything we see today
into a very small region requires an enormous amount of energy and
this is where the Big Bang theory is born. This hot, dense `soup' 
expanded, cooled and eventually formed all the
structures that we see today. 

\section{Symmetry breaking and inflation}
\label{symmetry}

In physics, as things get hotter they generally get simpler. At
a relatively low temperature ($\sim 10^{15}$~K), compared to the Big Bang, the
electromagnetic force and the weak
force, which binds the nucleus together, combine to 
form the electro--weak force.
At higher temperatures ($\sim 10^{28}$~K) the strong force (described
by Quantum Chromo-Dynamics), which keeps the
proton from splitting into its quarks, joins the electro--weak force
to become one force. This theory is called Grand Unification (GUT). 
It is hypothesised that the last force, the force of
gravity, joins the other forces in a theory of quantum gravity,
at even higher temperatures (corresponding to the very earliest times in the
Universe). At this stage everything in the Universe is
indistinguishable from everything else. Matter and energy do not exist
as separate entities and the Universe is very isotropic. 
To create structure in such a Universe 
there are two main theoretical models (or a combination of the two). 

Starting from Newtonian physics and considering the effect of
gravity on a unit mass at the edge of a sphere of radius $R$, then

\begin{equation}
\ddot{R}(t)=-{{GM}\over{R^2 (t)}},
\label{eq:newton}
\end{equation}

\noindent
where $t$ is the time since the Big Bang and 
$M$ is the mass inside the sphere, given by 

\begin{equation}
M={4\over 3} \pi R^3 (t) \rho (t).
\label{eq:mass}
\end{equation}

\noindent
In an expanding Universe, with no spontaneous particle creation, 
the amount of matter present does not change
and so $M$ is constant if we follow the motion of the edge of the
sphere. Thus, if we write the present density of the
Universe as $\rho_\circ$ then we have

\begin{equation}
\rho (t)={{R^3 (t_\circ)}\over{R^3 (t)}} \rho_\circ,
\label{eq:density}
\end{equation}

\noindent
where $t_\circ$ corresponds to now. The gravitational force per unit
mass in Equation~\ref{eq:newton} is therefore given by 

\begin{equation}
\ddot{R} (t)=-{4\over 3} \pi G \rho_\circ R^{-2} (t)
\label{eq:almost}
\end{equation}

\noindent
where $R(t_\circ )$, the radius of the sphere today, is taken as unity.
Integrating Equation~\ref{eq:almost} gives

\begin{equation}
\dot{R}^2 (t) = {8\over 3} \pi G \rho_\circ R^{-1} (t) - kc^2 .
\label{eq:almost2}
\end{equation}

\noindent
The constant of integration is found by including General
Relativistic considerations where $k$ is a measure of the curvature of space. 
So the equation of evolution of the Universe is 

\begin{equation}
\left({\dot{R}\over R}\right)^2 - {{8\pi G \rho}\over{3}} =
-{{kc^2}\over {R^2}}.
\label{eq:almost3}
\end{equation}

\noindent
We define Hubble's constant, the rate of expansion of the Universe, as
$H={\dot{R}\over R}$. In General Relativity the Universe is said to be
closed if the density is high enough to prevent it from
expanding forever. If the density is too low then the Universe will
continue to expand forever and is called open.  
The point at which the density becomes critical (in
between a closed and open Universe) corresponds to a flat space, or
$k=0$. This critical density is found from Equation~\ref{eq:almost3}
to be 

\begin{equation}
\rho_{crit}={{3H^2}\over{8\pi G}}
\label{eq:crit}
\end{equation}

\noindent
and we can define a new parameter called the closure parameter,
$\Omega={\rho\over{\rho_{crit}}}$. Using this definition, if $\Omega >
1$ then the Universe is heavier than the critical density (closed) and if
$\Omega < 1$ then it is lighter (open). 

If another constant is included in Equation~\ref{eq:newton} then an
effective zero energy (the scalar field) can be added to the General
Relativistic description of the Universe. This can be thought of as a 
`vacuum energy', the lowest possible state in the Universe. This alters 
Equation~\ref{eq:almost2} so that it includes the Cosmological constant 

\begin{equation}
\dot{R}^2 (t) = {8\over 3} \pi G \rho_\circ R^{-1} (t) +
{\Lambda\over 3} R^2 - kc^2
\label{eq:cosmoconst}
\end{equation}

\noindent
If this cosmological constant is the dominant term (as the scalar
field is expected to be at very high temperatures), and 
the other two terms in Equation~\ref{eq:cosmoconst} become negligible, 
it is possible to solve and find

\begin{equation}
R(t)\propto \exp \left[ (\Lambda/3)^{1\over2} t \right]
\label{eq:inflat}
\end{equation}

\noindent
which represents an exponential expansion. This expansion, in which
the Universe expands at speeds faster than light, is dubbed inflation
(Guth 1981 and Linde 1982). Figure~\ref{fig:scale} 
shows the effect of inflation on the size of the
Universe. 

\begin{figure}
\caption{The expansion of the Universe. The red line shows the Universe whose
size expands at the speed of light whereas the blue line includes an early
inflationary period.}
\label{fig:scale}
\end{figure}

With such an expansion, small quantum fluctuations
(produced by Heisenberg's uncertainty principle) would expand up
into large inhomogeneities in the Universe. These inhomogeneities are the
density fluctuations that then go on to form the structure that is present
in the Universe today. Exponential expansion stops when the
$\Lambda$ term becomes less dominant.

Another possible way to create fluctuations is through phase
transitions in the early Universe. The theory of phase transitions 
does not require inflation but it does not rule it out either. 
When one of the fundamental forces becomes
separated from the rest, the Universe is said to undergo 
a phase transition. If, during an early phase
transition, some of the energy of the Universe is trapped
between two regions undergoing the transition in a slightly different
way and is frozen out, then a
topological defect (Coulson \et 1994)
is formed. Depending on the original GUT the
Universe is described by we get different defects. 
There are four possible defects,
corresponding to zero, one, two and three dimensions, called
monopoles, strings (see Brandenberger 1989), 
domain walls and textures (Turok 1991) respectively. A string,
for example, can be thought of as a frozen one--dimensional region of
`early' Universe. It separates regions that went through the phase
transition at slightly different times so that the geometry around the
string is different from normal space--time geometry. In particular
the angle surrounding a string is less than $360\deg$.
These defects can act as seeds for structure formation through their
gravitational interaction. 

After inflation the Universe was still very hot and radiation
dominated so that no atoms could be formed. A thermal
equilibrium between matter and radiation was set up by continual scattering.
As the Universe cooled, processes not fully understood as yet 
caused an excess of matter over anti-matter
which started
to form basic nuclei (deuterium, helium and lithium). This
`soup' of interacting particles and radiation has a very high optical
depth and so the radiation could not escape. 

\section{Dark matter}
\label{cdm}

We have already defined the closure parameter of the Universe,
$\Omega$, as the ratio of the actual density of the Universe to the
critical density. For $\Omega > 1$ (a closed Universe)
gravity will dominate and the Universe will
collapse in on itself in a finite time. For $\Omega < 1$ (an open
Universe) the expansion will dominate
and the Universe will continue growing forever. We can weigh the
Universe by making observations of the stars and galaxies and
estimating how heavy the objects are that we can see. In
1978 observations were first reported of the rotation curve of galaxies
and it was calculated that there must be a lot more mass, unaccounted for by
light (see
Rubin, Ford \& Thonnard, 1978). Later, observations
were made of velocities of galaxies in clusters
and it was found that even more unseen mass was required to give the
galaxies their observed peculiar velocities. The Milky Way is in orbit
around the Virgo cluster with a peculiar velocity 
of $\sim 600$~${\rm kms}^{-1}$ (see Gorenstein
\& Smoot 1981 for the first measurement of this peculiar velocity,
determined from the dipole in the CMB). The luminous
mass in the Universe can  account for $\Omega_{lum}=0.003 h^{-1}$,
where $h=H_\circ$/100~km ${\rm s}^{-1}$~${\rm Mpc}^{-1}$ 
and so, taking into account the non--luminous mass, or dark
matter, this is a lower limit on $\Omega$  
(see White 1989 for a review on dark matter). 

It remains to be seen whether this dark matter can make $\Omega=1$ as
simple inflation predicts. The obvious question that comes to
mind is -- ``In what form does the dark matter come?''. The most
obvious candidate for dark matter is 
non--luminous baryon matter. This can not exist as free hydrogen or
dust clouds, otherwise we would expect to see large black objects
across the sky blocking out the starlight. If baryonic, the dark
matter must exist as
gravitationally bound matter, either in the form of brown dwarfs
(Carr 1990), planets, neutron stars or black holes. These may exist
in an extra--galactic halo around our galaxy and are called
Massive Compact Halo Objects or MACHO's for short (Alcock \et 1993). 
However, the amount
of baryonic matter present in the Universe is constrained by the
relative proportions 
of hydrogen, helium, deuterium, lithium and beryllium that 
are observed as these were formed together
in the early Universe. This gives us 
$0.009 \le \Omega_b h^2 \le 0.02$ (Copi {\em et al} 1995) 
and taking $h=0.5$ then $\Omega_b < 0.1$ which means that 90\%
of the Universe is made up of non--baryonic matter if it is spatially flat
and closed ($\Omega=1$). 

The non--baryonic matter must take the form of Weakly Interacting
Massive Particles (WIMPS; Turner 1991) which only interact with baryonic matter
through gravity (otherwise they would have been detected
already). The theory of WIMPS can be subdivided into two categories;
hot dark matter (HDM) and cold dark matter (CDM). Hot dark matter has
large thermal velocities (for example, heavy neutrinos) and will wipe
out structure on galactic scales in the early Universe 
due to streaming in the last
scattering surface. This is a `top--down' scenario. Cold dark matter
has low thermal velocities (for example, axions and supersymmetric
partners to the baryonic matter) and will enhance the gravitational
collapse of galactic size structures. This is a `bottom--up'
scenario. The two theories are not mutually exclusive and the combination
of CDM and HDM is called mixed dark matter (MDM).

Constraints are already possible on some of the dark matter
candidates. For example, in HDM if the dominant form of matter
consists of heavy neutrinos, then it can be shown (see Efstathiou
1989) that for a critical Universe (so $\Omega=1$) the neutrinos need to
be about 30~eV in mass. With better measurements of the power
spectrum of the
fluctuations seen in the CMB it will be possible to rule out or
confirm the existence of such particles. No candidate dark matter has yet
been detected. 

\section{Coming of age}
\label{age}

At a redshift of z $\sim 1100$, the Universe had cooled to a
temperature of $\sim 3000$~K. At this temperature electrons become
coupled with protons and form atoms. This essentially increases the
photon mean free path from close to zero to infinity in a very short time
($\Delta z\sim 80$). So the furthest we can look back is to this last
scattering surface.
This period is called the recombination era and is
the origin of the microwave background radiation studied in this thesis.
By observing this radiation the imprints of the
fluctuations from the early part of the Universe can be studied and hence
cosmological models can be tested. The temperature of the microwave
background has now cooled down through the effects of cosmic expansion and 
has been measured to a very high degree of accuracy. It is found (see
Mather, J.C. \et 1994) to be
at

\begin{equation}
T_\circ = 2.726 \pm 0.010 {\rm K} \quad (95\% \, {\rm confidence}).
\label{eq:tnought}
\end{equation}

The pattern of fluctuations in the radiation from the last scattering
surface will tell us a lot about the early Universe. There are two
models for how the matter fluctuations couple to the radiation
fluctuations. These are adiabatic and isocurvature
fluctuations. Inflation naturally produces the former but in special
conditions can produce the latter. 
Adiabatic fluctuations are perturbations in the density field which conserve
the photon entropy of each particle species (the number in a comoving 
volume is conserved). 
Isocurvature fluctuations are
fluctuations in the matter field with equal and opposite
fluctuations in the photon field, keeping the overall energy constant
and therefore a constant curvature of space--time. 

Due to the early coupling between matter and radiation, prior to the
last scattering surface, an almost perfect blackbody would exist
throughout the Universe at last scattering. 
For blackbody emission, the spectrum (i.e. the CMB) is given by the
differential Planck spectrum

\begin{equation}
\label{eq:TantCMB}
\Delta{T_A}={{\Delta T x^2 e^x}\over{({e^x}-1)^2}}.
\end{equation}

\noindent
Therefore the change in intensity is

\begin{equation}
\label{eq:intCMB}
\Delta{I(\nu)}={{\Delta T x^4 e^x}\over{({e^x}-1)^2}},
\end{equation}

\noindent
where $x={{h\nu}\over{kT}}$. The Universe progressively became more
transparent and so the fluctuations seen at the last scattering
surface are a superposition of fluctuations within a last scattering
volume (comprising of the region between a totally opaque and a
totally transparent Universe). 
This can be expressed in terms of the optical depth, $\tau$, as

\begin{equation}
{{\Delta T}\over{T}}=\int_{0}^{z} {{\delta T(z)}\over{T}} 
{e^{-\tau(z)}}{{d\tau}\over{dz}}dz,
\label{eq:fullcmb}
\end{equation}

\noindent
where $g(z)={e^{-\tau(z)}}{{d\tau}\over{dz}}$ is a Gaussian centred on
$z\sim 1100$ with $\Delta z\sim 80$, the width of
the last scattering surface (see, for example, Jones \& Wyse 1985). 
Radiation from 
fluctuations smaller than the width of last scattering will add
incoherently and therefore
the radiation pattern of fluctuations will be erased on these small scales. 
This corresponds to an angular size of 
$\theta=3.8\arcmin {\Omega_\circ}^{1\over2}$ on the sky  
today so all anisotropies smaller than this will be heavily suppressed. 

\subsection{The dipole}
\label{COBEdipole}

The main source of anisotropy in the CMB is not intrinsic to the
Universe. It is produced by the peculiar velocity of the observer. 
Moving towards an object causes emitted light to appear blueshifted.
As the Earth is not stationary
with respect to the CMB (it is moving around the sun, the sun around
the galaxy, the galaxy around the Virgo cluster etc.) there will be a
part of the CMB that the Earth moves towards and a part that it
moves away from. Therefore, we expect to see a large dipole created by
this Doppler effect. This dipole was clearly detected by the COBE satellite
(see Figure~\ref{fig:dipole}). It is necessary to remove this before
attempting to study the intrinsic fluctuations in the CMB. 

\begin{figure}
\caption{The dipole in the CMB as seen by the COBE satellite.}
\label{fig:dipole}
\end{figure}

\subsection{Sachs-Wolfe effect}
\label{sachs}

As the Universe grows older the observable Universe gets bigger,
due to the finite speed of light. The particle
horizon of an observer is the distance to the furthest object that
could have affected that observer. Any objects further than this point
are not, and never have been, in causal contact with the observer. At the last
scattering surface the particle horizon corresponds to $\theta \sim
2\deg$ as seen from Earth today. No physical processes will
act on scales larger than this. Therefore, at 
the epoch of recombination fluctuations must have been produced by
matter perturbations already present at this 
time. Inflation gives us a mechanism for the creation of these fluctuations. 
These matter perturbations give rise to perturbations in the
gravitational potential. Radiation will experience different
redshifts depending on the potential and hence
produce large angular scale anisotropies in the CMB. This process is
called the Sachs--Wolfe effect (Sachs \& Wolfe 1967). 

It can be shown (see for example Padmanabhan 1993) that the angular
dependence of the Sachs--Wolfe temperature fluctuations on scales
greater than the horizon size is given by 

\begin{equation}
{{\Delta T}\over T} \propto \theta^{{(1-n)}\over 2},
\label{eq:sachs}
\end{equation}

\noindent
where $n$ is the spectral index of the initial power spectrum of  
fluctuations ($P(k)=Ak^n$). In inflation the natural outcome is a spectral index
$n=1$ because the fluctuations originate from quantum fluctuations that
have no preferred scale (although recently it has been shown that
inflation does allow other possible values of $n$). 
This special case is called the
Harrison--Zel'dovich (Harrison 1970 and Zel'dovich 1972)
spectrum and leads to the ${{\Delta T}\over T}$
fluctuations being constant on all angular scales larger than the horizon size.
These fluctuations have been observed by the COBE satellite at an
angular scale of $7\deg$. The combined maps from the three observing frequencies
after two years of COBE measurements
at this angular scale are shown in Figure~\ref{fig:COBE}.

\begin{figure}
\caption{The combined COBE maps showing the CMB over the full sky.}
\label{fig:COBE}
\end{figure}

\subsection{Doppler peaks}
\label{doppler}

An overdensity in the early Universe does not collapse under the
effect of self-gravity until it enters its own particle horizon when
every point within it is in causal contact with every other point. The
perturbation will continue to collapse until it reaches the Jean's length, at
which time radiation pressure will oppose gravity and set up acoustic
oscillations. Since overdensities of the same size will pass the
horizon size at the same time
they will be oscillating in phase. These acoustic
oscillations occur in both the matter field and the photon field and so
will induce `Doppler peaks' in the photon spectrum.

The level of the Doppler
peaks in the power spectrum depend on the number of acoustic
oscillations that have taken place since entering the horizon. For
overdensities that have undergone half an oscillation there will be a
large Doppler peak (corresponding to an angular size of $\sim 1\deg$). Other
peaks occur at harmonics of this. As the amplitude and position of 
the primary and secondary peaks are
intrinsically determined by the number of electron scatterers and
by the geometry of the Universe, they can
be used as a test of the density parameter of baryons and dark matter, as
well as other cosmological constants.

\subsection{Defect anisotropies}
\label{defect}

The anisotropies produced by the various forms of defects arise from
the effect that the defect has on the surrounding space--time (for
example see Coulson {\em et al} 1994). Not only do they
leave imprints on the CMB at the time of last scattering but a large
proportion of the fluctuations due to defect anisotropies would be
produced at latter times. As an example, consider the
effect of cosmic strings (see Kaiser \& Stebbins 1984). 
A string moving with velocity $v$ will leave
behind it a `gap' in space--time. The angle around the string is not
$360\deg$ but is reduced by $8\pi G \mu$, where
$\mu$ is the energy density per unit length in the string. 
Therefore, photons travelling
through space behind the string will experience a Doppler
boost, with respect to photons in front of the string, as they have
less space to travel through. The value of this boosting is 

\begin{equation}
{{\Delta T}\over T}= 8\pi G \mu {v\over c}
\label{eq:string}
\end{equation}

\noindent
and is called the Kaiser--Stebbins effect. This results in a linear
discontinuity in the CMB when the string passes in front and is easily
discernible from the Gaussian anisotropies produced by the other
processes. The higher dimensional defects will produce more
complicated discontinuities. Recently however, Magueijo \et 1996 and Albrecht
\et 1996 have shown that this discontinuity effect may be masked by the defect
interaction with the CMB prior to recombination. 
On large angular scales the
discontinuities will also add together and mimic a Gaussian field (the central
limit theorem). Therefore, only a high resolution (on the arcmin
scale), high sensitivity, 
experiment will be able to distinguish between defect and
inflationary signatures on the CMB.

\subsection{Silk damping and free streaming}
\label{silk}

Prior to the last scattering surface the photons and matter interact
on scales smaller than the horizon size. Through diffusion the
photons will travel from high density regions to low density regions
`dragging' the electrons with them via Compton interaction. The
electrons are coupled to the protons through Coulomb interactions
and so the matter will move from high density regions to low density
regions. This
diffusion has the effect of damping out the fluctuations and is
more marked as the size of the fluctuation decreases. Therefore, 
we expect the Doppler peaks to vanish at very small angular
scales. This effect is known as Silk damping (Silk 1968).

Another possible diffusion process is free streaming. It occurs
when collisionless particles (e.g. neutrinos) 
move from high density to low density
regions. If these particles have a small mass then free streaming
causes a damping of the fluctuations. The exact scale this
occurs on depends on the mass and velocity of the particles
involved. Slow moving particles will have little effect on the
spectrum of fluctuations as Silk damping already wipes out the
fluctuations on these scales, but fast moving, heavy particles (e.g. a
neutrino with 30~eV mass), can wipe out fluctuations on larger scales
corresponding to 20~Mpc today (Efstathiou 1989). 

\subsection{Reionisation}
\label{reion}
 
Another process that will alter the power spectrum
of the CMB is reionisation. If, for some reason, the Universe
reheated to a temperature at which electrons and protons became ionised
after recombination, then the interaction 
with the photons would wipe out any small
scale anisotropies expected. Today, there is reionisation 
around quasars and high energy sources but this occurred too late in
the history of the Universe to have
any large effect on the CMB. Little is known about the period
between the last scattering surface and the furthest known quasar
(z$\sim 4$) so reionisation cannot be ruled out.
 
\subsection{The power spectrum}
\label{clpower}

The usual approach to presenting CMB observations is through spherical
harmonics. The expansion of the fluctuations over the sky can be
written as 

\begin{equation}
{\Delta T \over T} (\theta,\phi) = \sum_{\ell,m} a_{\ell m} Y_{\ell m}
(\theta,\phi)
\label{eq:spheric}
\end{equation}

\noindent
where $\theta$ and $\phi$ are polar coordinates. Here
$a_{\ell m}$ represent the coefficients of the expansion.
For a random Gaussian field all of
the statistical information can be obtained by considering the
two--point correlation function, given by

\begin{equation}
C(\beta)= \left< {{\Delta T\over T} ({\bf n_1}) {\Delta T\over T}
({\bf n_2})} \right >
\label{eq:twopoint}
\end{equation}

\noindent
for the unit vectors ${\bf n_1}$ and ${\bf n_2}$ that define the
directions such that ${\bf n_1}.{\bf n_2}=\cos(\beta)$. Substituting
Equation~\ref{eq:spheric} into Equation~\ref{eq:twopoint} gives

\begin{equation}
C(\beta)=\sum_{\ell m} \sum_{\ell^\prime m^\prime} < a_{\ell m} a^*_{\ell^\prime
m^\prime} > Y_{\ell m} (\theta, \phi) Y^*_{\ell^\prime m^\prime}
(\theta^\prime, \phi^\prime).
\label{eq:cofbeta}
\end{equation}

\noindent
If the CMB has no preferred direction, so that it is
statistically rotationally symmetric, then

\begin{equation}
C(\beta)={1\over{4\pi}} \sum_\ell (2\ell+1) C_\ell P_\ell (\cos\beta)
\label{eq:rotsym}
\end{equation}

\noindent
defining $<a_{\ell m} a_{\ell^\prime m^\prime}>=C_\ell \delta_{\ell \ell^\prime}
\delta_{mm^\prime}$ and the multiplication of 
spherical harmonics give the Legendre polynomials
$P_\ell (\cos\beta)$. If this is taken as a true indicator of the
CMB then the $C_\ell$ values can be used to give a complete statistical
description of the fluctuations. These $C_\ell$ values can be predicted
from theory (normalised to an arbitrary value) and constitute the
power spectrum of the CMB. For example, if a standard power law
($P(k)=Ak^n$) can be used to describe the fluctuations, as in the case
of the Sachs Wolfe effect, then $C_\ell$ is given by (see Bond \&
Efstathiou, 1987)

\begin{equation}
C_\ell = C_2 {{\Gamma\left[\ell+(n-1)/2\right]
\Gamma\left[(9-n)/2\right]}\over{\Gamma\left[\ell+(5-n)/2\right]
\Gamma\left[(3+n)/2\right]}} 
\label{eq:clinc2}
\end{equation}

\noindent
where $C_\ell$ is now normalised to the quadrupole term $C_2$ and
$\Gamma[x]$ are the Gamma functions. 

The power spectrum of the CMB is made up of a combination
of all the competing processes already described. At large angular
scales (corresponding to small $\ell$ values in the Fourier plane)
the level of fluctuations (and hence $\ell (\ell +1) C_\ell$) will be constant due to the
Sachs--Wolfe effect. At intermediate angular scales ($\sim 2\deg$) the level
will start to rise when the acoustic oscillations begin to act. At the
smallest angular scales the level will approach zero as the dissipative
processes take place. The
actual shape of the power spectrum can be calculated for all
inflationary scenarios but this is much more difficult for
defects. In all simulations of the CMB discussed in this thesis
the power spectrum used is that for a standard CDM inflationary 
model with $H_\circ = 50$~${\rm kms}^{-1}$, $\Omega_\circ =1$ 
and a baryon density parameter, $\Omega_b=0.05$. Figure~\ref{fig:speccmb}
shows the predicted spectrum for a standard CDM realisation of the
Universe. 

\begin{figure}
\caption{The predicted CMB spectrum (Sugiyama 1995) for a standard CDM
realisation of the Universe with $\Omega_b =0.06$ and $h=0.5$.}
\label{fig:speccmb}
\end{figure}

\section{The middle ages}
\label{middle}

With the seeds of fluctuations sown, gravity 
started to enhance the differences. Over--dense regions grew at the
expense of under--dense regions and new structures formed. Over
the next few billion years galaxies and clusters of galaxies would
decouple from the Hubble flow with the help of gravity.
These structures (both large and small) 
are the structures that are seen in 
the night sky. Unfortunately for a CMB astronomer, these structures
are also part of what they see when they point their telescopes at the
sky. The CMB emits in the microwave region of the spectrum but so do
extra--galactic sources and sources within the galaxy. This section
describes some of the foreground processes that have significant emission at
frequencies of interest to CMB astronomers, like extra--Galactic point
sources, as well as processes that interact with the CMB
photons to alter their spectra, like the Sunyaev--Zel'dovich effect. 

\subsection{Sunyaev--Zel'dovich effect}
\label{szeff}

Hot ionised gas interacts with the CMB photons to alter their power
spectrum. Such a hot region is found around clusters of galaxies. To
first order, the Doppler scattering of the photons from the electrons
in the hot gas averages to zero. However, to second order, the inverse
Compton effect will distort the power spectrum. A relatively cold
photon passing through a hot gas will gain a boost in its 
energy, moving its temperature up slightly leaving a hole in the
CMB. Therefore, there will be less CMB photons at lower frequencies while
at higher frequencies there will be an excess of CMB photons. This
results in a frequency dependence of the spectrum
given by (for derivation see Rephaeli \& Lahav 1991) 

\begin{equation}
\Delta I={{2(kT)^3}\over{(hc)^2}} y g(x),
\label{eq:thermal}
\end{equation}

\noindent
where $T$ is the temperature of the CMB, $\nu$ is the frequency, 
$y$ is the Comptonisation
parameter which is dependent on the electron interaction with the
photon, and $g(x)$ is given by 

\begin{equation}
g(x)={{x^4 e^x}\over{(e^x -1)^2}} \left[ x {\rm coth} \left({x\over
2}\right) -4 \right]
\label{eq:gx}
\end{equation}

\noindent
with $x={{h\nu}\over{kT}}$. 

If the cluster is moving with its own peculiar velocity (i.e. it is
not moving solely with the Hubble flow), then there will be an
extra Doppler shift in the spectrum. This boosts the spectrum up
slightly from the normal CMB spectrum but still preserves its
blackbody nature. The frequency dependence of this effect is 

\begin{equation}
\Delta I=-{{2(kT)^3}\over{(hc)^2}} {v_r \over c} \tau h(x),
\label{eq:kinetic}
\end{equation}

\noindent
where $v_r$ is the peculiar velocity of the cluster along the line of
sight, $\tau$ is the optical depth of the cluster and $h(x)$ is given by 

\begin{equation}
h(x)={{x^4 e^x}\over{(e^x -1)^2}}.
\label{eq:hx}
\end{equation}

\noindent
These two effects combine together to form the Sunyaev-Zel'dovich (SZ)
effect. They occur on the angular scale of clusters of galaxies, which
is generally below the scale where Silk damping has wiped out the
fluctuations intrinsic to the CMB. Figure~\ref{fig:szeffect} shows 
$g(x)$ and $h(x)$ as a function of frequency. As can be
seen the thermal SZ effect (arising from the inverse Compton
scattering) has a very characteristic spectrum which
makes it easy to identify (the zero point is at 217 GHz)
whereas the kinetic SZ effect (arising from the Doppler boost) 
has the same spectrum as the differential CMB blackbody and
is therefore harder to distinguish using statistical techniques.

\begin{figure}
\caption{The functional form of the thermal (solid line) and kinetic
(dashed line) SZ effect.}
\label{fig:szeffect}
\end{figure}

Since the SZ effect arises from the CMB interacting with cluster gas, the
spatial power spectrum of the anisotropies
will closely follow that of the cluster gas. The cluster
gas is gravitationally tied to galaxy clusters, which are distributed
in a Poissonian manner (white noise) across the sky. Therefore, the power
spectrum of the SZ effect ($C_\ell$), 
like that of the extra--galactic point sources, 
is expected to be constant with $\ell$.

\subsection{Extra--galactic sources}
\label{extragal}

One of the main foregrounds that is seen in CMB data originates from
extra-galactic sources. These are usually unresolved point sources
such as quasars and radio--loud galaxies. A study 
of the contribution by unresolved point sources to CMB
experiments has been produced by Franceschini \et (1989). 
They used numerous surveys, including VLA and IRAS data, to put limits on
the contribution to single beam CMB experiments 
by a random distribution of point sources. This
analysis assumes that there are no unknown sources that only emit
radiation in a frequency range between
$\sim 30$~GHz and $\sim 200$~GHz. This range of frequency has not been
properly surveyed and therefore there is still a cause of concern in the CMB
community. In spite of this, the analysis by Franceschini \et will be
used to put constraints on the contribution from unresolved point source to
the data discussed in this thesis. Figure~\ref{fig:franc} 
shows the expected fluctuation levels for a random
distribution of unresolved sources (it was assumed that all sources
above $5\sigma$ of the root mean square 
({\em rms}) point source contribution could be
resolved and subtracted effectively) as found in Franceschini \et.  
This analysis assumed that the point sources exhibit a Poissonian flux
distribution with no clustering and so this estimate is likely to be an
underestimate of the total {\em rms} signal 
expected (clustering enhances the level of fluctuations). More recent
studies of the contribution by point sources to CMB experiments
(e.g. De Zotti \et 1997) show similar results.

\begin{figure}
\caption{Curves of constant $log(\Delta T_A /T_A)$ for a random
distribution of point sources with a detection limit of
$5\sigma$. Taken from Franceschini {\em et al} 1989.}
\label{fig:franc}
\end{figure}

In CMB data it is often not easy to distinguish between a
resolved point source convolved with the beam and 
CMB fluctuations. Therefore, an estimate for these resolved sources,
as well as the expected level of unresolved sources, is needed. 
The survey carried out with the 300~ft Green Bank
telescope in 1987 (Condon, Broderick \& Seielstad, 1989) is used for
the estimates. This survey
consists of data between $0^\circ$ to $+75^\circ$ in declination at
1400~MHz and 4.85~GHz. At 1400~MHz the survey has a resolution of
$12\arcmin$ and is complete to $\sim 30$~mJy,
and at 4.85~GHz the resolution is $4\arcmin$ and is complete to $\sim 8$~mJy. 
To predict the point source levels for the various experiments
considered in this thesis the surveys must be converted to a common
resolution and gridding.

All fluxes in the two frequency surveys, above the
sensitivity levels, are then compared, pixel by pixel, to find the spectral
index of each pixel. The spectral index is then used to 
extrapolate the flux up to the frequency of the experiment being
considered. In this way a spatially varying spectral index for the
point sources is obtained.  
This has obvious disadvantages as it does not take into account
steepening spectral indices or the variability of sources but it is
the best simple estimate and will provide good constraints on the
data. The maps are then converted from flux, $S$, into antenna
temperature, $T_A$, using the Rayleigh--Jeans approximation to the differential
of the Planck spectrum given by 

\begin{equation}
T_A=\left( {\lambda^2 \over {2k\Omega_{b}}} \right) S,
\label{eq:fluxtoT}
\end{equation}

\noindent
where $\lambda$ is the wavelength of the experiment, $\Omega_{b}$ is
the area of the beam ($\Omega_b =2\pi \sigma^2$ where $\sigma$ is the beam
dispersion) and $k$ is the
Boltzmann constant. At $\sim 60$~GHz ${h\nu \over kT}$ becomes 1 and 
Equation~\ref{eq:fluxtoT} is no longer a good approximation
and the full spectrum should be used (e.g. for the Planck Surveyor satellite). 
For the 5~GHz Jodrell Bank
interferometer (see next chapter for description of the experiments)
the conversion using the Rayleigh--Jeans approximation is 

\begin{equation}
{{T_A}\over {S}}=64 \mu {\rm K} / Jy
\label{eq:conv10}
\end{equation}

\noindent
and for the 10~GHz, $8.3\deg$ FWHM  
beam switching experiment at Tenerife 

\begin{equation}
{{T_A}\over {S}}=14 \mu {\rm K} / Jy.
\label{eq:conv2}
\end{equation}

\noindent
The final maps of the expected point source temperatures are
then convolved with the experimental beam and
compared with the data from that particular experiment.

This prediction for the level of point source contamination is a good
first approximation but for accurate subtraction from the data more is
required. Many point sources are highly variable and so will
contribute to each data set differently. Without simultaneous
observations of each point source in the data this is very difficult
to account for. For example, one of the main contaminants to the
data sets, discussed in this thesis, is 3C345. The
variability of this source at various frequencies 
is shown as a function of time in
Figure~\ref{fig:3c345}. It is seen that, over the period that the
data discussed in this thesis was taken, 3C345 varied in flux by more
than a factor of two. This would have a large effect on the subtraction
of this source from the individual data scans and making a prediction
by averaging the data over ten years of data collection will give
incorrect results. It is noted that this is likely to be the most
variable point source in the region that is of interest in this thesis.
The data for the variability of the point sources has
only recently become available and so only the first approximation for
fitting point sources was used in the analysis presented in this
thesis. Therefore, care was taken to exclude any highly variable
point sources from regions that were used for CMB analysis. 

\begin{figure}
\caption{The variability of 3C345 as a function of time at various
frequencies. Taken from the Michigan and Metsahovi monitoring program.}
\label{fig:3c345}
\end{figure}

The distribution of extragalactic point sources across the sky is
Poissonian. This is just a simple white noise power
spectrum and, therefore, it is flat with varying angular scale. 
In contrast to the
varying spectrum of the CMB ($C_\ell \propto ( \ell (\ell +1) )^{-1}$)
we expect the power spectrum ($C_\ell$) to be 
constant for all values of $\ell$. This constant value is
determined through observational constraints. 

\subsection{Galactic foregrounds}
\label{foregrounds}

Galactic
emission processes, such as bremsstrahlung (free-free), synchrotron and
dust emission, are all important foregrounds in CMB experiments.
The following is a brief description of the
processes involved in the foreground emissions and gives the best estimate
of the spectral dependencies of each.

\subsubsection{\bf Dust emission}
\label{dust}

At the higher frequency range of the microwave background experiments,
dust emission starts to become dominant. This is the hardest galactic
foreground to estimate as it depends on the properties of the
individual dust grains and their environment. 

The emission from an ensemble of dust grains follows an opacity law.
For this process the intensity is given by
 
\begin{equation}
I(\nu)=\int\epsilon(\nu)dl,
\label{eq:inten}
\end{equation}
 
\noindent
where $\epsilon(\nu)$ is the emissivity at frequency $\nu$,
and the integral is along the
line of sight. The brightness temperature is found from the black body
equation and is therefore a solution of 
 
\begin{equation}
I(\nu)={{{2h\nu^3}\over{c^2}} {1\over{(e^{{{h\nu}\over{kT}}}-1)}}}.
\label{eq:Tbright}
\end{equation}
 
\noindent
In the Rayleigh-Jeans approximation (where $h\nu \ll kT$). 
 
\begin{equation}
{T_b}={{c^2I(\nu)}\over{2\nu^2k}}.
\label{eq:TbrightRJ}
\end{equation}
 
\noindent
This equation is used to convert between temperature and
flux for all of the foreground emissions. Considering a constant
line--of--site density of dust, it is possible to combine
Equations~\ref{eq:TbrightRJ} and \ref{eq:inten} to give
 
\begin{equation}
{T_b}\propto \epsilon(\nu)\nu^{-2}.
\label{eq:spec}
\end{equation}
 
\noindent
When modelling dust emission, it is therefore necessary to find the
emissivity as a function of frequency, as well as the flux level at a
particular frequency.

From surveys of the dust emission
(for example the COBE FIRAS results and the IRAS survey) it can be
shown that low galactic latitude dust (dust in the galactic plane) is
modelled well by a blackbody temperature of 21.3~K and an emissivity
proportional to $\nu^{1.4}$, while at high galactic latitudes
it is well modelled by a blackbody temperature of 18~K and an
emissivity proportional to $\nu^{2}$ (Bersanelli {\em et al}
1996). Since the observations
discussed in this thesis are all at high galactic latitudes
the dust models used have an assumed blackbody temperature of 18~K and 
a spectra which follows

\begin{equation}
\Delta{I(\nu)}\propto{{\Delta T x^6 e^x}\over{({e^x}-1)^2}},
\label{eq:dustspec}
\end{equation}

\noindent
where $x={{h\nu}\over{kT}}$. This equation is obtained by the
differentiation of Equation~\ref{eq:Tbright} with respect to $T$, 
multiplied by the dust emissivity. 

As the dust emission comes from regions of warm interstellar clouds it
is very likely that there will also be ionised clouds associated with
the neutral clouds (perhaps embedded within the neutral clouds or
surrounding them, see McKee \& Ostriker 1977 for an example of
correlated features) and so we should expect bremsstrahlung from
the same region (see below for description of bremsstrahlung). 
This results in an expected correlation between the
dust and bremsstrahlung. This correlation has been detected 
(see for example Kogut \et 1996a or Oliveira-Costa \et
1997 who find a cross--correlation between the DIRBE dust maps and 
the low frequency Saskatoon data which is contaminated by
bremsstrahlung). In a full analysis of any data
this correlation should be taken into account. 

From IRAS observations of dust emission (Gautier {\em et
al} 1992), it was found that the dust fluctuations have a power law that
decreases as the third power of $\ell$. This has also been confirmed at
larger angular scales by the COBE DIRBE satellite. This means that at
small angular scales (large $\ell$) there is less power in the dust
emission. 

\subsubsection{\bf Bremsstrahlung}
\label{freefree}

When a charged particle is accelerated in a Coulomb field it will emit
radiation to oppose this acceleration; a braking radiation or {\em
Bremsstrahlung} (also known as free--free emission). 
In ionised clouds of gas with no magnetic field this process will be
the dominant source of radiation. The expected spectrum of this
emission can be derived by considering the classical
non-relativistic case.
Also one can make the simplification that only the electron in an electron-ion
interaction will emit the radiation, as the 
acceleration is inversely proportional
to the mass of the particle and so the ion, being much heavier than the
electron, can be effectively thought of as stationary and, therefore,
does not emit. Since the ion is stationary  
the electron moves in a fixed Coulomb field.
 
\begin{figure}
\caption{Electron with charge $e$ passing through the Coulomb field of 
an ion with charge $Ze$.}
\label{fig:elecimp}
\end{figure}

First consider the radiation from one electron.
To derive the functional form of the radiation we assume that
the electron does not deviate a great deal from its original path while 
interacting with 
the Coulomb field (this is a good 
approximation if the electron is moving very fast so that the main change
in its momentum will be normal to the path and any change parallel is 
negligible). Figure~\ref{fig:elecimp} shows 
the path of the electron as it passes the ion. The 
parameter $b$ represents the electron's closest approach to the ion. 
 
The dipole moment of the electron is given by 

\begin{equation}
{\bf d}=-e{\bf R}
\label{eq:dipole}
\end{equation}

\noindent
and its second derivative with respect to time,
in terms of the velocity, ${\bf v}$, is

\begin{equation}
{\bf\ddot{d}}=-e{\bf\dot{v}}.
\label{eq:dipder}
\end{equation}

\noindent
The electric field from a dipole in the non-relativistic case is 
given by

\begin{equation}
{{\bf E}_{rad}}={q\over{4\pi{\epsilon_\circ}rc^2}} {\bf n}{\bf\times}
({\bf n}{\bf\times}{\bf \dot v}),
\label{eq:larmor}
\end{equation}

\noindent
where ${\bf n}$ is the line of sight from the observer to the particle
and $r$ is the distance.
In the case when the electron is not deviated a great deal from its
original path, so
that ${\bf\ddot d}$ is along the normal to ${\bf v}$, the 
electric field at a point $i$, distance $r$ away from the dipole, is given by

\begin{equation}
E(t)={1\over {4\pi\epsilon_\circ}}|{\bf\ddot{d}}(t)|{{\sin\theta}\over{rc^2}},
\label{eq:radmag}
\end{equation}

\noindent
where $E(t)$ and $|{\bf\ddot d(t)}|$ represent the magnitudes of ${\bf E}(t)$ and 
${\bf \ddot d}(t)$, and $\theta$ is the angle between the direction of 
${\bf\ddot d}$ and the point $i$. From this electric field the radiation 
energy per unit area per unit frequency is given by 

\begin{equation}
{dW\over{dAd\omega}}={{\epsilon_\circ}\over\pi}\left| \hat{E}(\omega)
\right|^2 ,
\label{eq:parseval}
\end{equation}

\noindent
where $\hat{E}(\omega)$ is the Fourier transform of $E(t)$.
This follows from Parseval's theorem for Fourier transforms. 
When integrated over $dA$, after substitution for the Fourier transform of 
Equation~\ref{eq:radmag}, it follows that

\begin{equation}
{dW\over{d\omega}}={{2{\mu_\circ}\omega^4}\over{3c}}\left|
\hat{d}(\omega) \right|^2 .
\label{eq:dWdw}
\end{equation}

The Fourier transform of Equation~\ref{eq:dipder} is given by

\begin{equation}
-\omega^2 {\bf \hat{d}}(\omega)=-{e\over{2\pi}}\int^{\infty}_{-\infty}
{\bf \dot{v}} e^{i\omega t} dt,
\label{eq:ftdip}
\end{equation}

\noindent
which will integrate to zero, because it oscillates, over long
integration times. For
short interaction times, however, the exponential is essentially unity and we 
have

\begin{equation}
{\bf \hat{d}}(\omega)\sim{e\over{2\pi\omega^2}}\Delta{\bf v},
\label{eq:dipft}
\end{equation}

\noindent
where $\Delta {\bf v}$ is the change in electron velocity during the collision.
With the assumption that the electron does not deviate from its 
path so that the change in velocity ($\Delta v$) is normal to the path, 

\begin{equation}
{\Delta v}={{Ze^2}\over{4\pi{\epsilon_\circ}m}}{\int^{\infty}_{-\infty}}{{b\over{{(b^2+v^2 t^2)}^{3\over2}}}dt}={{Ze^2}\over{2\pi{\epsilon_\circ}bmv}}.
\label{eq:deltav}
\end{equation}

\noindent
Using Equations~\ref{eq:dWdw}, \ref{eq:dipft} and \ref{eq:deltav} it
follows that

\begin{equation}
{dW\over{d\omega}}={{Z^2{\mu_\circ}e^6}\over{6\pi^4 c{\epsilon_\circ}^2m^2b^2v^2}},
\label{eq:findWdw}
\end{equation}

\noindent
remembering that this result is only valid for the short interaction times, or 
equivalently, interactions that are within a certain distance. 

Now 
expand this to include $n_e$ electrons per unit volume, interacting with $n_i$ 
ions per unit volume. 
Integrating over all interactions results in 

\begin{equation}
{{dW}\over{d\omega dVdt}}={2 \pi v n_e
n_i}{\int^{b_{max}}_{b_{min}}}{{dW}\over{d\omega}}bdb, 
\label{eq:flux}
\end{equation}

\noindent 
where we have taken the velocity of each electron to be $v$ so that the 
flux of electrons incident on an ion is $n_e v$ 
(the element of area around an ion is given by $2\pi bdb$). 
By substituting from Equation~\ref{eq:findWdw}
and integrating the final result it is found that 

\begin{equation}
{{dW}\over{d\omega dVdt}}={{Z^2{\mu_\circ}e^6 n_e n_i}\over{3\pi^2
c{\epsilon_\circ}^2m^2 v}} {\pi\over{\sqrt{3}}} g_{ff}(v,\omega), 
\label{eq:dWdwdVdt}
\end{equation}

\noindent
where the $b_{max}$ and $b_{min}$ parameters have been absorbed into the 
$g_{ff}(v,\omega)$ Gaunt factor. This factor depends on the energy of the 
interaction and includes quantum corrections when the 
full quantum analysis is considered. For the final part of the derivation 
consider a thermal ensemble of interacting pairs. 
Averaging over a Maxwellian distribution gives

\begin{equation}
{{dW(T,\omega)}\over{d\omega dVdt}}={{\int^{\infty}_{v_{min}}
{{dW}\over{d\omega dVdt}} v^2 exp \left({ -{{mv^2}\over{2kT}}}
\right)dv}\over{{\int^{\infty}_{0}}v^2 exp \left({
-{{mv^2}\over{2kT}}} \right) dv}}, 
\label{eq:finalres}
\end{equation}

\noindent
where $v_{min}$ is taking into account the photon discreteness as the electron's
kinetic energy must be at least as big as the photon energy that it is 
creating. The final result for the emissivity of bremsstrahlung is
therefore

\begin{equation}
\epsilon(\nu) \propto n_i n_e T^{-{1\over2}}
e^{-{{h\nu}\over{kT}}}g_{ff}(T,\nu), 
\label{eq:emisfree}
\end{equation}

\noindent
where $\nu={\omega\over{2\pi}}$. This result has been tabulated
on numerous occasions (see review article by Bressaard \& van de
Hulst, 1962). At the GHz frequency range 
of interest in this thesis, it is shown that 
$\epsilon(\nu) \propto \nu^{-0.1}$ which, from Equation~\ref{eq:spec},
in terms of the temperature fluctuations, gives $T_b \propto \nu^{-2.1}$.

Kogut {\em et al} (1995) have made fits to bremsstrahlung and
determined that its power spectrum decreases as the third power of
$\ell$. This agrees with the assumed correlation of bremsstrahlung and
dust, as the dust is found to have the same $\ell$ dependence.
To model this emission the IRAS templates, normalised to the
appropriate {\em rms}, were used as
described in the dust emission section.  

\subsubsection{\bf Synchrotron emission}
\label{synch}

When a relativistic particle interacts with a magnetic field ${\bf B}$, it will 
radiate. The equations of motion describing the motion of a charged
particle are

\begin{equation}
{d\over{dt}}(\gamma m{\bf v})={q\over c}{\bf v}{\bf\times}{\bf B}
\label{eq:motion1}
\end{equation}

\noindent
and

\begin{equation}
{d\over{dt}}(\gamma mc^2)=q{\bf v}{\bf \cdotp}{\bf E}=0.
\label{eq:motion2}
\end{equation}

\noindent
Equation~\ref{eq:motion2} implies that $\gamma$, the relativistic correction 
factor, is constant (the magnitude
of the velocity is constant) and only the particles direction is altered.
The velocity 
perpendicular to the field is therefore given by

\begin{equation}
{{d{\bf v}}\over{dt}}={q\over{\gamma m c}}{\bf v}{\bf\times}{\bf B}
\label{eq:vperp}
\end{equation}

\noindent
and its magnitude is constant. 
The velocity (magnitude and direction) parallel to the field must be constant. 
In the non-relativistic case the power from a charged particle is given by 
the surface integral of the Poynting flux

\begin{equation}
P = \int \int {1\over{\mu_o}} {\bf E} \times {\bf B} dS,
\label{eq:power}
\end{equation}

\noindent
where $\mu_o$ is the permeability of free space.
Using Equation~\ref{eq:larmor} and ${\bf B} =
{1\over c} [{\bf n}\times {\bf E} ]$ it is easily seen that 

\begin{equation}
P = \int \int {{\mu_o q^2 \dot{v}^2} \over {6\pi c}} \sin\theta dS,
\label{eq:power2}
\end{equation}

\noindent
where $\dot{v}$ is the acceleration of the particle and $\theta$ is
the angle between the line--of--sight from the observer 
to the particle and the acceleration. 
Transforming to the relativistic particle
($\dot{v}_{\parallel}'=\gamma^3 \dot{v}_{\parallel}$ which is
zero here and $\dot{v}_{\perp}'=\gamma^2 \dot{v}_{\perp}$) the 
power emitted in synchrotron radiation is given by 

\begin{equation}
P=\left< {{{\mu_\circ}q^4 \gamma^2 B^2 {v_\perp}^2}\over{6\pi cm^2}} \right>.
\label{eq:powerrel}
\end{equation}

\begin{figure}
\caption{The emission cones of synchrotron radiation.}
\label{fig:synch}
\end{figure}

Due to the relativistic speed at which the electron 
spirals through the ${\bf B}$ 
field, the radiation will be beamed into a small cone. The electron
emits radiation in 
one direction over the angle $\Delta \theta$ as seen in Figure~\ref{fig:synch}.
The frequency of rotation of the electron is given by

\begin{equation}
\omega_B={{qB}\over{\gamma mc}}
\label{eq:rotfreq}
\end{equation}

\noindent
(from Equation~\ref{eq:motion2}).
The radius of the circle shown in Figure~\ref{fig:synch} is 

\begin{equation}
a={v\over{\omega_B \sin\alpha}},
\label{eq:newa}
\end{equation}

\noindent
where the $\sin \alpha$ term is the projection of the circle into a plane
normal to the field and the angle $\Delta \theta$ is ${2\over\gamma}$. 
The distance that the particle has travelled between the 
start and finish of the pulse is 

\begin{equation}
\Delta S=a \Delta \theta={{2v}\over{\gamma \omega_B \sin\alpha}}
\label{eq:dsnew}
\end{equation}

\noindent
and the duration of the pulse is  

\begin{equation}
\Delta t={{\Delta S}\over v}={2\over{\gamma \omega_B \sin\alpha}}. 
\label{dttodsonv}
\end{equation}

\noindent
The time 
between the start and the finish of the pulse as seen by an observer is less
than $\Delta t$ by a factor $\Delta S\over c$, which is the time taken for 
the radiation to travel across $\Delta S$. So the observer sees a time
difference between pulses of 

\begin{equation}
\Delta t_\circ={2\over{\gamma \omega_B \sin\alpha}} \left( {1-{v\over
c}} \right)
\label{eq:dtcirc}
\end{equation}

\noindent
and as 

\begin{equation}
1-{v\over c} \sim {1\over{2\gamma^2}},
\label{eq:gammaapp}
\end{equation}

\noindent 
the time interval is proportional to the inverse third power of
$\gamma$. Let the inverse of the pulse delay be defined as the frequency 

\begin{equation}
\omega_c={3\over2}\gamma^3 \omega_B \sin\alpha.
\label{eq:omegacrit}
\end{equation}

\noindent
The synchrotron emission is dominated by this frequency. If the full
relativistic calculation is performed, then it can be shown that the
total power (from Equation~\ref{eq:powerrel}) is proportional to
$\omega\over{\omega_c}$. 

For an ensemble of electrons emitting synchrotron radiation,
assuming an energy distribution of the form 

\begin{equation}
N(E)dE=N_\circ E^{-p} dE,
\label{eq:engdist}
\end{equation}

\noindent
the total radiation power is given by 

\begin{equation}
P_{tot}(\omega)=N_\circ \int^{E_{max}}_{E_{min}} 
P\left({\omega\over{\omega_c}}\right)E^{-p}dE.
\label{eq:ptot}
\end{equation}

\noindent
$\omega_c$ depends on energy (through
$\gamma$ in Equation~\ref{eq:omegacrit}) and it is possible to substitute in
Equation~\ref{eq:ptot} to give 

\begin{equation}
\omega_c={{3E^2 qB\sin\alpha}\over{2m^3 c^4}}.
\label{eq:Edepwc}
\end{equation}

\noindent
Substituting for $x={\omega\over{\omega_c}}$ in Equation~\ref{eq:ptot}
the functional form of the power law is found to be

\begin{equation}
P_{tot}(\omega)\propto\omega^{-{{(p-1})\over2}}
\int^{x_2}_{x_1}P(x)x^{{(p-3)}\over2}dx. 
\label{eq:finalP}
\end{equation}

If the energy limits are sufficiently wide then the integral is a
constant and a power law in $\omega$ with spectral index
$\alpha={{p-1}\over2}$ is obtained, so that $P_{tot}(\omega)\propto
\omega^{-\alpha}$.
As the electrons radiate they lose energy and so in the full treatment
the energy distribution of the electron should be a function of time
($N(E,T)$). The electrons with the highest energy radiate faster than
the ones with
the lowest energy and so the spectrum is expected to steepen at higher
frequencies as the source grows older. In the galactic plane, where fairly
young sources are found, the spectral index for a typical synchrotron source is 
$\alpha\sim0.75$ resulting in the temperature fluctuations 
having a power law
($T_b\propto\nu^{-2.75}$). At higher galactic latitudes the spectral
index of the synchrotron steepens with energy (Lawson {\em et al}
1987) but due to the lack of full sky surveys at the GHz frequency range
it is very difficult to estimate the frequency dependence of this steepening. 

To estimate the effect of synchrotron radiation on experiments in the
GHz range low frequency maps are extrapolated using this form of power
law. However, there are inherent problems with this. Normally, the 
408~MHz (Haslam {\em et al} 1982) and 1420~MHz (Reich \& Reich 1986)
surveys are used to compute synchrotron radiation.
The 
408~MHz survey has a FWHM of $0.85^\circ$ and a 10\% error in
scale. The 1420~MHz survey has a FWHM of $0.58^\circ$ and a 5\% error
in scale. However, there are errors in the zero level (which would
not affect differencing experiments except if the maps are
consequently calibrated incorrectly) and there is also atmospheric
noise present in the final maps.
Figure~\ref{fig:sky1420} shows the high galactic latitude region of the
1420~MHz survey and it is obvious from the stripes that there are
a large amount of artefacts left in the survey, as well as a 
number of point sources, that will cause errors
in any extrapolation. The 408 MHz survey is slightly better but still 
contains some artefacts. Other than the errors inherent in the maps
themselves, there is also the extrapolation 
problem discussed above: we expect the
spectral index to steepen as we increase in frequency but we do not
know by how much. Therefore, using the low frequency surveys to
estimate the synchrotron emission at higher frequencies will lead to errors. 

\begin{figure}
\caption{The high galactic latitude region of the 1420MHz survey
showing scanning artefacts.}
\label{fig:sky1420}
\end{figure}

For the high Galactic latitude region, Bersanelli {\em et al} (1996)
have computed the power spectrum of low frequency surveys (the 408~MHz 
and the 1420~MHz). With the assumption that the major source of
radiation at these low frequencies is synchrotron, this power spectrum
should closely follow that of synchrotron. For $\ell >
100$ the power spectrum falls off roughly as the
third power of $\ell$, similar to both the dust emission and bremsstrahlung.

\section{Growing old}
\label{old}

What will happen to the Universe in the future? There are three main
possibilities. If the density of the Universe is large enough, so that $\Omega
> 1$, then it is gravitationally bound and will recollapse. The end
of this collapse is commonly referred to as the Big Crunch, but there is
a lot of debate on whether there really will be a Big Crunch or just a
whimper. If there is not enough mass to keep the Universe bound
($\Omega < 1$), then it will continue expanding forever and become
more and more sparse. If $\Omega = 1$, then there is just enough mass
to keep the Universe bound so that it will neither expand forever nor
will it collapse. At present, observations suggest that
$\Omega$ is very close to 1. It would be surprising to find $\Omega$
very close to 1 but not equal to 1 as expansion has the effect of 
moving $\Omega$ away from 1.
For example, if $\Omega=0.1$ today, then at the very early stages of
the Universe it
had to be $1-10^{-60}$. This constitutes the most accurately
determined number in physics and hence causes a problem of why
$\Omega$ was, originally, so close to unity.
Inflation gives us a solution to this
problem as it naturally predicts that $\Omega=1$ (if the cosmological
constant is zero).

\chapter{Microwave Background experiments}
\label{chap3}

In this chapter I describe the various considerations that go into
designing a Microwave Background experiment. The experiments used to
produce the data discussed in this thesis will also be
summarised.

When making measurements of the CMB fluctuations there are many
technical problems that need to be addressed. A basic CMB experiment
must be able to make high sensitivity observations while minimising
both foreground and atmospheric emissions (see Section~\ref{atmoseff}
for description of the atmospheric emission). 
An estimate of the level of Galactic free-free and synchrotron emissions can be
made by using experiments at lower frequencies (100~MHz to 10~GHz),
where these emissions are expected to be dominant, and then
extrapolating to higher frequencies. The 
Jodrell Bank 5~GHz experiment is used for this purpose. Dust emission
becomes important at frequencies higher than $\sim 200$~GHz and
so is not considered as a contaminant to the low
frequency experiments in this thesis. Between 10~GHz and 200~GHz the CMB
is expected to dominate over the Galactic foregrounds, although the
contamination from the atmosphere increases with
frequency. Therefore, a ground based experiment operating at
frequencies between 10~GHz and 100~GHz, or a space based experiment
operating at frequencies between 10~GHz and 200~GHz, 
should be used as a measure on the CMB. 
The ground based experiments chosen for this purpose are the Tenerife
experiments (10~GHz to 33~GHz). The
results from these are compared to the COBE satellite results (30~GHz
to 90~GHz) to check the consistency of the two results (they
both operate at similar angular scales)
and the primordial nature of the signal detected. 
As examples of the possible future CMB experiments both the
Planck Surveyor and MAP satellites are discussed. 

When designing any experiment systematic errors need
to be well understood so that useful constraints can be made on the data. Today
there are two types of receivers that can reach high sensitivity and
have well understood systematic errors at
the frequencies of interest to a CMB astronomer. At lower frequencies
($< 100$~GHz) High Electron Mobility Transistor (HEMT) devices are used,
whereas at higher frequencies Bolometer devices are used. The HEMT
devices work with an antenna receiver, the signal from which is then 
amplified with transistors. 
The bolometers are solid--state devices that increase in temperature
with incoming radiation. Both of these receiver
systems need to be cooled to lower the noise signal. 

\section{Atmospheric effect}
\label{atmoseff}

Another foreground that is seen with experiments looking at the
microwave background is closer to Earth than those already 
discussed. This is the atmosphere. Fluctuations in the atmosphere
are hard to distinguish from actual
extra--terrestrial fluctuations when limited frequency coverage is
available. There are
three ways to overcome this problem. The first method is to eliminate the
atmospheric effect completely.
Space missions are the best way to do this
but their main problem is cost. High altitude sites (either at the
top of a mountain or in a balloon) can reduce the atmospheric contribution, 
as can moving the experiment to a region with a stable atmosphere.
A cheaper alternative to physically moving the experiment
is to observe with the experiment for a long time. As the atmospheric effects
occur on a short time--scale, compared with the life-time of the
experiment (typically of order a few months for each data set taken with
ground based CMB experiments), and the extra-terrestrial fluctuations are
essentially constant, by
integrating over a long time the contribution from the extra--terrestrial 
fluctuations are increased with respect to the
atmospheric effects. 
Stacking together $n$ data points (taken from $n$ separate observations)
will reduce the variable atmospheric signal
with respect to the constant galactic or cosmological one by a factor of 
$\sqrt{n}$ (providing that they are independent with respect to the
atmospheric signal and any atmospheric effects on scales larger than
the beam which affect the gain have been
removed). The third way, which can also be combined with
both the first and second way, is to design the experiment to be as
insensitive as possible to 
atmospheric variations.

An obvious design
consideration is to make the telescope sensitive to frequencies at which
the atmospheric contribution is a minimum. By avoiding various bands in the
spectrum, where much emission is expected (for example water lines),
the atmosphere becomes less of a problem. Above a frequency of about
100~GHz the atmospheric effect is too large to allow useful
observations from a ground based telescope. Taken with the increasing
foreground contamination from the Galaxy at low frequencies (it is
expected that the Galaxy dominates over the CMB signal at frequencies
below 10 GHz) this reduces the observable frequencies for ground based
CMB experiments to between 10 and 100~GHz. This narrow observable range
results in the need for balloon or satellite experiments so that a
larger frequency coverage can be made to check the consistency of the
results and to check the contamination from the various foregrounds
that are expected. 

The largest atmospheric variations occur mainly on longer
time scales than the integration time of telescopes (typically
of order a few minutes), as the
variations are produced by pockets of air moving over the
telescope. If an experiment could be insensitive to these `long' term
variations then it should effectively see through the atmosphere. It
is noted that these `long' term variations are still on short
time scales compared to the lifetime of the experiment. 
An interferometer extracts a small range of Fourier coefficients from the
sky, reducing any incoherent signal (short time scale variations)
or any signal that is coherent
on large angular scales (long time scale variations), 
and so should see through the atmosphere very well. 
Similarly, an experiment that switches between two positions on
the sky relatively quickly will also reduce the long term atmospheric
variations. This technique is called beam switching. Church (1995)
modelled the atmosphere to predict the contribution that atmospheric
emission would make to interferometer and beam switching experiments
operating at GHz frequencies. Church found that the
level of atmospheric `snapshot' fluctuations expected was below 1~mK
in favourable conditions for an interferometer operating at sea level. 
After averaging over a relatively
short time scale (much shorter than the average lifetime of an
experiment), the atmospheric noise was 
well below the system noise and so negligible. A beam switching
experiment is less well able to eliminate the atmospheric emission but
operating at high altitudes, where the atmosphere is
drier, should allow good observations to be made with this type of
set up.

\section{General Observations}
\label{genobs}

Once the measurements of the CMB have been taken it is then necessary
to present data in a way that is consistent between all experiments. 
In this section I will attempt to summarise the way in
which most CMB data are presented. 

\subsection{Sky decomposition}
\label{clandal}

The usual method of presenting the results from a CMB experiment is
through the power spectrum of the spherical harmonic expansion
discussed in the previous chapter. Another value often
quoted is related to this analysis. The COBE group published their
data in terms of $Q_{rms-ps}$ which is given by 

\begin{equation}
Q_{rms-ps}=T \sqrt{{5C_2\over {4\pi}}}
\label{eq:qrms}
\end{equation}

\noindent
where $C_2$ is related to the $C_\ell$ values in the lower $\ell$ range
(where the Sachs Wolfe effect dominates) through Equation~\ref{eq:clinc2}. 
The relative values of the $C_\ell$s throughout the $\ell$ range 
depend mainly on the spectral
index, $n$, Hubble's constant, $H_\circ$, the density parameter, $\Omega_\circ$,
and, to a lesser extent, the other cosmological constants. 
For example, in the cold dark matter model of the Universe the height
of the Doppler peak depends mainly on Hubble's constant, whereas its
position depends mainly on the density parameter. 

\subsection{The effect of a beam}
\label{beameff}

One important thing to note is that the observations from a particular
experiment will not generally measure the $C_\ell$s directly. This is due
to the effect of the beam on the data. A different experiment will be
sensitive to different angular scales (and so a different range of $C_\ell$s). If
an experiment has a Gaussian beam (most experiments are not perfectly
Gaussian but can be approximated by one) then it will measure 

\begin{equation}
C_m (\beta) = {1\over{4\pi}} \sum_\ell (2\ell +1) C_\ell P_\ell (\cos\beta) \exp
[-(\ell+{1\over 2})^2 \sigma^2 ]
\label{eq:cmdep}
\end{equation}

\noindent
where $\sigma$ is the dispersion of the Gaussian (see Scaramella \&
Vittorio, 1988). This exponential
term follows through into the equation for temperature fluctuations
and must be taken into account when analysing the data if the results
are to be represented in this form.

\subsection{Sample and cosmic variance}
\label{samcosvar}

With the data presented like this and all systematic errors taken into
account there are still two errors which must be considered. In some
cases these errors will be larger than those caused by the
systematic or instrumental noise. The sample variance of the data arises from an
experiment that only measures a fraction of the sky. This is due to
the uncertainty that the part of sky measured was a `special'
part. As the distribution of the CMB is expected to follow a Gaussian
pattern there is a probability that the level of fluctuations in the 
fraction of sky that one
experiment measured is different to that in the fraction of sky
measured by another experiment. This is sample variance and is
inversely proportional to the sky area covered by the experiment. The
other error is sample variance on a cosmological
scale. The observable Universe is just one realisation of the parameters
(for example the $C_{\ell}$s in the case of a Gaussian field) taken from a
Gaussian distribution with the ensemble average described by the
underlying theory. Therefore, at large angular scales, where there are
less degrees of freedom (given by $2\ell +1$ for the Gaussian field), the
uncertainty caused by having just one realisation is greatest. There
is no way to reduce cosmic variance as this would require the study of
another observable Universe. Cosmic variance will dominate at large
angular scales while sample variance, if present, dominates at 
small angular scales.

\subsection{The likelihood function}
\label{likelihood}

If it is assumed that the CMB is described by a two--dimensional,
random Gaussian field then the properties of the fluctuations can be
described completely by their auto--correlation function $C(\beta)$
(see Equation \ref{eq:twopoint}). The data can then be used to find
the most likely variables that describe the auto--correlation function
(for example $n$ or $C_2$, and hence $Q_{RMS-PS}$, in
Equation~\ref{eq:rotsym} and \ref{eq:clinc2}). 

In the case of the Tenerife experiments care must be taken to account
for the switch beam and it is possible to write the covariance
matrix for two points $i$ and $j$ with coordinates
$(\alpha_i,\delta_i)$ and $(\alpha_j,\delta_j)$ as

\begin{displaymath}
M_{ij} = < \left\{ {\Delta T(\alpha_i,\delta_i)-{1\over 2} \left[ {\Delta
T(\alpha_i+\beta/\cos(\delta_i),\delta_i) + \Delta
T(\alpha_i-\beta/\cos(\delta_i),\delta_i) } \right] } \right\} 
\end{displaymath}

\begin{equation}
\;\;\;\;\;\;\;
\left\{ { \Delta T(\alpha_j,\delta_j)-{1\over 2} \left[ {\Delta
T(\alpha_j+\beta/\cos(\delta_j),\delta_j) + \Delta
T(\alpha_j-\beta/\cos(\delta_j),\delta_j) } \right] } \right\} >
\label{eq:covM}
\end{equation}

\noindent
where $\Delta T(\alpha_i,\delta_i)$ is the fluctuation in temperature
at point $(\alpha_i,\delta_i)$ after convolution with the Gaussian
beam pattern for a single antenna and $\beta$ is the switching angle. 
With the noise $\epsilon_i$ on point $i$ included the
total covariance matrix is given by

\begin{equation}
V_{ij} = M_{ij} + < \epsilon_i \epsilon_j >
\label{eq:fullcov}
\end{equation}

\noindent
where the noise term is non--zero only when $i=j$ if it is
uncorrelated from point to point. 

The likelihood function of this covariance matrix is defined as

\begin{equation}
L({\bf \Delta T} | p_i ) \propto {1\over {(\det {\bf V})^{1\over 2}}} \exp \left(
- {1\over 2} {\bf \Delta T}^T {\bf V}^{-1} {\bf \Delta T} \right) 
\label{eq:likelihood}
\end{equation}

\noindent
where $p_i$ are the parameters to be fitted in the covariance matrix
and ${\bf \Delta T}$ are the data. The maximum value of this function
corresponds to the most probable values of the parameters $p_i$ if we
interpret the likelihood curves in a Bayesian sense with a uniform a
priori probability distribution. As the likelihood function calculates
how probable a set of parameters are given a data set, rather than
trying to predict the parameters directly, both sample and cosmic
variance are taken into account in the analysis.

\section[Jodrell Interferometry]{The Jodrell Bank 5~GHz interferometer}
\label{jodintro}

The CMB is dominant over the Galactic foreground emissions at
frequencies higher than $\sim 10$~GHz (and below $\sim 200$~GHz). 
Therefore, to obtain a good
estimate of these foregrounds it is necessary to make observations at
lower frequencies. These observations can then be used to put
constraints on the foreground contribution to other CMB experiments. 

The 5~GHz interferometer located at Jodrell Bank, Manchester is a
twin horn, broad--band, multiplying interferometer 
(see Figure~\ref{fig:5ghz}). The horns are
corrugated and have a well defined beam with low side lobes to
minimise ground spill--over. They have an aperture diameter of 0.56~m
and a half power beam width of $8\deg$. The principle of interferometry 
ensures that uncorrelated
signals from the atmosphere are averaged down,
whilst the astronomical signals, which are correlated between antennae,
add coherently. Thus despite the relatively poor atmospheric conditions
prevalent at Jodrell Bank, an antenna temperature sensitivity of
$\sim 100$~$\mu $K per beam can be attained in one good day of observing.

\begin{figure}
\caption{The 5~GHz interferometer at Jodrell Bank.}
\label{fig:5ghz}
\end{figure}


The antennae are arranged in an East--West configuration with a
central frequency of 4.94~GHz and a variable baseline (the two
baselines to be discussed in this thesis are 1.79~m and 0.702~m). The
half power receiver bandwidth is 337 MHz.  The horns are mounted
horizontally and view the sky reflected through a plane mirror. The mirrors
can be tilted so that the centre of the beam is at a specific
declination. The rotation of the Earth sweeps the beam across the full
right ascension range every sidereal day (see Davies {\em et al}
1996a). Repeated 24 hour drift scans were taken at $2.5\deg$ intervals
in declination spanning the range $30\deg$ to $55\deg$
inclusive. Since the beam full width half maximum (FWHM) is $\sim 8\deg$ in
declination, this provides a fully sampled map of the sky.
The receivers are HEMTs
cooled to $\sim 15$~K by a closed cycle helium refrigerator. The
receiver noise temperature is $20 \pm 2$~K. 

The interferometer beam is made up of the convolution of two parts. 
Considering two infinitely thin horns separated by a distance $b$, 
as in Figure~\ref{fig:inter}, the path difference between the signals
arriving at the two horns can be shown to be $b\cos\theta $ by
simple geometry. The number of extra waves that propagate in
this path difference is given by 

\begin{equation}
a={b\cos\theta\over\lambda}
\label{eq:nowaves}
\end{equation}

\noindent
where $\lambda $ is the wavelength of the incoming radiation. Therefore, the
phase difference between the two horns is simply $2\pi a +
\gamma $. Here $\gamma $ is an artificially added phase after 
the data has been collected by the horns. Note that the path
difference compensation, shown in Figure~\ref{fig:inter}, is an
addition of phase that results in the two signals from the horns being coherent
but the $\lambda\over 4$ added phase results in the two signals being
$90\deg $ out of phase. Therefore, there are two output signals from
the interferometer which are orthogonal to each other (they will be
referred to as the cosine and sine channels).

\begin{figure}
\caption{The physical set up for an interferometer experiment.}
\label{fig:inter}
\end{figure}

The other part of the beam is a primary beam defined by the geometry
of the horns. This is usually modelled by a Gaussian beam (providing the
experiment has been well built). The beam response is 
multiplied with the infinitely thin horns' sinusoidal term, from the
correlator output, to give the
full response.

\begin{equation}
R(\alpha)=\exp\left[ -{{\theta^2}\over{2\sigma^2}} \right] \cos(2\pi a
+\gamma).
\label{eq:intres}
\end{equation}

\noindent 
However, in this case $\theta$ is needed in terms of the
Declination and Right Ascension (RA) of the source. If the beam
centre is pointed towards ($\alpha_1 ,\delta_1$), where $\delta_1$ is the
declination and $\alpha_1$ is the RA, and the source is located at 
($\alpha_2 ,\delta_2$), then the angle $\theta$ between the source and
beam axis in the Gaussian beam is given by

\begin{equation}
\theta=\cos^{-1}(\cos\delta_1 \cos\delta_2 \cos(\alpha_1 -\alpha_2)
+\sin\delta_1 \sin\delta_2).
\label{eq:alpha}
\end{equation}

\noindent
The interferometer baseline 
is changed due to projection effects and no longer depends solely on
$\theta $. As the interferometer is East--West then the projection will
only depend on the declination of the beam axis and not the
source. The path difference is now given by 

\begin{equation}
a={{b\cos\delta_1 \sin(\alpha_1 -\alpha_2)}\over\lambda}
\label{eq:anew}
\end{equation}

\noindent
and the full beam response is given by

\begin{equation}
R(\alpha_2,\delta_2)=\exp\left[ -{{\theta^2}\over{2\sigma^2}}
\right] \cos(2\pi{b\over\lambda}\cos\delta_1 \sin(\alpha_1
-\alpha_2)+\gamma)
\label{eq:fullres}
\end{equation}

\noindent
for a beam pointed at ($\alpha_1 ,\delta_1$) and $\theta$ is given in
Equation~\ref{eq:alpha}. This beam response and its
Fourier transform are plotted in Figure~\ref{fig:beamplot} for a beam
pointed at ($\alpha=0\deg,\delta=40\deg$). 

\begin{figure}
\caption{Beam response for the 5~GHz
interferometer cosine channel. 
The beam axis is at $40\deg$ Declination and $0\deg$ RA.}
\label{fig:beamplot}
\end{figure}

The complex correlator in the interferometer produces two
orthogonal sinusoidal outputs as already mentioned. In 
Equation~\ref{eq:fullres} this corresponds to two different expressions, one
with $\gamma = 0\deg$ and one with $\gamma=90\deg$. The outputs in the
two channels are binned into half degree pixels in Right Ascension.
Variations in the output levels occurring on long time scales, 
to which the experiment is not sensitive, are
removed from the data by smoothing it with a suitably large Gaussian
and subtracting the result. These baselines originate from calibration
errors (usually caused by atmospheric effects).
Data collected over a period of time at the same declination
is then stacked together to reduce the overall noise per pixel by a factor of
$\sqrt n$, where $n$ is the number of days of data collected at that pixel. 
For a more detailed
description of the experiment see Melhuish {\em et al} (1996).

\section{The Tenerife experiments}
\label{tenintro}

From the ground the best frequency window for making CMB observations
is between 10~GHz and 100~GHz. Python operates at 90~GHz and is located
at the Antarctic plateau which is both high in altitude and has a very
stable atmosphere, but most ground based experiments are confined to
frequencies between 10~GHz and 40~GHz. 
This minimises both the Galactic foregrounds
(free-free and synchrotron are dominant for frequencies less than
10~GHz) and the atmosphere (which becomes increasingly significant
with higher frequency). The main data used in this
thesis to put constraints on CMB emission are from the Tenerife
experiments which operate in this window. 

The Tenerife instruments
consist of three radiometers, each with two independent channels,
operating at frequencies of 10, 15 and 33~GHz. They are located at
2400~m altitude at the Teide Observatory in Tenerife. This area has a
smooth airflow which reduces spatial fluctuations in the water vapour
content. This ensures low
atmospheric contamination of the experiments. Approximately 70\% of
the time there is less than 3~mm of precipitable water vapour above the site.
Data taken during this period 
has very low atmospheric fluctuations and is regarded as `good' data. 
As in the case of the 5~GHz 
interferometer they are drift scanning experiments so that the
ground spill--over is constant and can be well accounted for. 
The instruments consist of two beams separated by an angle of $\theta_\circ = 
8.1\deg$ in the East-West direction. At 10~GHz
there are two experiments, one with $8.3\deg$ FWHM and another with
$4.9\deg$ FWHM, while at 15 GHz there is one with $5.2\deg$ FWHM and
at 33~GHz there is one with $5.0\deg$ FWHM. In each experiment the
difference between the two beams is calculated in real time. The
beams are then `wagged', by use of a tilting mirror (similar to the
stationary tilted mirrors in the interferometer) by one beam
separation ($8.1\deg$) so that the East beam is now in the position of
the old West beam. The new difference between the two beams is calculated in the
`wagged' position and the difference between the two beam--differences is
then calculated.
This double--difference between the two values
gives the experiments a triple beam form which is given in 
Equation~\ref{eq:tenbeam}. The beam is shown in
Figure~\ref{fig:tenbeam} and the functional form is given by

\begin{equation}
R(\alpha,\delta) = \exp{\left[-{{\theta^2}\over{2\sigma^2}}\right]} - {1\over
2}\left( \exp{\left[ -{{(\theta-\theta_\circ)^2}\over{2\sigma^2}}\right]} +
\exp{\left[ -{{(\theta-\theta_\circ)^2}\over{2\sigma^2}}\right]} \right) 
\label{eq:tenbeam}
\end{equation}

\noindent
where $\sigma$ is the beam dispersion and 
$\alpha$ is the angular separation between the source and the beam
centre. In terms of Declination and Right Ascension $\theta$ is given
by Equation~\ref{eq:alpha}.

\begin{figure}
\caption{The triple beam pattern of the Tenerife beam--switching
experiment. The beam shown is for the $4.9\deg$ FWHM primary beam.}
\label{fig:tenbeam}
\end{figure}

The mirror is tilted every $\sim 4$ seconds and data is taken from
each beam by use of a Dicke switch at 32~Hz. Over a period of 82
seconds (consisting of 8 difference pairs), the double--difference and
its standard deviation, as well as a calibration signal, are
recorded. The final data set consists of $1\deg$ bins in Right
Ascension with the average variance of the data taken from the 82
second cycles that contribute to that bin. 
The bandwidth of the receivers is 470~MHz at 10~GHz, 1.2~GHz
at 15~GHz and 3~GHz at 33~GHz. The data is recorded over a continuous period
of $4\times 24$ hours and thus a single scan contains a maximum of
four full coverages in Right Ascension, with data being taken over a
period of up to 5 calendar days. 
The full data set, once collected
(over a period of years), is then stacked together,
as discussed in the next Chapter, after the removal of long period baseline
drifts caused by slow variations in the atmosphere. This removal can
be performed in a similar way to the 5 GHz interferometer but is
actually done using the Maximum Entropy algorithm that will be
described latter (see Chapter~\ref{chap5}). The experiment is
described in more detail in Davies \et (1992). 

\section{The COBE satellite}
\label{cobeintro}

The first detection of fluctuations in the CMB was made by a NASA satellite in 1992
and the data from this experiment has always been central to CMB
work. Therefore, a comparison between this data set and that from the
Tenerife experiment will provide a very useful check on the
consistency of the two data sets. 

The NASA Cosmic Microwave Background Explorer (COBE) satellite,
launched on November $18^{th}$ 1989, had three experiments on
board. The Diffuse Infrared Background Experiment (DIRBE) measured
fluctuations in high frequency emission mainly produced by dust in our
galaxy. The Far Infrared Absolute Spectrophotometer (FIRAS) measured
the CMB spectrum between 1 cm and 100 $\mu$m. The results from FIRAS
showed the CMB to have a black body spectrum that was correct to 1
part in $10^4$, the most accurate black body known to
science. From this experiment the temperature of the CMB was shown to
be $T_{CMB} = 2.726\pm 0.010$~K at 95\% confidence (see Mather {\em et
al} 1994). The Differential Microwave Background Radiometer (DMR) maps
the fluctuations in the CMB over the full sky. The satellite is shown
in Figure~\ref{fig:COBEsat}.

\begin{figure}
\caption{The COBE satellite showing the location of the main
experiments on board.}
\label{fig:COBEsat}
\end{figure}

The DMR experiment is made up of six radiometers, two at each
frequency of 31.5~GHz, 53~GHz and 90~GHz. The frequencies were chosen
to overlap the expected minimum in Galactic foreground emission. Each
radiometer pair have two independent receivers (denoted by A and B) 
that measure the
difference in the level of CMB in beams of FWHM $7\deg$ separated by
$60\deg$ in the sky. After removal of the dipole effect (see 
Chapter~\ref{chap2}, Section~\ref{COBEdipole}) 
the sum and difference between the A and B are calculated. The sum
maps, which enhance any signal present in both A and B, 
give an estimate of the fluctuations present in the CMB while the
difference maps, which remove any consistent signal, 
gives a measure of the instrumental noise. The data
was calibrated by using a combination of an on--board calibration
source, microwave emission from the Moon and the level of the dipole
in the CMB (see Bennett {\em et al} 1992b). 
The first detection of fluctuations in the CMB was made by
COBE using data from one year of flight (Smoot {\em et al} 1992). The dipole and
the combined maps of the A+B channels 
from all three frequencies were shown in Chapter~\ref{chap2}. 
The data used in this thesis are the publicly available
processed data after four years of flight. 

\section{The Planck Surveyor satellite}
\label{cobrasintro}

The Planck Surveyor satellite is
due to be launched by ESA in 2006. The goal of Planck is to
make full--sky maps of the CMB fluctuations on all angular scales
greater than 4 arc minutes with an accuracy set by astrophysical
limits. The satellite and its proposed operation is described in
detail in Bersanelli \et (1996). Since the publication of
Bersanelli \et (1996) substantial improvements have been made to the
telescope and the latest specifications (G. Efstathiou, private
communication) will be used in the simulations performed in this
thesis. The main improvements include a change in frequencies and a
substantial improvement in noise sensitivities for the lower frequency
channels and an additional 100~GHz channel which can be used as a
cross--check between the two different types of receiver technologies
used. The satellite consists of ten frequency
channels between 30 and 900~GHz which are summarised in Table~\ref{ta:cobras}. 
The four lowest frequency channels consist of HEMT
radio receivers while the six highest frequency channels are
bolometer arrays. This difference in detector technology was chosen to
achieve the best sensitivity to the signal and accounts for the
apparent discontinuity in the table between the two 100~GHz 
channel sensitivities. 

\begin{table}
\begin{center}
\begin{tabular}{|c|c|c|c|c|c|c|c|c|c|c|} \hline
Frequency & 30 & 44 & 70 & 100 & 100 & 143 & 217 & 353 & 545 & 857 \\ 
(GHz) &&&&&&&&&& \\ \hline
Number of & 4 & 6 & 12 & 34 & 4 & 12 & 12 & 6 & 6 & 6 \\
detectors &&&&&&&&&& \\
Angular & $33\arcmin$ & $23\arcmin$ & $14\arcmin$ & $10\arcmin$ & $10.6\arcmin$ 
& $7.4\arcmin$ & $4.9\arcmin$ & $4.5\arcmin$ & $4.5\arcmin$ & $4.5\arcmin$ \\
resolution &&&&&&&&&& \\
Bandwidth & 0.2 & 0.2 & 0.2 & 0.2 & 0.37
& 0.37 & 0.37 & 0.37 & 0.37 & 0.37 \\
(${\Delta \nu}\over \nu$) &&&&&&&&&& \\
Transmission & 1.0 & 1.0 & 1.0 & 1.0 & 0.3 & 0.3 & 0.3 & 0.3 & 0.3 & 0.3 \\
${\Delta T}\over T$ sensitivity & 1.6 & 2.4 & 3.6 & 4.3 &
1.81 & 2.1 & 4.6 & 15.0 & 144.0 & 4630 \\
($10^{-6}$) &&&&&&&&&& \\
\hline
\end{tabular}
\end{center}
\caption{Summary of the Planck Surveyor satellite frequency
channels (G. Efstathiou, private communication). 
The sensitivity is for a beam pixel after 14 months of observations.}
\label{ta:cobras}
\end{table}

Figure~\ref{fig:cobras} shows an artist's impression of the
Planck Surveyor satellite. The input data presented in this thesis are
simulations of the observations that will be taken by the satellite with the
characteristics shown in the above table. These simulations were
produced by Francois Bouchet of the Institut d'Astrophysique de Paris in a
collaboration with the MRAO. 

\begin{figure}
\caption{An artists impression of the Planck Surveyor satellite. Produced
for the Bersanelli {\em et al} 1996 phase A study.}
\label{fig:cobras}
\end{figure}

\section{The MAP satellite}
\label{mapsat}

NASA are also due to launch a new satellite called the Microwave Anisotropy 
Probe (MAP) in 2000. It is intended to be a follow up of the COBE satellite
with full sky coverage but at higher resolution. It has five frequency channels 
from 20~GHz to 90~GHz which are summarised in Table~\ref{ta:map}
(improvements have also been made to the MAP satellite design but the
values quoted here are those available on the NASA MAP web
site\footnote{Since the simulations presented here were performed
the resolution of the MAP satellite has improved to 12 arc minutes}).
The expected results quoted by the MAP team in their publications
usually assume that dust emission will be negligible at 
these frequencies (this assumption was also made when the COBE data were 
analysed). It is cheaper to build than the Planck Surveyor satellite and is 
due to be launched up to six years earlier
but the resolution will not be as good. 
An introduction to the data analysis that is proposed for this satellite
can be seen in Wright, Hinshaw and Bennett (1996).

\begin{table}
\begin{center}
\begin{tabular}{|c|c|c|c|c|c|} \hline
Frequency (GHz) & 22 & 30 & 40 & 60 & 90 \\ \hline
Number of detectors & 4 & 4 & 8 & 8 & 16 \\
Angular resolution  & $54\arcmin$ & $39\arcmin$ & $32\arcmin$ & $23\arcmin$ & $17\arcmin$ \\
${\Delta T}\over T$ sensitivity ($10^{-6}$) & 12 & 12 & 12 & 12 & 12 \\
\hline
\end{tabular}
\end{center}
\caption{Summary of the MAP satellite frequency channels.}
\label{ta:map}
\end{table}

Figure~\ref{fig:map} shows an artist's impression of the MAP satellite. Again,
the input data presented in this thesis are simulations of the
observations that will be taken by MAP with the
characteristics shown in the above table. The same input actual sky
simulations as used for the 
Planck Surveyor satellite will be used, to allow a comparison of 
the results from the two experiments.

\begin{figure}
\caption{Artists impression of the MAP satellite. Produced for the MAP Internet
home pages.}
\label{fig:map}
\end{figure}

\chapter[The data]{The data from the 5 GHz interferometer and Tenerife experiments}
\label{chap4}

In this chapter I will present the raw and processed data from the
four Tenerife experiments and the Jodrell Bank interferometer. 
Brief analyses done on the raw data itself and comparison with
previous surveys are also presented. A further analysis technique that
obtains the best information from the data will be discussed in the
next chapter and results from this are presented in Chapter~\ref{chap7}.

\section[Jodrell Interferometry]{The Jodrell Bank 5GHz interferometer}
\label{jodrell}

\subsection{Pre--processing}
\label{5prepro}
 
From the complete data set a data subset
that is useful must be selected and any data that is obviously not due to
real sky fluctuations discarded. The first process is to use an automatic
program (developed by Simon Melhuish at Jodrell Bank, see Melhuish
{\em et al} 1997) that excises the
regions in the data contaminated by the Sun and the Moon. The program
looks at the time of observations and the position of the Sun and the
Moon at that time. It deletes data that is within $47\deg$ and
$17\deg$ of the Sun and Moon respectively. Any abnormal signals that
deviate from the mean of the data by more than $3\sigma$ are also deleted as
these probably arise from noise within the telescope or bad atmospheric
conditions. 

The two output channels from the interferometer should be in
quadrature - however, this is not always the case. Errors in the
correlator output lead to an output that can be $80\deg$ out of
phase rather than $90\deg$ and a difference in amplitude between the
two channels of 1\% was commonly observed. 
This is easily corrected by multiplying the
channels with a suitably chosen matrix and the amplitude and
quadrature is restored. Long term baseline drifts caused by
instrumental drifts were removed by applying a high--pass filter of
one hour ($15\deg$ Gaussian) in Right Ascension. The final results,
binned in $0.5\deg$ pixels are stacked together to produce one low noise
data scan per declination. The typical stack contains about 30 days of
data. 

\subsection{Calibration}
\label{5calib}

Before any analysis can be performed on the data it is
necessary to calibrate the level of the signal and accurately measure
the beam shape (the theoretical beam shape will not be achieved unless
the experiment is ideal). For both
of these purposes the Moon crossings in the data can be used. The
position and flux of the Moon is known very accurately and so it is
possible to
accurately model its contribution to the data. Figure~\ref{fig:calibrat5} 
shows a typical Moon crossing. In analysing the
moon care must be taken as it is slightly resolved by the
interferometer and so appears with smaller amplitude than
theoretically predicted (but this can be
easily modelled). It also moves across the sky during measurements
and so appears slightly extended. 
The Gaussian beam is fitted to the amplitude of the moon crossing and
the cosine and sine channels are fitted to the phase of the moon
crossing. 

\begin{figure}
\caption{The moon crossing of a typical scan. This is used to
calibrate the interferometer.}
\label{fig:calibrat5}
\end{figure}

To calibrate the interferometer with known sources (of which the moon
is an example) it is necessary to convert the flux units normally used
for the calibration sources to antenna temperature. The effective
area, $A_e$, of a telescope is given by 

\begin{equation}
A_e = \epsilon A_p
\label{eq:aeff}
\end{equation}

\noindent
where $A_p$ is the actual physical area of the telescope and
$\epsilon$ is the aperture efficiency. The aperture efficiency is
given by (Kraus 1982)

\begin{equation}
\epsilon={{\lambda^2 D}\over{4\pi A_p}}
\label{eq:effic}
\end{equation}

\noindent
where $\lambda$ is the wavelength of the experiment and $D$ is the
antenna directivity. In general the antenna directivity depends on the
exact beam pattern, $P_n(\alpha,\delta)$ by

\begin{equation}
D={{4\pi}\over{\int\int_{4\pi} P_n (\alpha,\delta) d\Omega}}
\label{eq:direct}
\end{equation}

\noindent
but we can estimate it by using the model Gaussian beam. From this it
is found that $\epsilon=0.72$. The antenna temperature is then related to the
flux by

\begin{equation}
T_A = {{\epsilon A_p S} \over {2k}}
\label{eq:antenT}
\end{equation}

\noindent
which gives the flux to temperature conversion of $64$~$\mu$K/Jy (this is
consistent with the result using the beam area in Equation~\ref{eq:conv2}).

The raw output units of the data from 
the interferometer are referred to as calibration
units. To convert these into antenna temperature we can now compare
with predictions of known point sources as well as the moon. The two
main point sources seen in the data are in the galactic plane. Casiopia A
has a flux density of 670~Jy and is at Declination $58.5\deg$ and Cygnus
A has a flux density of 375~Jy and is at Declination $40.6\deg$. Bright
sources in high Galactic latitudes away from the Galactic plane, 
like those shown in Table~\ref{ta:sources}, 
are easily seen in the data scans and can also be used to check the
calibration of the data. 

\begin{table}
\begin{center}
\begin{tabular}{|c|c|c|c|} \hline
Source & Flux (Jy) & Declination & Right Ascension \\ \hline
3C$84^\dagger$ & 34 & $41.3\deg$ & $49.1\deg$ \\
3C$345^\dagger$ & 6 & $39.9\deg$ & $250.3\deg$ \\
4C$39.25^\dagger$ & 9 & $39.25\deg$ & $141.0\deg$ \\
3C147 & 10.2 & $49.8\deg$ & $84.7\deg$ \\
3C286 & 7.3 & $30.8\deg$ & $202.2\deg$ \\
3C48 & 5.2 & $32.9\deg$ & $23.7\deg$ \\
\hline
\end{tabular}
\end{center}
\caption{Some of the sources used in the calibration of the
data. Sources marked with a $\dagger$ are highly variable and flux
data from a survey carried out by the University of Michigan
simultaneously with the 5 GHz survey was used in the calibration.}
\label{ta:sources}
\end{table}

The application of this calibration gives a value of $T_{cal} =
3.0 \pm 0.2$~K/CAL for the conversion from the interferometer output
units (CAL) to degrees Kelvin. 

\subsection{Data processing}
\label{jodproc}

The final stacked scans for one of the output channels in 
the wide spacing interferometer data (1.79
m baseline) are shown in Figure~\ref{fig:raw5ghz}. The principle
Galactic plane crossing at RA $\sim 20.5h$ ($308\deg$) and the weak anti--centre
crossing at RA $\sim 4h$ ($60\deg$) are clearly seen at each declination. 
Over the full 24 hour ($360\deg$) range the errors obtained per pixel and the
number of days of data at each declination are summarised in
Table~\ref{ta:5noise}.

\begin{table}
\begin{center}
\begin{tabular}{|c|c|c|} \hline
Declination & Mean noise RMS ($\mu$K) & Mean number of days \\ \hline
$30.0\deg$ & 25 & 42.0 \\
$32.5\deg$ & 53 & 15.8 \\
$35.0\deg$ & 23 & 80.8 \\
$37.5\deg$ & 33 & 30.2 \\
$40.0\deg$ & 18 & 162.8 \\
$42.5\deg$ & 36 & 53.9 \\
$45.0\deg$ & 22 & 84.5 \\
$47.5\deg$ & 69 & 8.3 \\
$50.0\deg$ & 24 & 72.9 \\
$52.5\deg$ & 32 & 33.5 \\
$55.5\deg$ & 36 & 39.0 \\
\hline
\end{tabular}
\end{center}
\caption{Noise levels and number of days in each stack for the wide
spacing interferometer setup.}
\label{ta:5noise}
\end{table}

\begin{figure}
\caption{The raw data from the cosine channel at 5 GHz for all eleven declinations. The galactic plane crossing is easily seen on this plot.}
\label{fig:raw5ghz}
\end{figure}

At this frequency
and resolution, the dominant contributor to the principal crossing in
the central sky area is the discrete 
radio source Cygnus A (S(5GHz) $\simeq 200$~Jy), with
an additional contribution from diffuse emission in the Galactic plane.
Other discrete radio sources contribute to the data and for comparison
with previous surveys these are assumed to remain at a constant flux
level with time. Radio sources at these frequencies are expected to 
be variable at the $\sim 30$\% level and consequently radio source 
variability rather than random noise is the major source of
uncertainty in the data. 
A measure of the uncertainty involved in the assumption that the
sources are constant rather than variable can be
obtained by testing the method on the discrete sources 3C345 (7.8~Jy) and 4C39
(7.6~Jy), which
are clearly detected in all three scans that surround Declination $40\deg$
where the sources are located. 
3C345 lies at RA $250.3\deg$ and Dec. $39\deg 54\arcmin
11\arcsec$ whilst 4C39 is located at RA $141.0\deg$
and Dec. $39\deg 15\arcmin 24\arcsec$. 
Figure~\ref{fig:sources} shows the data (black line) and predicted 
discrete source
emission (red line) from the Green Bank catalogue 
for the region RA$ 120\deg -270\deg$ at Dec $37.5\deg$
on an expanded scale; for clarity the $\sim 33$~$\mu$K error bars 
have not been shown.
The position of the sources 3C345 and 4C39 in the data
agree well with the prediction from the Green Bank 
catalogue and the amplitudes agree to within the expected source
variability. Any discrepancies
are consistent with the presence of noise and signal due to the
Galaxy.

\begin{figure}
\caption{Comparison between the raw data (black line) and the
predicted point source contribution (red line) from the Green Bank catalogue at
Dec. $37.5\deg$ for the cosine channel of the interferometer.}
\label{fig:sources}
\end{figure}

Figure~\ref{fig:GBraw} shows the amplitude data ($\sqrt{C^2 + S^2}$
where $C$ is the cosine data and $S$ is the sine data) for 
all eleven declinations compared with the prediction from the Green
Bank catalogue. Except for a few variable sources (notably 3C345 at
Dec. $37.5\deg$ and $35.0\deg$, and the source at the centre of
Dec. $50\deg$), this comparison shows a very good agreement between
the data and that of the catalogue. Since the Green Bank experiment is
only sensitive to sources with an angular size of less than 10.5
arcmin it is fairly safe to say that there is little Galactic emission
present in the data. The wide-spacing interferometer data can,
therefore, be used as a point source estimation for the other higher
frequency experiments at Tenerife. The narrow-spacing data is more
sensitive to larger angular scales and, taken together with the wide
spacing data, can be used as a good estimate for the Galactic
emission. Further analysis of the data is presented in Chapter~\ref{chap7}. 

\begin{figure}
\caption{Comparison between the amplitude of the data and the prediction from the GB catalogue at 4.85 GHz.}
\label{fig:GBraw}
\end{figure}

\section[The Tenerife scans]{Structure in the Tenerife switched-beam scans}
\label{tenerife}

\subsection{Pre--processing}
\label{tenpre}

Data within $50\deg$ of the Sun and $30\deg$ of the Moon were removed
from the raw data of all the Tenerife experiments with an automated
process, similar to the
interferometer. Data taken in poor atmospheric conditions (about 30\%
of all data) and any individual pixel that deviated by more
than $3\sigma$ from the average were also removed. These individual
pixels correspond to technical failures in the instrumental
system or anomalous sources (like butterflies flying into the horns). 
The data taken in poor atmospheric conditions was removed by looking
at each scan individually. If data appeared to be affected by
atmospheric fluctuations then a portion of data around the affected
region was removed. Portions of bad data were eliminated by eye because an
automated analysis would prove too complicated due to the complexity
of the data. 

A similar technique
to the pre--processing of data from the Jodrell Bank interferometer, 
by smoothing the data, could have been used to remove long term
baseline drifts caused by atmospheric offsets, but it was decided to
leave the baselines in and remove them simultaneously with the Maximum
Entropy reconstruction described in the next chapter. This method
basically finds the best astronomical signal consistent with all the
scans, subtracts this from each scan and then performs the smoothing on
the residual signal. The final
stacked results shown below are therefore after Maximum Entropy
processing. However, problems arise when the baseline variations in the raw
data are so extreme that they prevent their successful removal in the MEM
deconvolution analysis. As noted in Davies \et 1996, this problem
is exaggerated at the higher frequencies where the water vapour
emission is higher. At these higher frequencies (the 33~GHz experiment
in the case of Tenerife observations)
it is clear that the variations in baseline are, in certain cases, too 
extreme for removal 
and will therefore result in artefacts in the
final stacked scan. These artefacts result from
poor observing conditions rather than being intrinsic to the astronomy,
because such problems occur only for days with severe baselines and
appear in a randomly distributed fashion for different days.  Removal
of such data is essential if the necessary sensitivity
to detect CMB fluctuations is to be obtained. This involves the task of
examining each raw scan (the best covered declination for the $5\deg$ FWHM
data sets contains over 200
days of data) and its baseline and deciding if the data are
usable. In such cases where the data is un-salvageable, then the data
for the full $360^{\circ}$ observation are discarded. This ensures that
there is no bias introduced by selectively removing features in the
scans. After this final stage of editing, the baseline fitting must be
repeated for the full remaining data set. The MEM process will now be
able to search for a more accurate solution and will produce a new set
of more accurate baselines.
The coverages of a given declination can now be stacked together and
the process repeated until all artefacts of this type are removed. This
process is carried out for each of the Tenerife data sets discussed in
this thesis but will not be mentioned again.

\subsection{Stacking the data}
\label{tenstack}

The data set consists of $1\deg$ bins in right ascension covering a
range of declinations and taken over a large number of days (the best
scans have nearly 200 days of data). To reduce the amount of data it
is necessary to stack all the data for each point that were taken at
different days together. This is done after baseline subtraction to
avoid addition of atmospheric effects. By taking a weighted mean over
the $ns$ scans (where $ns$ is the number of days) the final data scan
is obtained. The number of days of data at each declination and
frequency for the $5\deg$ FWHM Tenerife experiment 
is summarised in Table~\ref{ta:daysten}.
If the data in each $1\deg$ bin is given by $y_{ir}$,
where $i$ is the bin and $r$ is the day, and the error on $y_{ir}$ is given
by $\sigma_{ir}$ then the final stacked data is given by

\begin{equation}
Y_{i} = {{\sum_{r=1}^{ns} w_{ir} \left( y_{ir} - b_{ir} \right)} \over
{\sum_{r=1}^{ns} w_{ir} }}
\label{eq:finaly}
\end{equation}

\noindent
where $b_{ir}$ represents any long term baselines that have to be
subtracted from the data before stacking and 
the weighting factor $w_{ir}$ is given by

\begin{equation}
w_{ir} = {1\over \sigma_{ir}^2 }.
\label{eq:weight}
\end{equation}

\noindent
The error on the final stack is given by the scatter over the days
contributing to each point

\begin{equation}
\sigma_i^2 = \left( {{\sum_{r=1}^{ns} w_{ir} \left[ (y_{ir} - b_{ir}) - Y_{i}
\right]^2 }\over{\sum_{r=1}^{ns} w_{ir} (ns-1)}} \right).
\label{eq:finalsig}
\end{equation} 

\begin{table}
\begin{center}
\begin{tabular}{|c|c|c|c|} \hline
Declination & 10~GHz & 15~GHz & 33~GHz \\ \hline
30.0 & - & 104 & - \\
32.5 & 85 & 74 & - \\
35.0 & 82 & 191 & 40 \\
37.5 & 129 & 103 & - \\
40.0 & 134 & 134 & 90 \\
42.5 & 50 & 85 & - \\
45.0 & 51 & 67 & - \\  
\hline
\end{tabular}
\end{center}
\caption{Number of independent measurements for the $5\deg$ FWHM
 Tenerife experiments.}
\label{ta:daysten}
\end{table}

\subsection{Calibration}
\label{tencal}

The data was calibrated using a continuous online noise injection
diode so that the data amplitude response remains constant throughout
the day's observing. The output is then in units of CAL (the
calibration level). Moon observations and the Galactic plane crossing
were then used to calibrate the signal and convert the output into degrees
Kelvin. Previous studies of the moon (e.g. Gorenstein \& Smoot 1981)
were used to calculate the expected level in each scan. Other,
large amplitude, point sources can also be used as a check on the
calibration of the data. The different calibration methods are
consistent (Davies \et 1992) and it is believed that the final
calibration of each instrument is accurate to 5\%.

\subsection{Error bar enhancement}
\label{errenh}

Each of the Tenerife experiments contain two independent
receivers. These operate simultaneously and look at the same region of
sky so as to have a check on the consistency of each
receiver (and to reduce the noise by a factor of $\sqrt{2}$). 
However, the two receivers will also be looking through the
same atmospheric signals and so the noise on the two channels will be
correlated. This noise is equivalent to a Gaussian noise common to
both channels with a coherence time smaller than the binning time, the
net effect of which is an enhancement of the error bars (see Gutierrez
1997 or Davies \et 1996a). 

The correlated noise is seen in the two channels in all three
frequency experiments. The maximum effect is at 33~GHz where an
atmospheric signal of $\sim 30\mu$K is seen (Davies \et 1996a). The
level of the signal can be calculated by looking at the correlation
and cross-correlation between the channels (Gutierrez
1997). Gutierrez (1997) showed that the correlation between the
channels only existed on timescales less than 4 minutes (the bin size)
and so no significant correlation is found between adjacent positions
in RA. The enhancement required for each of the declinations and
frequencies is summarised in Table~\ref{ta:noisenh} (Gutierrez private
communication). 

\begin{table}
\begin{center}
\begin{tabular}{|c|c|c|c|} \hline
Declination & 10~GHz & 15~GHz & 33~GHz \\ \hline
$30.0\deg$ & -     & 1.045 & - \\
$32.5\deg$ & 1.023 & 1.067 & - \\
$35.0\deg$ & 1.092 & 1.039 & - \\
$37.5\deg$ & 1.057 & 1.055 & - \\
$40.0\deg$ & 1.042 & 1.042 & 1.166 \\
$42.5\deg$ & 1.068 & 1.070 & - \\
$45.0\deg$ & 1.038 & 1.040 & - \\
\hline 
\end{tabular}
\end{center}
\caption{Noise enhancement for each declination and frequency of the
Tenerife experiment. This extra multiplication factor should be
included to account for atmospheric correlations between channels.}
\label{ta:noisenh}
\end{table}

\subsection[$8.3\deg$ FWHM experiment]{The $8.3\deg$ FWHM 10~GHz experiment}
\label{8degsec}

Between 1984 and 1985 the Tenerife experiment consisted of a
double--switching telescope with $8.3\deg$
FWHM at 10.45~GHz. Table~\ref{ta:8deg} summarises the observations
taken during this period. One third of the total sky was covered but the
sensitivity reached was not very high as the integration time at each
declination was limited. Figure~\ref{fig:typical15} shows the full
data set for the $46.6\deg$ Declination data. The long term baseline
drifts are still apparent at this stage. After the baseline has been
removed using the maximum entropy technique the data was stacked
together and compared with the expected signals (see
Chapter~\ref{chap7}). At this frequency and angular scale the
experiment is more sensitive to Galactic emission than to the CMB
radiation and so this data will be used to put constraints on the
Galactic contamination to the $5\deg$ Tenerife experiments. 

\begin{table}
\begin{center}
\begin{tabular}{|c|c|c|c|} \hline
Declination & Number & Mean scan & RMS length \\ 
 & of scans & length (hours) & (hours) \\ \hline
$+46.6^{\circ}$ & 15 & 14.1 & 3.1 \\
$+42.6^{\circ}$ & 16 & 22.0 & 9.7 \\
$+39.4^{\circ}$ & 42 & 14.1 & 6.3 \\
$+37.2^{\circ}$ & 18 & 14.6 & 7.6 \\
$+27.2^{\circ}$ & 17 & 11.3 & 3.9 \\
$+17.5^{\circ}$ & 16 & 21.1 & 10.3 \\
$+07.3^{\circ}$ & 13 & 9.7 & 3.3 \\
$+01.1^{\circ}$ & 52 & 15.8 & 11.1 \\
$-02.4^{\circ}$ & 6 & 10.7 & 4.9 \\
$-17.3^{\circ}$ & 20 & 8.4 & 4.2 \\
\hline
\end{tabular}
\end{center}
\caption{Observations with the $8.3^{\circ}$ FWHM 10.4 GHz
 experiment.}
\label{ta:8deg}
\end{table}

\begin{figure}
\caption{The 15 scans obtained at Dec $=46.6\dg $
displayed as a function of right ascension. Each plot shows
the second difference in mK after binning into $1\dg $
bins. A running mean has been subtracted from each scan.
Long scans are displayed modulo $360\dg $.}
\label{fig:typical15}
\end{figure}

\subsubsection{\bf The Likelihood results}
\label{stat8deg}

The statistical properties of the signals present in the data have
been analysed using the likelihood function and a Bayesian analysis.
This method has been widely
used in the past by the Tenerife group (see {\it e.g.} Davies \et 1987) and
incorporates all the relevant parameters of the observations:
experimental configuration, sampling, correlation between measurements,
etc. The analysis assumes that both the noise and the signal follow a
Gaussian distribution fully determined by their respective
auto-correlation function. 
The source of dominant noise in the 
data is thermal noise in the receivers which is independent in each
data-point (Davies \et 1996) and therefore it only contributes to
the terms in the diagonal of the auto-correlation matrix. For this analysis the
section of data away from the Galactic plane has been selected
otherwise the local contribution would dominate by orders of magnitude
over the CMB fluctuations. 
Also, the analysis has been restricted
to data in which there is a minimum number of 10 independent
measurements for the full RA range (Dec. $7.3\deg$ does not have
enough data) and to data for which we have a point source prediction
(Dec. $-17.3\deg$ is not covered by the Green Bank survey).
This region represents approximately 3000 square degrees on
the sky. Table~\ref{ta:tab2} presents the sensitivity per
beam in the RA range used
in this analysis. Also column 4 gives the mean number of
independent measurements which contribute to each point. This
statistical analysis has been performed directly on the scan data, and
not on the MEM deconvolved sky map produced during the baseline
subtraction process. Thus, for this section, any effects of using a
MEM approach are restricted to the baselines subtracted from the raw
data, which will not contain, or affect, any of the astronomical
information to which the likelihood analysis is sensitive. 
 
\begin{table}
\centering
\begin{tabular}{|ccccc|} \hline
Dec. & RA  & $\sigma$ ($\mu$K)& Indep. &
 $(\frac{\ell (\ell +1)}{2\pi}C_\ell )^{1/2}\, (10^{-5}$) \\
 \hline
$+46.6\deg$ & $161\deg-250\deg$ & 116 & 15 & $\le 8.5$  \\
$+42.6\deg$ & $161\deg-250\deg$ & 117 & 14 & $\le 10.3$\\
$+39.4\deg$ & $176\deg-250\deg$ &  81 & 42 & $1.8^{+2.3}_{-2.0}$ \\
$+37.2\deg$ & $161\deg-250\deg$ & 113 & 17 & $5.7^{+3.2}_{-2.9}$ \\
$+27.2\deg$ & $161\deg-240\deg$ & 139 & 11 & $4.5^{+3.0}_{-3.9}$ \\
$+17.5\deg$ & $171\deg-240\deg$ & 144 & 12 & $4.3^{+3.0}_{-4.4}$ \\
$+1.1\deg$ &  $171\deg-230\deg$ &  96 & 58 & $5.0^{+3.3}_{-3.0}$\\ \hline
\end{tabular}
\caption{Statistics of the data used in the analysis. 95\% confidence
limits are shown.}
\label{ta:tab2}
\end{table}
 
Two different likelihood analyses were made: the first considers the data of
each declination independently, and the
second considers the full two-dimensional data set. 
A likelihood analysis was performed assuming a
Harrison-Zel'dovich spectrum (so that the covariance matrix is given
by Equation \ref{eq:clinc2} with $n$=1), thus the parameter fitted for was
$Q_{RMS-PS}$. Since the $\ell$ range sampled is small, Equation
\ref{eq:qrms} can be used to give an estimate for $C_\ell$ and this is what is shown
in the Table \ref{ta:tab2}. The experiment has a peak sensitivity to an $\ell$ of
$14^{+7}_{-6}$. The fifth column of
Table \ref{ta:tab2} gives, for the one-dimensional analysis, the amplitude of the
signal detected with the one-sigma confidence level. The confidence
limits on these signals were found by integration over a uniform prior
for the likelihood function.
These analyses ignore correlations
between measurements at adjacent declinations. Therefore a full
likelihood analysis, taking this correlation into account,
should constrain the signal more efficiently. It should be noted that
the two dimensional analysis assumes that the signal has the same
origin over the full sky coverage but this may not be the case because
of the differing levels of Galactic signal between declinations and
across the RA range. In
Figure \ref{fig:likel} the likelihood function resulting from this
analysis is shown. It shows a clear, well defined peak at
$(\frac{\ell (\ell +1)}{2\pi}C_\ell )^{1/2}\, (10^{-5}) = 2.6^{+1.3}_{-1.6}$
(95 \% confidence level). This value would
correspond to a value of $Q_{RMS-PS} = 45.2^{+23.8}_{-27.2}$~$\mu$K.
The results are compatible with the constraints on the signal
in each declination considered separately but it is clear that the two
dimensional likelihood analysis improves the  constraints on the amplitude
of the astronomical signal.
 
\begin{figure}
\caption{The likelihood function from the analysis of the full data set. There
is a clearly defined peak at
$(\frac{\ell (\ell +1)}{2\pi}C_\ell )^{1/2}\, (10^{-5}) = {2.6^{+1.3}_{-1.6}}$
($95\% $ confidence level).
}
\label{fig:likel}
\end{figure}
 
A comparison between the results obtained here and the amplitude of the CMB
structure found in Hancock \et (1994) at higher frequencies can be
made. They found $Q_{RMS-PS}
\sim 21$~$\mu$K in an $5\dg$ FWHM switched beam and taking into account
the extra dilution a slightly lower level is expected in an $8\dg$ FWHM
switched beam, assuming a $n=1$ power spectrum.
It is seen that the majority of the
signal in the 10 GHz, FWHM=$8\deg$ data is most likely due to Galactic
sources. Assuming that the majority of the signal found here
is Galactic and using a spatial spectrum of $C_\ell \propto \ell^{-3}$ (estimated
from the Haslam \et 1982 maps) to predict the
galactic contamination in a $5\dg$ FWHM beam at 10 GHz, then using
a full likelihood analysis, it is found
that an {\em rms} signal of 
$\Delta T_{rms} = 55^{+32}_{-26}$~$\mu$K is expected. It
should be noted that this is an upper limit on the Galactic contribution to the
$5\dg$ data as the variability of the sources has been ignored when
the subtraction was performed (this results in a residual signal from
the point sources in the data during the likelihood analysis) and the
analysis also includes regions where the Galactic signal is expected
to be higher (for example the North Polar Spur). The $5\deg$ FWHM
Tenerife scans are centred on Dec. $40\deg$ and it can be seen from
Table \ref{ta:tab2} that this is the region with the lowest Galactic
contamination. The results
reported in Gutierrez {\em et al} (1997), for the $5\deg$ FWHM, 10~GHz
Tenerife experiment, show that the signal found was $Q_{RMS-PS} <
33.8$~$\mu$K (corresponding to a signal of $\Delta T < 53$~$\mu$K) 
which is consistent with the prediction here (also taking into
account the more significant contribution from the CMB at $5\deg$).
This comparison allows a restriction on the {\em maximum}
Galactic contribution to the signal found in Hancock \et 1994 to be
$\Delta T_{rms} \sim 18 - 23$~$\mu$K at 15~GHz and $\Delta T_{rms} \sim 2
- 4$~$\mu$K at 33~GHz depending on whether the contamination is
dominated by synchrotron or free-free emission.

\subsection[$5\deg$ experiments]{The $5\deg$ FWHM Tenerife experiments}
\label{5degex}

The aim of the $5\deg$ FWHM experiments is to produce consistent maps
at three operating frequencies (10.45, 14.90 and 33.00~GHz) of the CMB
fluctuations and using
the frequency coverage to put constraints on the levels of galactic
foregrounds (namely synchrotron and free--free which are expected to
be the most important at these frequencies). The experiments cover a range
between Declinations $27.5\deg$ and $45\deg$. The 10.45~GHz,
$4.9\deg$ FWHM experiment, the 14.9~GHz, $5.2\deg$ FWHM experiment and
the 33~GHz, $5.0\deg$ FWHM experiment have been taking data since 1985,
1990 and 1992 respectively. Table~\ref{ta:indepmeas} summarises the
results up to July 1996 (the data presented in this thesis is taken up
to August 1997 which includes a new 15~GHz declination scan at
$27.5\deg$). It is seen that the full area is not covered
by all three experiments but they are continually taking new data to
build up a well sampled, high sensitivity data set. Table~\ref{ta:noise} 
shows the noise levels per $1\deg$ bin achieved in this data set. 

\begin{table}
\centering
\begin{tabular}{|c|c|c|c|} \hline
Declination & 10~GHz & 15~GHz & 33~GHz \\ \hline
$30.0\deg$ & - & 104 & - \\
$32.5\deg$ & 85 & 74 & 28 \\ 
$35.0\deg$ & 82 & 191 & 40 \\
$37.5\deg$ & 129 & 103 & 36 \\
$40.0\deg$ & 134 & 134 & 90 \\
$42.5\deg$ & 50 & 85 & 19 \\
$45.0\deg$ & 51 & 67 & - \\ \hline
\end{tabular}
\caption{The number of independent measurements in the RA range
 $161\deg - 250\deg$ in each of the Tenerife data sets.}
\label{ta:indepmeas}
\end{table}

\begin{table}
\centering
\begin{tabular}{|c|c|c|c|} \hline
Declination & 10~GHz & 15~GHz & 33~GHz \\ \hline
$30.0\deg$ & - & 20 & - \\
$32.5\deg$ & 54 & 24 & 42 \\ 
$35.0\deg$ & 52 & 17 & 33 \\
$37.5\deg$ & 41 & 19 & 33 \\
$40.0\deg$ & 44 & 19 & 21 \\
$42.5\deg$ & 63 & 22 & 50 \\
$45.0\deg$ & 80 & 26 & - \\ \hline
\end{tabular}
\caption{The noise per beam in $\mu$K for each of the Tenerife final
stacked scans.}
\label{ta:noise}
\end{table}

\subsubsection[Dec. $35\deg$]{\bf Dec. $35\deg$ at 10 and 15~GHz}
\label{dec35}

The data set consisting of the Declination $35\deg$ scan at 10~GHz and
15~GHz has been analysed separately to the remainder of the data as an
example of the analysis techniques possible on the individual scans
(see Gutierrez \et 1997). The declination $40\deg$ was analysed
previously (see Hancock \et 1994) but new data has been added since
then. This
region was chosen as it contains one of the best coverage of high
Galactic latitudes (see Table~\ref{ta:indepmeas}). 
The remaining data (including the $35\deg$ data)
will be analysed with Maximum entropy in a full two--dimensional
analysis in Chapter~\ref{chap7}. 
The analysis is concentrated in the region RA=$161\deg-250\deg$
which is at Galactic latitude $b\gta 40\deg$ where the CMB signal is
not buried beneath foreground contributions. Figure~\ref{fig:dec35_10_15} 
shows the stacked data in the region of
interest. First the possible sources of non-CMB foregrounds in the data
have been considered; the
contribution due to known point-sources has been calculated using the
Kuhr \et (1981) catalogue and the Green Bank sky survey (Condon \&
Broderick 1986); this was complemented with the Michigan and Metsahovi
monitoring programme. In this band of the sky the most intense
radio-source at high Galactic latitude is 1611+34
(RA=16$^h$11$^m$48$^s$, Dec.=$34\deg20^\prime20^{\prime \prime}$) 
with a flux density $\sim
2$~Jy at 10 and 15~GHz. After convolution with the triple beam of the
experiment this represents peak amplitudes of $\sim$60 and $\sim
$30~$\mu$K at 10 and 15~GHz respectively, which are in the limit of the
detection of each data-set. The predicted {\em rms} of the point-source
contribution from the remaining unresolved sources, using the above
surveys, along the scan is 26 and 11 $\mu$K at 10 and 15~GHz.
Therefore, even considering uncertainties in the prediction of this
order, this can only be responsible for a small fraction of the
detected signals as will be discussed below. In the following analysis
this point source contribution has been subtracted from the data scans.
The contamination by
a foreground of unresolved radio-sources has been predicted to have an {\em rms}
$\lta 30$ and $\lta $15~$\mu$K at 10 and 15~GHz respectively
(Franceschini \et 1989) consistent with the findings here. 
In previous papers (Davies, Watson and Guti\'errez 1996) the unreliability of the
predictions of the diffuse Galactic foreground using the low frequency
surveys at 408~MHz (Haslam \et 1982) and 1420~MHz (Reich \& Reich 1988)
has been demonstrated.
It is possible to estimate the magnitude of such a
contribution from the $8\deg$ FWHM Tenerife experiment. At 15~GHz it
was shown that the maximum contribution from the Galactic foregrounds
was $\sim 20$~$\mu$K. 

\begin{figure}
\caption{The Dec $35\deg$ data at 10 GHz and 15 GHz in the region
RA=$161\deg -250\deg$.}
\label{fig:dec35_10_15}
\end{figure}

\subsubsection{\bf The Likelihood results}
\label{5deglike}

The likelihood analysis has been applied to the data at 10~GHz and
15~GHz in the range
RA=$161\deg-250\deg$. Assuming a Harrison-Zel'dovich spectrum for the
primordial fluctuations the likelihood curve for the 15~GHz data shows
a clear peak with a likelihood $L/L_0=5.5\times $10$^4$ in a signal
of $\Delta T_{RMS}\sim 32$ $\mu$K which 
corresponds to an expected power spectrum normalisation
$Q_{RMS-PS}=20.1$ $\mu$K at an $\ell$ of $18^{+9}_{-7}$. 
The maximum Galactic contribution to this
data set was shown to be $\Delta T_{rms}=23\mu$K (from the analysis of
the $8\deg$ data) which cannot account for the signal seen here. 
Analysing the curve in a Bayesian sense
with a uniform prior, a value of $Q_{RMS-PS}=20.1^{+7.1}_{-5.4}$~$\mu$K 
(68 \% confidence limit)    
was obtained. These results do not depend greatly on the precise region
analysed; for instance analysing the sections RA=$161\deg-230\deg$ or
RA=$181\deg-250\deg$, the results show $Q_{RMS-PS}=20.3^{+8.7}_{-7.3}$
$\mu$K and $Q_{RMS-PS}=20.0^{+8.0}_{-6.0}$~$\mu$K respectively. As a
consequence of the noisy character of the 10~GHz data, there is no
evidence of signal in the likelihood analysis of the data at this
frequency; instead a limit on $Q_{RMS-PS}\le 33.8$~$\mu$K
(95 \% confidence limit) is obtained which is 
compatible with the amplitude of the signal
detected at 15 GHz. The signal at 15~GHz is only slightly larger than
the level of the signal present in the COBE DMR data ($Q_{RMS-PS}=18\pm
1.5$~$\mu$K, Bennett \et 1996) and this indicates that, if the CMB
fluctuations do indeed correspond to the standard COBE DMR normalised
Harrison-Zel'dovich model, the possible foreground contamination would
contribute with $\Delta T_{RMS}\lta 16$ $\mu$K at 15~GHz.

A joint likelihood analysis (Guti\'errez \et 1995) was run between
the data presented here at 15 GHz and those at Dec.$=40\deg$ presented
in Hancock \et 1994. The angular separation between both strips is
similar to the beam size of the individual antennas so a partial correlation 
between the Dec.=$+35\deg$ and Dec.=$+40\deg$ scans is expected.
The region between RA=$161\deg-230\deg$ at
Dec.$=+40\deg$ was chosen so as to exclude the variable radio-source 3C 345
(RA$\sim 250\deg$) that was clearly detected in the data. 
Analyses of models with a power spectrum $P\propto k^n$ for
the primordial fluctuations (tilted models) were performed independently. 
Figure \ref{fig:amp_n} shows the amplitude
of the signal for each model and the one-sigma level bounds in the
spectral index versus $Q_{RMS-PS}$ plane. The relation between the
expected quadrupole and the spectral index can be parametrised by
$Q_{RMS-PS}=25.8^{+8.0}_{-6.5}\exp\{-(n-0.81)\}$ $\mu$K. In the case of
a flat spectrum ($n=1$) $Q_{RMS-PS}=21.0^{+6.5}_{-5.5}$ $\mu$K was
obtained in agreement with the results obtained analysing 
the strips at Dec.=$+35\deg$ (see above) and
Dec.=$+40\deg$ ($Q_{RMS-PS}=22^{+10}_{-6}$ $\mu$K, Hancock \et
1997) independently.

\begin{figure}
\caption{The constraints on $Q_{RMS-PS}$ given by the Dec. $35\deg$
data.}
\label{fig:amp_n}
\end{figure}

\subsubsection{\bf The full data set}
\label{fulldata}

The remaining data taken with the $5\deg$ FWHM experiments will be analysed
simultaneously to give a better constraint on underlying signals. As the
sampling rate of the Tenerife declinations scans is less than a beam
width, the adjacent scans will have strong correlations which can be
used to put a stronger constraint on the underlying sky signal. As an
example of the quality of the 
new data taken recently, the declination $27.5\deg$
data is shown in Figure~\ref{fig:dec27}. The Galactic plane
is clearly seen at RA $\sim 300\deg$ whereas the Galactic anti-centre
has not been observed. At the positions where more data was
take (bottom figure) it is easily seen that the errors on the final
stacked scan are lowest (middle figure) as expected. Due to the level
of noise in this scan at present no obvious point sources are visible
but more data is due to be taken over the next year to increase the
sensitivity. 

Figures~\ref{fig:tharr10} to
\ref{fig:tharr33} show the stacked scans (with errors) for each of the
declinations at each of the frequencies. The stacking was done after
baseline removal with MEM. The accuracy of the data is easily seen
through the observation of point sources. For example, the two main point
sources shown in this region of the scans are 3C345 at RA $250.3\deg$
and Dec. $39\deg 54\arcmin 11\arcsec$ and 4C39 at RA $141.0\deg$
and Dec. $39\deg 15\arcmin 24\arcsec$. These are both clearly seen
(4C39 is on the very left of the scans) in
the 10~GHz and 15~GHz scans. Unfortunately due to the increased level
of atmospheric contribution to the scans at 33~GHz it has been
impossible to complete the analysis of the declinations at this
frequency in time for this thesis. Therefore, only the old declination
$40\deg$ data (which is the highest sensitivity data at this
frequency and was presented in Hancock 1994) 
is included here. It is expected that the final set of 33~GHz data
will be available by Easter of 1998.

\begin{figure}
\caption{The declination $27.5\deg$ data at 15~GHz. The top figure
shows the stacked data. The principal Galactic plane crossing is
clearly visible at RA $\sim 300\deg$. The middle figure shows the
typical errors across the scan. It is seen that in the best region a
sensitivity of $\sim 60\mu$K has been achieved. The bottom scan shows
the number of independent data points that have gone into the scan. It
is seen that no data was taken of the Galactic anti-centre. }
\label{fig:dec27}
\end{figure}

\begin{figure}
\caption{The stacked scans at 10~GHz for the Tenerife $5\deg$ FWHM
experiments.}
\label{fig:tharr10}
\end{figure}

\begin{figure}
\caption{The stacked scans at 15~GHz for the Tenerife $5\deg$ FWHM
experiments.}
\label{fig:tharr15}
\end{figure}

\begin{figure}
\caption{The stacked scan at 33~GHz for the Tenerife $5\deg$ FWHM
experiments.}
\label{fig:tharr33}
\end{figure}

\subsubsection{\bf The Likelihood results}
\label{5deglikefull}

The likelihood analysis that was applied to the Dec. $35\deg$ above
was then applied to the full data set. It has already been seen that
the 15~GHz and 33~GHz data sets should be relatively free from Galactic
contamination and so the results from the likelihood analysis to these
two data sets should give very good constraints on the level of the
CMB. The 10~GHz data set is more sensitive to Galactic free-free and
synchrotron and so a higher level of fluctuations is expected in this
data set (although if the CMB and Galactic fluctuations are exactly in
anti-phase then a lower level would be achieved but this is highly
unlikely to be the case). 

The results from the likelihood analyses are shown in
Table~\ref{ta:awjres} and Figures~\ref{fig:10ghzinlike} and
\ref{fig:15ghzinlike}. As is seen the level of all of the 15~GHz and 33~GHz
data sets are $\sim 20\mu$K which is consistent with a CMB origin for
the fluctuations. The higher level in the 10~GHz channel is also
seen. Unfortunately due to the noisy level of the 10~GHz channel only
95\% upper limits were possible on most of the declination data
sets. 

It is possible to analyse the data sets simultaneously at each
frequency as each declination is correlated with its neighbours (the
beam is wider than the declination strip separation). A two dimensional
likelihood for the 10~GHz data set (consisting of the 6 declinations)
and the 15~GHz data set (consisting of the 8 declinations) was performed.
As in the Dec. $35\deg$ case the likelihood curve for the
full two dimensional 15~GHz data set is a ridge in the 
$n$/$Q_{RMS-PS}$ plot and so no
constraints on $n$ are possible with the Tenerife data alone. However,
for $n=1$ the full two dimensional analysis gives $Q_{RMS-PS} =
22^{+5}_{-3} \mu$K. 
The COBE results showed a level of $18\pm 2\mu$K for the
CMB anisotropy (Bennett \et 1996) and this is consistent with the data
presented here for a Harrison-Zel'dovich spectrum (scale invariant). A
likelihood analysis of the two data sets together can also be
performed to put constraints on the slope of the primordial
spectrum and this is done below. 

\begin{table}
\begin{center}
\begin{tabular}{|c|c|c|c|} \hline
Dec. & 10~GHz & 15~GHz & 33~GHz \\ \hline
$27.5\deg$ & - & $25^{+14}_{-10}$ & - \\
$30.0\deg$ & - & $24^{+9}_{-7}$ & - \\
$32.5\deg$ & $\le 57$ & $22^{+14}_{-12}$ & - \\
$35.0\deg$ & $\le 38$ & $20^{+9}_{-7}$ & - \\
$37.5\deg$ & $\le 33$ & $15^{+7}_{-7}$ & - \\
$40.0\deg$ & $\le 33$ & $20^{+13}_{-8}$ & $24^{+10}_{-8}$ \\
$42.5\deg$ & $\le 47$ & $19^{+11}_{-7}$ & - \\
$45.0\deg$ & $46^{+28}_{-22}$ & $\le 26$ & - \\
\hline
\end{tabular}
\end{center}
\caption{Results from the one dimensional likelihood analysis of the
Tenerife $5\deg$ FWHM experiments in the RA range $160\deg$ to
$230\deg$. All are for $Q_{RMS-PS}$ in $\mu$K and show the 68\%
confidence limits.}
\label{ta:awjres}
\end{table}

\begin{figure}
\caption{The likelihood results for the individual declination scans
at 10~GHz for the Tenerife $5\deg$ FWHM experiment.}
\label{fig:10ghzinlike}
\end{figure}

\begin{figure}
\caption{The likelihood results for the individual declination scans
at 15~GHz for the Tenerife $5\deg$ FWHM experiment.}
\label{fig:15ghzinlike}
\end{figure}

A separate analysis of the same data set was made in parallel by
Carlos Gutierrez (Gutierrez, private communication) and the results
from this analysis are shown in Table~\ref{ta:gutres}. The main
difference between the two analysis was a different point source
catalogue was used for the predictions but the results are consistent
with those presented here. The two point source catalogues used were taken at
different times and so a difference is expected between the two predictions
due to the variability of some of the sources. 
To eliminate even this small discrepancy a 
new point source data set has become available. This includes observations
taken of each of the point sources within the Tenerife field over the 
last ten years and so a simultaneous fit for the point source flux is 
possible and should allow better subtraction (see Figure~\ref{fig:3c345} for
an example of the continuous monitoring of one of the variable sources). As 
this data has only just become available there has not been time to 
incorporate the analysis into the results presented here so it should be 
noted that these are preliminary results.  

\begin{table}
\begin{center}
\begin{tabular}{|c|c|c|c|} \hline
Dec. & 10~GHz & 15~GHz & 33~GHz \\ \hline
$27.5\deg$ & - & - & - \\
$30.0\deg$ & - & $22^{+10}_{-6}$ & - \\
$32.5\deg$ & $33^{+18}_{-13}$ & $25^{+12}_{-11}$ & - \\
$35.0\deg$ & $\le 33$ & $20^{+8}_{-6}$ & - \\
$37.5\deg$ & $35^{+16}_{-13}$ & $19^{+8}_{-7}$ & - \\
$40.0\deg$ & $\le 31$ & $23^{+11}_{-9}$ & $23^{+11}_{-8}$ \\
$42.5\deg$ & $\le 44$ & $29^{+12}_{-10}$ & - \\
$45.0\deg$ & $59^{+22}_{-17}$ & $21^{+14}_{-10}$ & - \\
\hline
\end{tabular}
\end{center}
\caption{Results from the one dimensional likelihood analysis of the
Tenerife $5\deg$ FWHM experiments in the RA range $160\deg$ to
$230\deg$, done in parallel to those presented
in this thesis (Gutierrez, private communication). All are for
$Q_{RMS-PS}$ in $\mu$K and show the 68\% confidence limits.}
\label{ta:gutres}
\end{table}

In terms of the {\em rms} temperature fluctuations the results from the full
two dimensional likelihood analysis of the Tenerife 10~GHz and 15~GHz
$5\deg$ FWHM data set is shown in Figure~\ref{fig:2dimdt}. The peak of
the likelihood for the 10~GHz data gives $\Delta
T_{rms}=53^{+13}_{-13}$~$\mu$K and that for the 15~GHz gives $\Delta
T_{rms}=39^{+8}_{-7}$~$\mu$K. When comparing this with the prediction
from the $8\deg$ FWHM Tenerife experiment for the upper limit to the
Galactic contamination ($\Delta T_{rms} = 55^{+32}_{-26}$~$\mu$K at
10~GHz and $< 23\mu$K at 15~GHz) it is seen that the majority of the
signal at 10~GHz is likely to be Galactic in origin whereas the
majority of the 15~GHz signal must come from other sources
(namely the CMB). A full analysis of the Tenerife data would include a
simultaneous analysis of all data from the full frequency range of the
Tenerife experiments as well as data from other experiments more
sensitive to the foregrounds (e.g. the Galactic surveys at lower
frequencies or the DIRBE dust map at a higher frequency). This is work
that is in progress utilising the new multifrequency Maximum Entropy
Method (see Chapter~\ref{chap5}) but for present purposes it is useful
to make the assumption that the 15~GHz signal is CMB alone (which has
been shown to be a good assumption).

\begin{figure}
\caption{The likelihood results for the full data sets at 10~GHz and
15~GHz for a Gaussian autocorrelation function. This gives
constraints on the level of fluctuations in temperature.}
\label{fig:2dimdt}
\end{figure}

Assuming that the majority of the signal in the
15~GHz data set is CMB and not Galactic in origin it is possible to
use this data set as a constraint on models. 
The Dec $+40\deg$ data presented in Hancock {\em et al} (1997) was used to put
constraints on the value of $n$, the spectral index of the power
spectrum of the initial fluctuations. This data set is now larger and
so the errors have been reduced. The data was compared with the COBE
data so as to put a greater constraint on the spectral index.
In the likelihood analysis performed
it was found that $n=1.10^{+0.25}_{-0.30}$ (at 68\% confidence)
which is consistent with
the result from COBE alone (Tegmark \& Bunn 1995). 

The same method can be applied to the full data set. To simplify the
analysis it is assumed
that there are no correlations between the COBE and Tenerife data
sets. There is a large overlap between the filter functions of the two
experiments and the region of sky observed by Tenerife is also
observed by COBE. Therefore, there should be significant correlations
between the two data sets for these pixels. However, for
the majority of the COBE pixels there is no correlation with the
Tenerife data and so, as a first attempt, this approximation is valid. The
assumption of no cross-correlation between the two data sets allows
separate inversions of the covariance matrices and so all that is
required in the joint likelihood analysis 
is a multiplication of the COBE likelihood curve (Tegmark, private
communication) by the
Tenerife likelihood curve produced above. The result from this joint
likelihood analysis is shown in
Figure~\ref{fig:cobetenlike}. The peak of the likelihood curve is at
$n=1.19$ and $Q_{RMS-PS} = 19.3\mu$K. 

It is also possible to integrate over
one of the parameters to put a constraint on the other. Integrating
over $Q_{RMS-PS}$ it was found that $n=1.1^{+0.2}_{-0.2}$ (68\%
confidence limits) and
integrating over $n$ it was found that $Q_{RMS-PS} = 19.9^{+3.5}_{-3.2}
\mu$K (68\% confidence limits). It is noted that the slope
between the Tenerife and COBE scales is expected to be slightly greater than one
(up to 10\%) even for the scale invariant spectrum because of the
contribution to the Tenerife data set from the tail of the Doppler
peak. However, even without this enhanced slope the joint analysis is
consistent with a Harrison-Zel'dovich spectrum. 
For $n=1$ it was found that $Q_{RMS-PS} = 22.2^{+4.4}_{-4.2}
\mu$K. This can also be compared with the joint likelihood analysis
for the 15~GHz and 33~GHz Tenerife Dec. $40\deg$ data which gives
$Q_{RMS-PS} = 22.7^{+8.3}_{-5.7} \mu$K for $n=1$ which is consistent
with the joint COBE and 15~GHz Tenerife likelihood analysis. Currently
work is being done into calculating the full likelihood results for
the joint analysis incorporating all of the cross-correlations between
the COBE and Tenerife data sets. 

\begin{figure}
\caption{The likelihood results for the combination of COBE and
Tenerife data sets. This is very preliminary analysis as no
cross-correlations between the two data sets have been taken into
account.}
\label{fig:cobetenlike}
\end{figure}

By placing the results from Tenerife for the level of the CMB onto the power
spectrum graph, along with the results from other experiments, it can
be seen that the standard CDM model (with inflation) is consistent
with the findings. Figure \ref{fig:specres} shows the levels of
fluctuations found by various CMB experiments. The solid line is a
polynomial fit to the points that minimises $\chi^2$ and the two
dashed lines correspond to altering the Saskatoon experiment
calibration by 14\% (the quoted uncertainty in the results). It is
clearly seen that there is evidence for the Doppler peak in the
spectrum (see Hancock \et 1996).

\begin{figure}
\caption{The levels of fluctuations found from various CMB
experiments. The solid line is a $\chi^2$ minimisation for the
points. The two dashed lines correspond to altering the Saskatoon
results by $\pm 14\%$. Evidence for a Doppler peak is clearly visible
and it is seen that the Tenerife data is in agreement with the results
from other experiments. The vertical bars on each point correspond to
the error on the levels calculated and the horizontal bars correspond
to the window function of the experiments.}
\label{fig:specres}
\end{figure}


\chapter{Producing sky maps}
\label{chap5}

The purpose of any Microwave Background experiment is to produce a map
of the fluctuations present on the last scattering surface. However,
the methods used to obtain the sensitivity required 
involve scanning, beam switching
and interferometry and so a way of working back from the
data to the underlying real sky is needed. Similar problems occur when
more than one component is present in the data and a method of
separating out the different signals is required. 

In this Chapter analysis techniques that can be used to 
find the best astronomical signals from a given data set are presented. The four 
methods to be described are 
CLEAN (Section~\ref{clean}), Singular Value Decomposition 
(Section~\ref{SVD}), a new Maximum Entropy Method (Section~\ref{memalg}) 
and the Wiener filter (Section~\ref{wiener}).
In the following Chapters the methods are used to analyse the data from
each of the experiments described in Chapter~\ref{chap3}.

The Maximum Entropy Method has been introduced previously in Hancock
(1994). A fuller account of the method and the choice of parameters 
is given in this chapter, as
well as new error calculations and 
new applications to interferometry and multiple frequency
observations. Therefore, the equations that appear in Chapter~4 of Hancock (1994)
are repeated here (Equations~\ref{eq:entropy} to \ref{eq:chisqu}) so
that the full method can be followed. 

In general, the data from a CMB experiment will take the form of the true 
sky convolved with the instrumental response matrix with any baseline 
variations or noise terms added on. 
It is assumed that the observations obtained from a particular experiment 
have been integrated into discrete bins. For the $i$-th row
and the $j$-th column,
the data, $y_i^{(j)}$, recorded by the instrument can be
expressed as the instrumental response matrix $R_i^{(j)}(i',j')$ acting
on the true sky $x(i',j')$:

\begin{equation}
\label{eq:data}
y_i^{(j)}= \sum_{(i',j')} R_i^{(j)}(i',j') x(i',j') +\epsilon_i^{(j)},
\end{equation}

\noindent
where $i'$ and $j'$ label the true sky row and column position respectively.
The $\epsilon_i^{(j)}$ term represents a noise term, assumed to be 
random, uncorrelated Gaussian noise.

It is immediately clear from Equation~\ref{eq:data} that the inversion
of the data $y_i^{(j)}$ to obtain the two-dimensional sky distribution
$x(i',j')$, is singular. The inverse, $R^{-1}$, of the instrumental
response function does not exist, unless the telescope samples all of the
modes on the sky. Consequently there is a set of signals,
comprising the annihilator of $R$, which when convolved with $R$ gives
zero. Furthermore, the presence of the noise term $\epsilon$ will
effectively enlarge the annihilator of $R$ allowing small changes in
the data to produce large changes in the estimated sky signal. It is, 
therefore, necessary to use a technique that will approximate this inversion. 

\section{CLEAN}
\label{clean}

The CLEAN technique is an operation which reduces the data into
a set of point source responses. The point source responses are then
convolved with the beam to obtain the `CLEAN' map. For the two
dimensional data a two dimensional CLEAN was implemented. Firstly the
algorithm looks for the peak value in the data amplitude (in the case
of the interferometer this is a complex amplitude) and notes its
position, $(i,j)$. 
It then subtracts the normalised beam centred on $(i,j)$
multiplied by $\gamma$ times the peak value, where $\gamma$ is the
loop gain ($\gamma < 1$) set by the user, 
from the data. This process is then repeated
iteratively. If this is allowed to continue ad--infinitum all the data
would be fitted by the point source beam responses but this has no
advantage over the original data so a criterion must be set as to when
to halt the iterations. Two methods can be used to halt the
iterations. Either the number of iterations can be set ($N_{iter}$) or
the peak amplitude of the residual map can be set so that no points
below this will be fitted. The final responses data set is then
convolved with the beam to produce a cleaned map. The obvious problem
with this method is that it eliminates all fluctuations in the data
below the threshold level (or below the level set by $N_{iter}$). 
However, it does reconstruct all large sources very quickly.

\section{Singular Value Decomposition}
\label{SVD}

In singular value decomposition (SVD) the solution to Equation~\ref{eq:data}
can be approximated by solving 

\begin{equation}
y_i^{(j)}= \sum_{(i',j')} R_i^{(j)}(i',j') \hat{x}(i',j')
\label{eq:svddata}
\end{equation}

\noindent
where $\hat{x}$ is the best estimate (given the noise and beam
convolution) of the underlying true sky $x$. The
solution is found by minimising the residual

\begin{equation}
r=\sum_{(i,j)} \left| y_i^{(j)}- 
\sum_{(i',j')} R_i^{(j)}(i',j') \hat{x}(i',j') \right|^2.
\label{eq:ressvd}
\end{equation}

\noindent
To find an approximate inverse of the ${\bf R}$ matrix, 
so that the best $\hat{x}$ can
be found, it is written as the product of three matrices.

\begin{equation}
{\bf R} = {\bf U} {\bf W} {\bf V}^T
\label{eq:svddecomp}
\end{equation}

\noindent
where ${\bf U}$ and ${\bf V}$ are both orthogonal matrices and ${\bf W}$ is purely
diagonal with elements $w_k$. The formal inverse of ${\bf R}$ is therefore easily
found and is equal to 

\begin{equation}
{\bf R}^{-1} = {\bf V} \left[ {\rm diag}\left(1\over w_k\right)
\right] {\bf U}^T.
\label{eq:svdinverse}
\end{equation}

\noindent
The singular values of ${\bf R}$ (comprising the annihilator)
are easily seen as the values of $w_k$ that are zero. To overcome this
problem the SVD analysis sets a condition number which is defined as
the ratio of the largest $w_k$ to the smallest $w_k$. Any $w_k$
below the minimum value set by the 
condition number (so that it is close to zero and
therefore represents a possible singular value) has the corresponding
$1/w_k$ set to zero. The condition number  
is usually set to machine precision and so
rounding errors and singularities are removed making the approximate inverse
possible to calculate. 

The SVD analysis does not take into account the level of noise in the
data vector and so may have the tendency to amplify the noise. A
method of regularising this is required. 

\section[Bayes' theorem]{Bayes' theorem and the entropic prior}
\label{bayes}

There are many degenerate sky solutions that are consistent with the 
data but it is desired to assume as little as possible about the true sky 
so as to avoid bias towards any particular model.  By considering the problem 
from a purely Bayesian viewpoint, the most probable sky distribution 
given our data set and some prior information is desired. Given a
hypothesis $H$, data $D$ and background information $I$, then Bayes'
theorem states that the posterior probability distribution $\Pr(H| DI)$
is the product of the prior probability $\Pr(H | I)$ and the likelihood
$\Pr(D | HI)$, with some overall normalising factor called the evidence $\Pr(D
| I)$:

\begin{equation}
\label{eq:bayes}
\Pr(H | DI)=\frac{\Pr(H | I) \Pr(D | HI)}{\Pr(D | I)} .
\end{equation}

\noindent
The likelihood $\Pr(D|HI)$ is fully defined by the data, but there is
freedom to choose the prior $\Pr(H|I)$. The most conservative prior
assumption is to simply take the sky fluctuations to be small at some level. 
It is therefore desired
to find the sky that contains the least amount of information, 
the one closest to being smooth, that
is still consistent with the data set. This forms the 
basis of the so called ``maximum
entropy methods'' (MEM) (Gull 1989, Gull \& Skilling
1984) used in the reconstruction of images. If
the class of sky models is restricted to those with this property then
this leads to a prior of form (Skilling 1989),

\begin{equation}
\label{eq:entprior}
\Pr(H | I) \propto \exp(\alpha S(f,m))
\end{equation}

\noindent
for an image $f(\xi)$ and
prior information $m(\xi)$, where $\xi$ is the position on the sky.
The regularising parameter, $\alpha$, is a
dimensional constant that depends on the scaling of the problem and
$S(f,m)$ is the cross entropy.  For a positive, additive distribution (PAD),
the cross
entropy is given by (Skilling 1988):

\begin{equation}
S(f,m)= \int d \xi \left [f(\xi)-m(\xi)-f(\xi)
\ln \left( \frac{f(\xi)}{m(\xi)} \right) \right ],
\end{equation}

\noindent
where $m(\xi)$ can also be considered a constraint on $f(\xi)$ such that 
when the entropic
prior dominates, the aposteriori sky map $f(\xi)$ does not deviate
significantly from $m(\xi)$. This form is chosen
for the entropy to ensure that the global
maximum of the entropy at $f(\xi)=m(\xi)$ is zero and it is the only 
form of entropy consistent with coordinate and scaling invariance
(Skilling 1988) 
which does not introduce any biases away from a small fluctuation model. 

\subsection[Pos/neg reconstruction]{Positive and negative data reconstruction}
\label{posneg}

Due to the log entropy term, this standard form is
inapplicable in the more general case of an image $x(\xi)$ that can
take both positive and negative values. 
Various methods in the past have been tried to reconstruct data 
containing negative peaks. Laue, Skilling and Staunton (1985) proposed 
a two channel MEM, which involved splitting the data into positive and
negative features and then reconstructing each separately but not taking into
account any continuity constraint between the two. This method
is inappropriate for differencing experiments as the positive and negative 
features originate from the beam shape and not from separate sources. 
White and Bunn (1995) have proposed adding a constant onto the data
to make it wholly positive. They use the Millimetre-wave Anisotropy 
Experiment (MAX) data to reconstruct a $5\dg \times 2\dg$ region of sky. 
As simulations performed show, this method introduces a bias 
towards positive (or negative if the data are inverted) 
reconstruction. The reason for this is that the added constant 
has to be small enough so that numerical errors are not introduced into
the calculations but a lower constant will give less range for the 
reconstruction of negative features and so the most probable sky will 
be a more positive one. A method to overcome both of these 
problems is proposed here (see Jones {\em et al} 1998 and Maisinger
{\em et al} 1997).

Consider the image to be the difference between two PADS:
\[
x(\xi)=u(\xi)-v(\xi),
\]

\noindent
so that the expression for the cross entropy becomes:

\begin{displaymath}
S(u,v,m)= \int d\xi \left [u(\xi)-{m_u}(\xi)-u(\xi) \ln \left( \frac{u(\xi)}{{m_u}(\xi)} \right)
\right]
+ 
\end{displaymath}

\begin{equation}
\;\;\;\;\;\;\;\;\;\;\;\;\;\;\;\;\;\;\;\;
\int d\xi \left [v(\xi)-{m_v}(\xi)-v(\xi) \ln \left( \frac{v(\xi)}{{m_v}(\xi)} \right) \right].
\label{eq:entropy}
\end{equation}

\noindent
With the prior thus defined, it is possible to calculate the probability of
obtaining the reconstructed sky $x$ (the hypothesis $H$) given $y$ 
(the data $D$):
\begin{equation}
\Pr(x | y) \propto \Pr(x) \Pr(y | x),
\end{equation}
and then maximise $\Pr(x | y)$ to obtain the most likely 2-D image of the
microwave sky.

\section[MEM in real space]{Maximum Entropy deconvolution in real space}
\label{memalg}

The application of Maximum Entropy to experiments with large sky
coverage will now be discussed. 

\subsection{Long period baseline drifts.}
\label{longbase}

In Figure~\ref{fig:scan5}, one of
the Dec $=+46.6\deg$ scans from the $8\deg$ FWHM Tenerife experiment, 
a slow variation
in baselevel (with a peak to peak amplitude of $\sim 2$ mK) is distinctly
evident. Most of the scans obtained from this and the other Tenerife
experiments, show these
variations, to a greater or lesser degree, and therefore their removal
is an important part of the analysis. These long period baselines vary
both along a given scan and from day to day, clearly indicating that
they are due, in the main, to atmospheric effects, with a possible
contribution from diurnal variations in the ambient conditions. The
remainder of this section concentrates on the $8\deg$ FWHM Tenerife
experiment but the other Tenerife experiments are easily analysed in
the same way with minor modification to the equations derived (note
that the baseline variations are removed from the Jodrell Bank
interferometer in the pre--processing stage of analysis). 
The time scale for these baselevel variations appears to be several
hours. With variations of this kind included, Equation~\ref{eq:data} can be written as

\begin{equation}
y_i^{(j)}= \sum_{(i',j')} R_i^{(j)}(i',j') x(i',j')
+\epsilon_i^{(j)}+b_i^{(j)}
\label{eq:dataandb}
\end{equation}

\noindent
where $b_i^{(j)}$ represents the long term baselevel variation.
It is possible to examine this quantitatively by calculating the transfer
function of the experiment, which defines the scales of real structures
on the sky to which the telescope is sensitive. Variations produced on
scales other than these will be entirely the result of non-astronomical
(principally atmospheric) processes and should be removed.

\begin{figure}
\caption{The data from scan 5 of Figure~\ref{fig:typical15}, Chapter~\ref{chap4},
displayed on an expanded temperature scale against RA bin number. Long 
time scale variations in the mean level are evident in the RAW scan
(bottom panel). The middle panel shows the 
baseline fit found by the method of Section~\ref{memalg}.
The top panel shows the
baseline corrected scan. The bin numbers exceed 360 since 
the scan begins near the end of an LST day, and the data are not folded modulo 
$360\dg $.}
\label{fig:scan5}
\end{figure}

As the Earth rotates the beams are swept in RA over a band of sky
centred at a constant declination.  For present illustrative
purposes, it is sufficient to approximate the beams as one-dimensional
in RA, with the beam centre and the East and West throw positions lying
at the same declination.  The beam shape for each individual horn is
well represented by a Gaussian with dispersion $\sigma=$FWHM$/2\sqrt{2\ln 2}=3.57\dg$:

\begin{equation}
B(\theta) = \exp \left(- \frac{\theta^2}{2 \sigma^2} \right),
\end{equation}

\noindent
and the beam switching in right ascension 
may be expressed as a combination of delta
functions:

\begin{equation} \label{eq:switch} S(\theta)= \delta(0)- \frac{1}{2}
(\delta(\theta_b) + \delta(- \theta_b)), 
\end{equation}

\noindent
for a switch angle $\theta_b=8.3\dg$ in RA. So, the beam pattern is
\[R(\theta)=B(\theta)*S(\theta). \] Thus, the transfer function, (\ie
the Fourier transform of the beam pattern) is just:

\begin{equation}
g(k)=2 \sqrt{2 \pi}\sigma \exp \left( \frac{- k^2 \sigma^2}{2} \right) \sin^2 \left( \frac{k \theta_b}{2} \right) .
\end{equation}

In Figure~\ref{fig:tranfn}, the transfer function for waves of period
$\theta= 2 \pi/k$ is plotted. As a function of declination the $\theta$
co-ordinate must be multiplied by a factor of $\sec \delta$ because a
true angle $\theta$ on the sky covers $\sim \theta / \cos \delta$ in right
ascension.  The peak response of the instrument is to plane waves of
period $\sim 22\dg \sec \delta$, \ie individual peaks/troughs with FWHM
$\sim 7\dg$.  The response drops by a factor 10 for structures with
periods greater than $\sim 7$ hours and less than $\sim 40$ minutes.
The long period cutoff is due to cancellation of the large-scale
structures in the beam differencing pattern, while the short period
cutoff is simply due to dilution of structures within a single beam.
The cutoff on large scales in particular is significant for the
analysis, since it tells us that variations in the data on time scales
$\simgt 7^{h}\sec\delta$ are almost certainly due to long time scale
atmospheric effects, or terrestrial and environmental effects, rather
than being intrinsic to the astronomical sky.  Thus identification and
removal of such `baseline' effects is important. 
By using the whole data set to calculate the most probable
astronomical sky signal with maximum entropy deconvolution, it is
possible to simultaneously fit a long-time scale Fourier component baseline to 
each scan. Then a stack of 
$n$ days of data at a given declination to obtain a final scan with a
$\sim \sqrt{n}$ improvement in sensitivity to true astronomical
features is possible.

\begin{figure}
\caption{The transfer function for the Tenerife experiments.}
\label{fig:tranfn}
\end{figure}

\subsection{The beam}
\label{beam}

Note that in the
case of the Tenerife and Jodrell experiments, it is not
necessary to re-define the beam matrix for each position in RA (which
requires a matrix $R_i^{(j)}$ for the $i$-th bin in RA and $j$-th bin
in declination) since the beam is translationally
invariant in the RA direction. However, it is clear that the beam shape
projected into RA and Dec co-ordinates will be a function of
declination. The beam matrix in Equation~\ref{eq:data}
can be written as $R^{(j)}(i',j')$, 
for declination $j$. This simplifies the problem slightly but does not
invalidate the use of MEM for a more general experiment. 
 
For example, in the case of the Tenerife experiment 

\begin{equation}
R^{(j)}(i',j')=
C \left[ \exp \left(-\frac{\theta_C^2}{2 \sigma^2} \right )-
\frac{1}{2} \left( \exp \left(-\frac{\theta_E^2}{2 \sigma^2} \right )
+ \exp \left(-\frac{\theta_W^2}{2 \sigma^2} \right ) \right )  \right ]
\label{eq:beam}
\end{equation}

\noindent 
where $\theta_C$, $\theta_E$, and $\theta_W$ represent the true angular
separation of the point $(i',j')$ from the beam centre and the East and
West throw positions respectively. The normalisation of the beam matrix
is determined by $C$ and 
is implemented with respect to a single beam.
The angles  $\theta_C$, $\theta_E$, and $\theta_W$ can be calculated
using spherical geometry. If the beam is centred at
Dec. $\delta^{(j)}$ and 
$\alpha_0$ is the (arbitrary) RA origin for the definition of all the
beams, for a source at a general $(\alpha,\delta)$ corresponding to the
grid point $(i',j')$ the distance from the main beam centre is

\begin{equation}
\theta_{C} = \arccos \left( \sin \delta^{(j)} \sin \delta + \cos (\alpha_{0} -
\alpha) \cos \delta^{(j)} \cos \delta \right).
\label{eq:thetac}
\end{equation}
 
There are also two other beams (with half the amplitude of the central 
beam), due to
the beamswitching and mirror
wagging, a distance $\theta_b$ (the beamthrow) either side of the
central peak. These have RA centres given by $\theta_{E}$ or
$\theta_{W}$ as in Equation~\ref{eq:beam} defined as 
\[ \cos ( \theta_{E {\rm \, or \, } W} - \alpha_{0}) =
\frac{\cos \theta_{b} - \sin^{2} \delta ^{(j)}} { \cos^{2}
\delta^{(j)}}, \] 
and (fairly accurately for the beamthrow used in
practice) Dec's of $\delta^{(j)}$ still. Thus their angular distances
from the source at $(i',j')$ can be worked out from Equation~\ref{eq:thetac}, yielding
$\theta_{E {\rm \, or \, } W}$, and the final $R^{(j)}$ entry
computed from Equation~\ref{eq:beam}.

\subsection{Implementation of MEM}
\label{implem}

If the assumption of random Gaussian noise if made,
the likelihood is given by:

\begin{equation}
\label{eq:chilike}
\Pr(y | x) \propto \exp \left( \frac{- \chi^2}{2} \right),
\end{equation}

\noindent
where $\chi^2$ is the misfit statistic (Equation~\ref{eq:chisqu}).
Thus in order to maximise $\Pr(x | y)$ it is simply necessary to minimise the
function

\begin{equation}
\label{eq:f}
F=\chi^2 - \alpha S
\end{equation}

\noindent
where a factor of two has been absorbed into the constant $\alpha$.
Thus the process is to iterate to a minimum in $F$ by consistently
updating the 2-D reconstructed sky $x(i',j')$. For the Tenerife data set 
long term baselines for each scan are also built up simultaneously. 
The baselines are represented by a Fourier series
\begin{equation}
\label{eq:baseline}
b_{ir}^{(j)}=C_{0r}^{(j)} + \sum_{n=1}^{nmax} \left[ C_{nr}^{(j)} \cos(\frac{2n \pi i}{l}) + D_{nr}^{(j)} \sin(\frac{2n \pi i}{l}) \right] 
\end{equation}
for the $r$-th scan at the $j$-th declination with $nmax$ baseline
coefficients to be fitted. The basis
data vector index $i$ runs from $1$ to $l$, the length of the data sets 
($3 \times 24^{h}$ for the Tenerife $8\deg$ FWHM data set). 
Thus to obtain a minimum period solution \simgt
${7^{h}} \sec \delta$ (Section~\ref{longbase}) 
we must limit the number of baseline coefficients, $nmax$ in 
Equation~\ref{eq:baseline} to less than 9 (for the case
$\delta=40\deg$) for the Tenerife $8\deg$ FWHM data set and less than
16 for the Tenerife $5\deg$ FWHM data set.

For the $r$th Tenerife scan, Equation~\ref{eq:dataandb} may be written as

\begin{equation}
y_{ir}^{(j)}=y_i^{pred(j)}+b_{ir}^{(j)}+\epsilon_{ir}^{(j)},
\end{equation}

\noindent
with the baseline variation $b$ included and 

\begin{equation}
\label{eq:pred}
y_i^{pred(j)}= \sum_{i',j'} R^{(j)}(i',j')x(i'+i,j')
\end{equation}

\noindent
is the predicted signal produced by the telescope in the
absence of noise and baseline offsets. The beam matrix $R$ is now defined
with respect to an origin (which was chosen to be zero) 
in the $i'$ direction as it is
translationally invariant in RA. This gives us the term
$x(i'+i,j')$ instead of $x(i',j')$. So, the $\chi^2$ for the problem is
\begin{equation}
\label{eq:chisqu}
\chi^2=\sum_{j=1}^{ndecs}\sum_{i=1}^{nra} w_{i}^{(j)} (y_i^{pred(j)}-y_{i}^{obs(j)})^2,
\end{equation}
for a total number of declinations $ndecs$,
total number of RA bins $nra$ and observed stacked data values $y_{i}^{obs(j)}$
with weighting factor $w_{i}^{(j)}$, which are a weighted average over the
$ns$ scans with the baseline $b_{ir}^{(j)}$
subtracted from each of the scans $y_{ir}^{(j)}$ ($r$ is an index running over the $ns$ scans).
In the absence of data
$w_{ir}^{(j)}$, for each individual scan, is set to zero and when data is present it is given by
the inverse of the variance for the data point. 
It is a fairly simple task to compute $y_i^{pred(j)}$, since
it is possible to use prior knowledge of
the geometry of the instrument to calculate the expected response
function $R^{(j)}(i',j')$ for each $i'$, $j'$, at RA $i$ and 
declination $j$, thus $\chi^2$ is fully defined.

For the Jodrell Bank interferometer there are two orthogonal
channels that should be included.
As the entropy term depends on an unconvolved map this
does not change with the introduction of these two orthogonal
terms. However, for the $i-$th bin in right ascension and the
$j-$th bin in declination the data, the $\chi^2$ term becomes
 
\begin{equation}
\label{eq:chisqu2}
\chi^2=\sum_{j=1}^{ndecs}\sum_{i=1}^{nra} w_{ic}^{(j)} (y_{ic}^{pred(j)}-y_{ic}^{
obs( j)})^2+
\sum_{j=1}^{ndecs}\sum_{i=1}^{nra} w_{is}^{(j)} (y_{is}^{pred(j)}-y_{is}^{obs( j)
})^2,
\end{equation}
 
\noindent
for a total number of declinations $ndecs$,
total number of RA bins $nra$ and observed data value $y_{ic/s}^{obs(j)}$
with weighting factor $w_{ic/s}^{(j)}$ taken to be the inverse of the variance
of the pixel.
The cosine and sine channels are given by
 
\begin{equation}
\label{eq:cosdata}
y_{ic}^{(j)}= \sum_{(i',j')} C^{(j)}(i',j') x(i'+i,j') +\epsilon_{ic}^{(j)},
\end{equation}
 
\noindent
and
 
\begin{equation}
\label{eq:sindata}
y_{is}^{(j)}= \sum_{(i',j')} S^{(j)}(i',j') x(i'+i,j') +\epsilon_{is}^{(j)},
\end{equation}
 
\noindent
where $i'$ and $j'$ label the true sky right ascension bin and declination
bin respectively, and $C$ and $S$ are the beam matrices for the cosine
and sine channel respectively. The beams are once more defined with respect to an
arbitrarily chosen origin of $i=0$ as they are translationally
invariant in RA.
The $\epsilon_{ic/s}^{(j)}$ term represents a noise term,
assumed to be random Gaussian noise.
 
Applying the requirements of continuity in entropy with respect to $x$ 
(so that the partial gradients
with respect to $u$ and $v$ are equal and opposite, which leads to $uv=m^2$)
to Equation~\ref{eq:entropy} implies
an entropy term, $S(x,m)$, of form:
\begin{equation}
\sum_{i',j'}  \left[\psi_{i',j'}-2m_{i',j'}-x_{i',j'} \ln \left( \frac{\psi_{i',j'}+ x_{i',j'}}{2m_{i',j'}} \right) \right],
\label{eq:entropy2}
\end{equation}
where $\psi_{i',j'}=u_{i',j'}+v_{i',j'}=(x_{i',j'}^2+4m_{i',j'}^2)^{1/2}$. 
Using our prior
information that the sky fluctuations have zero mean,
${m_{i',j'}}={m_u}={m_v}$ (from Equation~\ref{eq:entropy}) was chosen, and 
the minima of the large entropy case corresponds to 
$x=\left(1{-{{m_u}\over{m_v}}}\right) u=0$;  
$m_{i',j'}$ can therefore be considered as a level of `damping' imposed 
on $x_{i',j'}$ rather than a default model as in the positive-only
MEM. A large value of $m$ allows large
noisy features to be reconstructed whereas a small value of $m$ will not
allow the final sky to deviate strongly from the zero mean.
 
Thus, if the value of the regularising parameter $\alpha$ and the
`damping' term $m$ is known then
$F$ is determined and the best sky reconstruction is that for which $\partial
F /\partial x_{ij}=0$, $\forall x_{ij}$.  This is most easily
implemented by applying one-dimensional Newton-Raphson iteration
simultaneously to each of the $x_{ij}$ to find the zero of the function
$G(x)=\partial F  / \partial x$. This means that $x$ is updated from the
$n$-th to the $(n+1)$-th iteration by

\begin{equation}
x_{lm}^{n+1} = x_{lm}^{n} -\gamma \left( \frac{G(x_{lm}^n)}{\left . \frac{\partial G}{\partial x_{lm}} \right |_{x_{lm}^n} } \right ).
\label{eq:newtonraph}
\end{equation}

\noindent
Convergence towards a global minimum is ensured by setting a suitable
value for the loop gain $\gamma$ and updating $x_{lm}$ only if $\left .
\frac{\partial G}{\partial x_{lm}} \right |_{x_{lm}^n}$ is positive.
The differential with respect to the sky model is easily found
analytically. For $\chi^2$ it is found that

\begin{equation}
{{\partial\chi^2}\over{\partial x_{lm}}}=2 \sum_k \sum_i w_i^{(k)} \left(
R^{(k)} (l-i,m) (y_i^{pred(k)}-y_i^{obs(k)}) \right)
\label{eq:chigrad}
\end{equation}

\noindent
and, considering only the diagonal terms that are of concern here,

\begin{equation}
{{\partial^2\chi^2}\over{\partial {x_{lm}}^2}}=2 \sum_k \sum_i w_i^{(k)}
\left( R^{(k)} (l-i,m) \right)^2 
\label{eq:chicurv}
\end{equation}

\noindent
and for the entropy term, $S$, it is found that

\begin{equation}
{{\partial S}\over{\partial x_{lm}}}=-\ln \left({u_{lm} \over m_{lm}}
\right) = -\ln \left( {{\psi_{lm} + x_{lm}}\over{2 m_{lm}}} \right)
\label{eq:entgrad}
\end{equation}

\noindent
and

\begin{equation}
{{\partial^2 S}\over{\partial {x_{lm}}^2}}=-{1\over{u_{lm} +v_{lm}}}=-{1\over{\psi_{lm}}}.
\label{eq:entcurv}
\end{equation}

In the case of the Tenerife data where the long-term baseline
fluctuations are left in the data to be analysed by MEM, it is
possible to
simultaneously fit for the parameters of the baselines.
The best reconstruction of the microwave sky is calculated
and subtracted from each individual scan and then an 
atmospheric baseline is fitted to each scan. To fit for the
baseline parameters $C_{0r}^{(j)}$, $C_{nr}^{(j)}$ and $D_{nr}^{(j)}$ as
expressed in Equation~\ref{eq:baseline} it is sufficient to implement a
simultaneous but independent
$\chi^2$ minimisation on each of these to obtain the baseline for the
$r$-th scan. From the Bayesian viewpoint minimising $\chi^2$ is just
finding the maximum posterior probability by using a uniform prior. This is 
also done with a Newton-Raphson iterative technique with a new loop gain,
$\gamma_b$.

\subsection{Errors on the MEM reconstruction}
\label{memerror}

The posterior probability of the MEM reconstruction can be written as

\begin{equation}
{\rm Pr}(H|DI) \propto {\rm exp}\left[ -\left( \chi^2-\alpha S \right)
\right]
\label{eq:postprob}
\end{equation}

\noindent
and expanding the exponent around its minimum gives

\begin{equation}
\chi^2-\alpha S= \left( \chi^2-\alpha S \right)_{min} + x \nabla \left(
\chi^2-\alpha S \right)_{min} + {1 \over 2} x^\dagger 
\nabla^2 \left( \chi^2-\alpha S \right)_{min} x.
\label{eq:minent}
\end{equation}

\noindent
At the minima (the maximum probability) $\nabla \left( \chi^2-\alpha S
\right)=0$ and so, to second order, Equation~\ref{eq:postprob} can be written as 

\begin{equation}
{\rm Pr}(H|DI) = {\rm Pr}(H_{MP}|DI) {\rm exp} \left[ - {1\over 2}
{x^\dagger \nabla^2 \left( \chi^2-\alpha S \right)_{min} x} \right]
\label{eq:expent}
\end{equation}

\noindent
where $H_{MP}$ is the most probable reconstruction of $H$ given the
data, $D$, and information, $I$ (i.e. it is the MEM solution). 
This is a Gaussian in $x$ with
covariance matrix given by the inverse Hessian, $M^{-1}=\left( \nabla^2 \left(
\chi^2-\alpha S \right)_{min} \right)^{-1}$. For the reconstruction 
of data from CMB experiments the Hessian is given by

\begin{equation}
M_{lk}= R_{il} R_{ik} + {\alpha\over x_{lk}} \delta_{lk}
\label{eq:hessian}
\end{equation}

\noindent
where $R$ is the beam and $x$ is the sky reconstruction. This follows
directly from considering all of the terms, on and off the diagonal,
in Equations~\ref{eq:chicurv} and \ref{eq:entcurv}. The second
differential of the entropy is a diagonal matrix, however, the second
differential of the $\chi^2$ is not diagonal. The errors on the
reconstruction, therefore, involve the inversion of a large matrix that
is computationally intense. For the case of the Planck surveyor this would
involve the inversion of a $(400\times 400\times 5)^2$ 
matrix. For this reason the simpler method of
performing Monte Carlo simulations was used to obtain an estimate on
the errors. However, in the Fourier plane the problem is dramatically
simplified (see Section~\ref{memalgfs}).

\subsection{Choosing $\alpha$ and m}
\label{alpha}

In this MEM approach the entropic regularising parameter,
$\alpha$, controls the competition between the requirement for a
smoothly varying sky and the noisy sky imposed by the data. The larger
the value of $\alpha$ the more the data are ignored. The smaller the value of
$\alpha$ the more noise is reconstructed.  A choice of 
$\alpha$ that will take maximum
notice of the data vectors containing information on the true sky
distribution, while using the beam response shape to reject the noisy
data vectors is required. In some sense, the entropy term may be
thought of as using the
prior information that the sky is continuous at some level to fill in
for the information not sampled by the response function, thereby
allowing the inversion process to be implemented.

Here, $m$ is chosen to be of similar size to 
the {\em rms} of the fluctuations so that the
algorithm has enough freedom to reconstruct the expected features. 
Increasing/decreasing $m$ by an order of magnitude 
from this value does not alter the final 
result significantly so that the absolute value of $m$ is not important.
This is different to a positive--only MEM because in that 
case $m$ is chosen to be the default model (the value of the sky 
reconstruction in the absence of data) and is therefore more 
constrained by the problem itself. In the case of positive/negative
MEM, as $m$ is the default 
model on the two channels and not on the final sky, there is a greater
freedom in its choice. 

For a data set with independent data points, $\alpha$ is completely
defined in a Bayesian sense (see Section~\ref{bayesalpha}). For data sets
containing a large number of non-independent points (as in the case of
real sky coverages that contain beam convolutions so that the data
set is too large to allow inversion of the Hessian; see 
Section~\ref{bayesalpha}), 
the optimum choice of $\alpha$ is somewhat controversial in the
Bayesian community and while several methods exist (Gull 1989, 
Skilling 1989) it is difficult
to select one above the others that is superior. The 
criterion that $\chi^2-\alpha S=N$, where $N$ is the number of data points 
that are fitted in the convolved sky, is used here. If any of the data 
points are weighted to zero, as the galactic plane crossing is in the
cases considered here, these points should not be included in $N$. 
Increasing/decreasing
$\alpha$ by a factor of ten decreases/increases the amplitude of the
fluctuations derived in the final analysis by \simlt 5 \%. The value
of $\alpha$ is decreased
in stages until $\chi^2-\alpha S=N$; experience has
shown that a convergent solution, for the 10~GHz, $5\deg$ FWHM Tenerife
experiment, is best obtained with the typical
parameter values given in Table~\ref{ta:params}. Below this value for 
$\alpha$ the noisy features in the data have a large effect and the 
scans are poorly fitted. This can be seen in Figure~\ref{fig:funcalp}
where $\chi^2-\alpha S$ begins to flatten out as $\alpha$ is lowered
further than the chosen value (shown as a cross). From this figure it
is easy to chose the value of $\alpha$. Note that 
any significance cannot be attached to the absolute value of
$\alpha$, since it is a parameter that depends on the scaling of the
problem. Also shown in Figure~\ref{fig:funcalp} is the effect of
varying $m$, the `damping' term. It is seen that a larger value of $m$
requires a larger value of $\alpha$ to produce the same result. This
is because the increased freedom in the reconstruction due to $m$ must
be compensated for by a larger restriction due to $\alpha$. 

\begin{table}
\begin{center}
\begin{tabular}{|c|c|} \hline
Parameter & Value\\ \hline
$\alpha$ & $4\times 10^{-2}$\\
$m$ & $50\mu$K\\
$\gamma$ & $0.01$\\
$\gamma_b$ & $0.05$\\
\hline
\end{tabular}
\end{center}
\caption{The parameters used in the MEM inversion.}
\label{ta:params}
\end{table}

\begin{figure}
\caption{The effect of varying $m$ and $\alpha$ on the final
$\chi^2-\alpha S$ of the MEM algorithm. The cross shows the values
chosen for the final reconstruction presented in Chapter~\ref{chap7}
of the 10~GHz $5\deg$ Tenerife experiment.}
\label{fig:funcalp}
\end{figure}

\subsection{Galactic extraction}
\label{extract}
 
By including different frequency information
(for example data from the 10, 15 and 33~GHz beam switching Tenerife experiments,
the 5~GHz interferometer at Jodrell Bank and the 33~GHz interferometer at
Tenerife, or the multifrequency channels of the Planck surveyor satellite)
it should also be possible to separate Galactic foreground emission
from the intrinsic CMB signals utilising the different power law dependencies.
It is possible to rewrite Equation~\ref{eq:data} for the full
multi--frequency data set. For the $k$-th frequency channel of an
experiment with $m$ frequencies

\begin{equation}
y_k({\bf r})= \sum_{l=1}^n \sum_{\bf r\arcmin} Q_{kl} ({\bf r,r\arcmin})
 x_l ({\bf r\arcmin}) +\epsilon_k ({\bf r})
\label{eq:newy}
\end{equation}

\noindent
where $x_l ({\bf r\arcmin})$ is the $l$-th component 
(e.g. CMB, dust emission, point source emission {\em etc.}).
$Q_{kl} ({\bf r,r\arcmin})$ is therefore a
combination of the beam response matrix and the frequency spectrum of
each component. The drawback in using this method is that 
the frequency spectrum of the individual components
needs to be known before the method can be implemented. However, in
most cases, the spectrum is known to a good approximation and, if
desired, MEM can be used to find the most probable spectrum for any
unknown components although this requires much longer computational
time. The MEM
treatment follows directly from this equation similarly to the single
frequency MEM, except that there are $n$ entropy terms (one for each
component) and $m$ $\chi^2$ terms (one for each frequency channel).

%
%
%

\section[MEM in Fourier space]{Maximum entropy in Fourier space}
\label{memalgfs}

If the sky coverage of a particular experiment is not too large, so 
that the sphericity is negligible and 
a Fourier transform can be used, then the problem becomes
simplified. This may also be possible with larger sky coverages but
will require a more sophisticated decomposition of the sky harmonics
(e.g. full sky spherical harmonic decomposition, for which the main
limitation is computational time). Therefore, in Fourier space it is
useful to look at the full Maximum Entropy problem again. 

If the Fourier transform of Equation~\ref{eq:newy} is taken then we
find

\begin{equation}
{\hat{y}_k} ({\bf k}) = \sum_l \hat{Q}_{kl} ({\bf k,k^\prime})
\hat{x}_l ({\bf k^\prime}) +
\hat{\epsilon}_l ({\bf k})
\label{eq:dataft}
\end{equation}

\noindent
where ${\bf k}$ is the Fourier mode. The convolution
between $Q$ and $x$ has been replaced by a multiplication. This in
itself represents a substantial simplification of all of the
calculations involved in implementing the MEM approach. It is seen
that each Fourier mode is independent of each other Fourier mode and
so the Maximum Entropy algorithm can be implemented on a pixel by
pixel basis (where each pixel is now a separate Fourier mode). However, it is
also possible to include further information to MEM using this
method. 

If the correlation of $x$ is known then it may be useful to
use this extra information to further constrain the image
reconstruction. We define the Fourier transform of the 
signal covariance matrix as 

\begin{equation}
\hat{C}_{ll^\prime} ({\bf r}) = \left< \hat{x}_l ({\bf r}) \hat{x}_{l^\prime} ({\bf r})
\right>
\label{eq:cll}
\end{equation}

\noindent
where $\hat{C}_{ll^\prime}$ is real with dimensions $n\times n$ (the number
of components). The diagonal elements of $\hat{\bf C}$ contain the ensemble
average power spectra of the different components at a reference
frequency $\nu_\circ$ and the off-diagonal terms contain the cross
power spectra between components. The known correlations can be
incorporated into MEM through the use of an intrinsic correlation
function (ICF) and a set of independent hidden variables (Gull \&
Skilling 1990 and Hobson {\em et al} 1998a). This is most easily
achieved by the inclusion of lower triangular matrix $\hat{\bf L}$
such that

\begin{equation}
\hat{\bf x} = \hat{\bf L} \hat{\bf h}.
\label{eq:cholesky}
\end{equation}

\noindent
$\hat{\bf L}$ is obtained by performing a Cholesky decomposition
(Press {\em et al} 1992) of the matrix $\hat{\bf C}$ such that 
$\hat{\bf L} \hat{\bf L}^T = \hat{\bf C}$. Thus, the components of
$\hat{\bf h}$ are uncorrelated and of unit variance and so 

\begin{equation}
\left< \hat{\bf x} \hat{\bf x}^\dagger \right> = \left< \hat{\bf L}
\hat{\bf h} \hat{\bf h}^\dagger \hat{\bf L}^T \right> = \hat{\bf L}
\left< \hat{\bf h} \hat{\bf h}^\dagger \right> \hat{\bf L}^T = \hat{\bf
L} \hat{\bf L}^T = \hat{\bf C}
\label{eq:longone}
\end{equation}

\noindent
as required. Now the `default' model in the MEM is a measure on
$\hat{\bf h}$ and not $\hat{\bf x}$ and so the {\em rms} expected
level is simply unity. 

Equation~\ref{eq:dataft} can now be written as 

\begin{equation}
{\hat{y}_k} ({\bf k}) = \sum_{l,m} \hat{Q}_{kl} ({\bf k,k^\prime})
\hat{L}_{lm} ({\bf k^\prime}) \hat{h}_m ({\bf k^\prime}) +
\hat{\epsilon}_k ({\bf k})
\label{eq:dataftwL}
\end{equation}

\noindent
and $\hat{\bf L}$ can be absorbed into $\hat{\bf Q}$ 
so that it now contains frequency and
spatial information for each of the foregrounds and the CMB. 
Now the MEM problem becomes the
reconstruction of $\hat{h}$. Equation~\ref{eq:entropy} follows directly
except that now the difference between $u$ and $v$ is used to
reconstruct $\hat{h}$. 

\subsection{Implementation of MEM in Fourier space}
\label{implemft}

The problem is now exactly the same as before except all the terms are
complex. The log term in the entropy does not allow for complex
reconstructions but it is noted that the real and imaginary parts of
the sky are independent of each other and so it is possible to split
them up into two different PADS. Therefore, one complex reconstruction
is split into a total of four channels; $u_{real}$, $v_{real}$,
$u_{imag}$ and $v_{imag}$. 
The $\chi^2$ term is easily calculable from the reconstruction
$\hat{h}$ as it depends linearly upon it (through the
multiplication). Therefore, the reconstruction for each $k$-mode can
be found separately. 

If there is an initial guess for the distribution of $\hat{h}$ then it
is possible to incorporate this into the MEM algorithm in two
different ways. The first way is to incorporate the information into
the covariance matrix and allow MEM to reconstruct the best hidden
variables given the covariance matrix. In the absence of any initial
guess at the shape of the covariance matrix, it is set to a flat model
with the total power chosen to be the same as that in the input maps. If it
is noted that the difference between $m_u$ and $m_v$ is the `default'
level for $\hat{h}$ (the real and imaginary parts of $\hat{h}$ need to be
considered separately here) then another method for taking into
account an initial guess is possible. Taking a uniform background common to both
default values (this is taken as the {\em rms} of $\hat{h}$) the
initial guess at $\hat{h}$ is split into a positive part (added onto
$m_u$) and a negative part (added onto $m_v$) so that $m_u-m_v$ is
equal to the guess. Without the initial guess then it can be seen that
$\hat{h}$ represents phase information and so both $m_u$ and $m_v$ are
assigned with unit amplitude. In the absence of data $\hat{h}$ will
default to the original guess. 

The function $F$ is now fully defined and the maximum of the posterior
probability is required. To find the minimum of $F$, which is now
dependent on a very small number of variables, a quick minimisation
routine is required. The variable metric minimisation (Press {\em et
al} 1992), which uses the first derivative of the function and
estimates the second derivative, approaches the minimum
quadratically. This proved to be the most efficient method of
minimising in Fourier space (other methods were tried and comparison
between the various minimisations showed that each method was
consistent). 

\subsection{Updating the model}
\label{update}

After the posterior probability has been maximised an estimate of the
underlying sky signal $\hat{x}$ is obtained. This could be used as the
best sky reconstruction but it is also possible to use this estimate
as the initial guess to MEM to find a better reconstruction. The guess
can either be used to update the covariance matrix or the default sky
models as described above. It is then possible to iterate until no
significant change is observed between iterations (a 3\% change in any
pixel flux was used as a measure of whether the image had converged or
not in the simulations in the following Chapter). 

\subsection{Bayesian $\alpha$ calculation}
\label{bayesalpha}

A common criticism of MEM has been the arbitrary choice of $\alpha$,
the Lagrange multiplier. However, it is possible to completely define
$\alpha$ using a Bayesian approach. If the problem is thought of as
trying to find the maximum probability with respect to both $\hat{h}$
and $\alpha$, given some background information $I$, 
it is possible to first integrate over all possible $\hat{h}$ values
to find that

\begin{equation}
\Pr(\alpha | \hat{y},I) = {1\over{\Pr(\alpha | I)}} \int \Pr(\hat{h},\hat{y},\alpha | I)
{d^N \hat{h} \over {\rm det} [\psi]^{1\over 2}}
\label{eq:bayespr}
\end{equation}

\noindent
and maximise this in a similar way to maximising over $\hat{h}$
before. The metric on the space of positive/negative distributions is
given by $-\nabla_h \nabla_h S(h,m)$ which leads to the invariant
volume, ${\rm det}[\psi]^{-1/2}$, in the integral (this can be derived
by considering the difference between two Poisson distributions; see
Hobson \& Lasenby 1998).

The full posterior probability can be written as 

\begin{equation}
\Pr(\hat{h},\hat{y},\alpha | I) = \Pr(\hat{h}|\alpha,I) \Pr(\hat{y}|\hat{h},\alpha,I) \Pr (\alpha|I)
\label{eq:bayesrp}
\end{equation}

\noindent
and expanding this one finds

\begin{equation}
\Pr(\hat{h},\hat{y},\alpha | I) = {\Pr(\alpha|I) \over {Z_s Z_L}} \exp \left(\alpha
(S_1 + S_2 + .... S_N) - {1\over 2} \chi^2 \right)
\label{eq:byfull}
\end{equation}

\noindent
where $Z_s$ is the normalisation constant for the entropy term and
$Z_L$ is that for the likelihood term. The total entropy is the sum of
the entropy for each of the separate maps that go into $\hat{h}$. 

If one Taylor expands each term about its minimum (e.g. $S= S_{min} + {1\over
2} \hat{h} \nabla^2 S_{min} \hat{h}+ ...$) then it is found that, to
second order,

\begin{equation}
Z_s = (2\pi)^{N\over 2} \alpha^{N\over 2}
\label{eq:zs}
\end{equation}

\noindent
where $N$ is the number of independent variables to be found. For
example, $N=6$ in
the case of a data set made from three input maps (i.e. $\hat{h}$
consists of three channels) if the
calculation is performed on each Fourier mode separately 
(remembering that each of the three channels is made
up of a real and an imaginary channel) and $N=6 n_p$, where $n_p$ is
the number of pixels in each map, if the calculation is performed on
all Fourier modes simultaneously. The function to be minimised
is given by

\begin{equation}
F(h)=F(h)_{min} - {1 \over 2} \hat{h} [\psi^{-{1\over 2}} ] A [\psi^{-{1\over
2}} ] \hat{h}
\label{eq:fexp}
\end{equation}

\noindent
where $A=[\psi^{1\over 2}] M [\psi^{1\over 2}]$ and $M$ is the Hessian
and $\psi=- \nabla^2 S$. The probability
distribution for $\alpha$ is given by Equation~\ref{eq:bayespr}
and so substituting in the expansions it is found that

\begin{equation}
\Pr(\alpha| \hat{y},I) = {1\over Z_L} { {(2\pi)^{N\over 2} (\det A)^{-{1\over
2}}} \over{(2\pi)^{N\over 2} \alpha^{-{N\over 2}}}} \exp \left(\alpha
(S_1 + S_2 + ... S_N) - {1\over 2} \chi^2 \right)
\label{eq:nearba}
\end{equation}

The most probable $\alpha$ can now be calculated by taking the
derivative of the log of Equation~\ref{eq:nearba} and equating to zero.

\begin{equation}
(S_1 + S_2 + ... S_N) +{N\over 2\alpha} -{1\over 2} {d\over d\alpha}
\log \det A =0
\label{eq:alba}
\end{equation}

\noindent
and noting that the differential of $A$ with respect to $\alpha$ is just
the identity matrix it is found that 

\begin{equation}
S_{tot} + {N\over 2\alpha} - {1\over 2} {\rm Tr} A^{-1}=0
\label{eq:onemore}
\end{equation}

\noindent
Rearranging Equation~\ref{eq:onemore}, the most probable value for $\alpha$ at
each Fourier mode is the solution to 

\begin{equation}
2\alpha S_{tot} + N = \alpha {\rm Tr} A^{-1} (\alpha)
\label{eq:bayesalpha}
\end{equation}

\noindent
It should be noted that $A$ is also a function of
$\alpha$, so the solution of Equation~\ref{eq:bayesalpha} is
non-trivial. An iterative approach to the problem is necessary.
First solve

\begin{equation}
\alpha_{new} = {n \over {{\rm Tr} A^{-1} (\alpha_{old})  + 2 S_{tot}}}
\label{eq:bayesaln}
\end{equation}

\noindent
and then use this new $\alpha$ to perform a new minimisation until
convergence is reached for $\alpha$. Figure~\ref{fig:convalp} shows
the typical convergence around the minima for $\alpha$. 
This method for calculating
$\alpha$ can only be used when one minimisation of $F(\hat{h})$ has
been performed. Thus an initial guess for $\alpha$ is required. This will
be described in the next section. It is noted that this can be
incorporated with the update procedure described in 
Section~\ref{update} until global convergence is reached. In real space this
method is infeasible, as it involves the inversion of the
Hessian, which, with convolutions, is a very large matrix. 

\begin{figure}
\caption{The convergence of $\alpha$ for a typical experiment (in this
case the analysis of Planck surveyor data). The vertical axis is 
$2\alpha S_{tot} + N - \alpha {\rm Tr} A^{-1} (\alpha)$ and the
Bayesian $\alpha$ is found on the point of intersection.}
\label{fig:convalp}
\end{figure}

\subsection{Errors on the reconstruction}
\label{eq:errorft}

In the Fourier domain the inverse of the Hessian (the inverse of the 
Fourier transform of Equation~\ref{eq:hessian}) is much easier to find
than in real space. Instead of the large matrix inversions that were necessary
with the convolution it is now possible to do each $k$-mode separately.
This is used to put errors on the power spectrum (or the full
two dimensional Fourier transform of the underlying sky). 
 
\vspace{1cm}
The full multi-channel MEM has been derived from information theoretic
considerations within the context of Bayes' theorem. A different
approach to the choice of prior probability will now be discussed and
its connection to MEM will be highlighted. 

\section{The Wiener filter}
\label{wiener}

The most conservative prior, ${\em Pr}(H|I)$, in 
Equation~\ref{eq:bayes} may not the best choice in the presence of added
information (see Zaroubi {\em et al} 1995). If the form of the 
hypothesis $H$ is known then this information can also be used 
to further constrain its reconstruction. Taking $m_u =m_v =m =1$ and 
assuming that the
levels of the fluctuations are small, so that $\hat{h}$ is small,
the entropy can be rewritten, to second order, as

\begin{equation}
\alpha S(\hat{h})=-\sum_{\bf k} \alpha { \hat{h}^2 ({\bf k}) \over 4}
\label{eq:wienent}
\end{equation}
 
\noindent
where the sum is over the Fourier modes. Using
$\hat{x}=\hat{L}\hat{h}$, the prior probability is therefore found to be 

\begin{equation}
\Pr(\hat{x} | I) = \exp \left(-{\alpha\over 4} \hat{x}^{\dagger}
(\hat{L}^{T} \hat{L})^{-1}  \hat{x} \right).
\label{eq:wienpri}
\end{equation}

\noindent
Noting that $\hat{L}^{T} \hat{L}=\hat{C}$ this is equivalent to a
Gaussian prior if $\alpha=2$. 
Therefore, in the limit of small fluctuations
it is seen that Maximum Entropy reduces to a quadratic form
fully defined by a Gaussian covariance probability. This is the Wiener
filter. Figure~\ref{fig:alpdiff} shows the
difference in the range of amplitude in the reconstruction that Wiener and MEM
allow. It is seen that there is very little difference between the
$\alpha=2$ MEM and the Wiener filter out to about three standard
deviations. If the fluctuations are not necessarily Gaussian then it 
is possible to take $\alpha=2$ as our starting
guess in the full Maximum Entropy approach and then use
Bayesian alpha calculations to properly define the Lagrangian
multiplier in subsequent iterations. 

\begin{figure}
\caption{The short dashed line shows the value of $\alpha S$ for the
case in which $\alpha=2$. The horizontal scale is the standard
deviation of the reconstruction away from the model value (in this
case taken to be unity). It is seen that the maximum value of the
entropy is zero and the algorithm will default to this value in the
absence of data. Given the data, the algorithm will try to maximise the
entropy and so will reconstruct data with smaller amplitudes (the
range in $x$). The equivalent plot for the Wiener filter is shown as
the solid line and it can be seen that MEM approaches Wiener for small
values of $x$, whereas at large amplitudes, 
a greater range of $x$ can be reconstructed by
MEM. The long dashed line shows the results for a Bayesian $\alpha$
calculation on the Planck surveyor analysis and it can be seen that
the Wiener filter falls far short of the required dynamic range for
reconstruction.}
\label{fig:alpdiff}
\end{figure}

For the Gaussian case the prior probability has become

\begin{equation}
{\rm Pr}(\hat{x}|I) \propto {\rm exp} \left({-{1\over 2} \hat{x}^\dagger
\hat{C}^{-1} \hat{x}}
\right)
\label{eq:wienprob}
\end{equation}

\noindent
and the matrix $\hat{C}$ is the covariance matrix of the model given by

\begin{equation}
\hat{C}=<\hat{x}^\dagger \hat{x}>.
\label{eq:covar}
\end{equation}

The problem then becomes the minimisation of 

\begin{equation}
F=\chi^2 + \hat{x}^\dagger \hat{C}^{-1} \hat{x}
\label{eq:wienmin}
\end{equation}

\noindent 
This minimisation can be done in the
same way as MEM through an iterative algorithm or it can be solved
analytically. In full, 

\begin{equation}
F=(\hat{y}-\hat{Q}\hat{x})^\dagger \hat{N}^{-1}
(\hat{y}-\hat{Q}\hat{x}) + \hat{x}^\dagger \hat{C}^{-1} \hat{x}
\label{eq:wienfull}
\end{equation}

\noindent
where $\hat{Q}$ is the instrumental response matrix, from Equation~\ref{eq:newy} 
and $\hat{N}$ is the variance matrix of the data. Minimising
with respect to $\hat{x}$, the solution is

\begin{equation}
\hat{x}\arcmin =\hat{C} \hat{Q}^\dagger (\hat{Q}\hat{C}\hat{Q}^\dagger
+\hat{N})^{-1} \hat{y}.
\label{equation}
\end{equation}

The Wiener filter, $W$, that finds the best linear approximation to
$\hat{x}_i ({\bf r})$ has resulted: 

\begin{equation}
\hat{x}_i\arcmin ({\bf r}) = W_{ij} ({\bf r,r\arcmin}) \hat{y}_j ({\bf r\arcmin})
\label{eq:yinverse}
\end{equation}

\noindent
where $i$ is now running over all the pixels in the component maps,
$j$ is over all the pixels in the data maps and the Wiener filter is
given by 

\begin{equation}
W=\hat{C} \hat{Q}^\dagger (\hat{Q}\hat{C}\hat{Q}^\dagger
+\hat{N})^{-1}.
\label{eq:wienerfilter}
\end{equation}

\noindent
This can be easily
written in the hidden variable space, as for MEM, but for illustrative
purposes it is left in Fourier space. 
The Bayesian probability can now be written in terms of this filter
(completing the square of Equation~\ref{eq:wienfull} can also be used to
find the Wiener filter):

\begin{equation}
{\rm Pr}(\hat{x}|\hat{y}I) \propto {\rm exp} \left( {-{1\over 2} \left[ \hat{x} -
W \hat{y} \right]^\dagger (\hat{C}^{-1} +
\hat{Q}^\dagger \hat{N}^{-1} \hat{Q}) \left[ \hat{x}-W \hat{y}
\right] } \right)
\label{eq:wiencomsq}
\end{equation}

\noindent
which is seen to have the quadratic form of the simplified MEM
approach. From this it is easily seen that the covariance matrix on this
reconstruction is given by 

\begin{equation}
H_w^{-1} =  (\hat{C}^{-1} + \hat{Q}^\dagger \hat{N}^{-1} \hat{Q})^{-1}.
\label{eq:sigwien}
\end{equation}

The Wiener filter can now be considered as minimising the
variance of the reconstruction errors of the Fourier components given by 

\begin{equation}
\left< \hat{\Delta}_i ({\bf k})^2 \right> = \left< \left| \hat{x}_i\arcmin ({\bf
k}) - \hat{x}_i ({\bf k}) \right|^2 \right> .
\label{eq:varian}
\end{equation}

It is well known that the Wiener filter smoothes out the fluctuations
at low flux levels (corresponding to levels below the noise). This has
the effect of reducing the power in a map. Therefore, the Wiener
filter cannot be used in an iterative fashion as it will tend to
zero. 

Both MEM and the Wiener filter (the quadratic approximation to MEM) in
the Fourier plane have a disadvantage over the real space MEM (Wiener
in real space involves the inversion of very large matrices and so is
infeasible to run). As the calculations are done on a pixel by pixel
basis in Fourier space, the number of pixels at each frequency has to
be the same. Due to the different pixelisation usually seen at
different frequency experiments this requires an additional first step
of re-pixelisation. In real space the full MEM analysis does not
require this extra pixelisation as all the separate pixelisations can
be taken into account in the matrix $Q$ which does not have to be
square. Also, the Wiener filter requires the same
number of pixels in the output maps as in the input maps 
and so cannot be used to reconstruct a map in irregularly
sampled data (as in the case of the Tenerife or Jodrell Bank data). 

\subsection{Errors on the Wiener reconstruction}
\label{wienerror}

It has already been shown that the covariance matrix (the inverse
Hessian) for the Wiener
reconstruction is given by Equation~\ref{eq:sigwien}. Therefore, the
assignment of errors on the Wiener filter reconstruction is
straightforward. It should be noted however, that the simple
propagation of errors in the Wiener filter are a direct result of the
Gaussian assumption of the initial covariance structure. This differs
from the MEM error calculation as the MEM approximates the peak of the
probability to be Gaussian to calculate the errors and not the whole
probability distribution as in the Wiener filter case. 

\subsection{Improvements to Wiener}
\label{wienimprov}

An `improvement' to the Wiener filter has been proposed by a number of
authors (see, for example, Rybicki \& Press 1992,
Bouchet \et 1997, Tegmark 1997). They propose a rescaling of the power
spectrum (either by dividing the power spectrum by a
quality factor or by the introduction of a Lagrange multiplier into
the prior probability) so that the total reconstructed power is equal
to the total input power of the maps. It has been shown that the power
spectrum reconstructed by the standard Wiener filter is a biased
estimation of the true power spectrum (for example Bouchet \et
1997) in that it weights the reconstructed power spectrum by
signal/(signal+noise). This bias can be quantified by introducing a
quality factor for each component $l$, $S_l ({\bf k})$, given by

\begin{equation}
S_l ({\bf k}) = \Sigma_m W_{lm} ({\bf k}) Q_{ml} ({\bf k})
\label{eq:qualityfac}
\end{equation}

\noindent
where $W_{lm} ({\bf k})$ is the Wiener matrix at Fourier mode ${\bf
k}$, $Q_{lm} ({\bf k})$ is the instrumental response matrix and the
sum is over the frequencies of the experiment. The quality factor
varies between unity (in the absence of noise) and zero. If $\hat{x}_l
\arcmin ({\bf k})$ is Wiener reconstruction of the $l$th component and $k$th
Fourier mode and $\hat{x}_l ({\bf k})$ is the exact value it can be shown
(Bouchet \et 1997) that 

\begin{equation}
\left< |\hat{x}_l\arcmin ({\bf k})|^2 \right> = S_l ({\bf k}) 
\left< |\hat{x}_l ({\bf k})|^2 \right>.
\label{eq:newpower}
\end{equation}

\noindent
It is seen that the power spectrum of the reconstruction is an
underestimate of the actual power spectrum by the quality factor. By
rescaling the Wiener matrix it can easily be seen that the resulting
power spectrum will be unbiased. However, it is found that the
variance of the reconstructed maps increases as they are noisier. 

The other proposed variant on Wiener filter is the introduction of a
Lagrange multiplier into the prior probability that scales the input
power spectrum. It has been suggested (Tegmark 1997) that this
parameter should be chosen to obtain a desired signal to noise ratio
in the final reconstruction. However, if it is noted that this
Lagrange multiplier has exactly the same role as the MEM $\alpha$ it
is seen that this parameter is completely defined in a Bayesian sense
similarly to $\alpha$. The inclusion of this Lagrange multiplier
simply results in the Wiener filter becoming the quadratic
approximation to MEM without the automatic setting of $\alpha$ due to
the absolute value of the Gaussian probability covariance matrix. 

In Wiener filtering the introduction of the quality factor results in
noisier maps. The addition of the Lagrange multiplier
just results in the quadratic approximation to MEM and as
MEM and Wiener are indistinguishable in the Gaussian case there is no
need to perform both methods. Therefore, as the final product of the
analysis presented in this thesis is intended to be the real sky maps, 
only classic Wiener and the full MEM will be tested in
the next chapter. 

\vspace{1cm}
The methods discussed in this chapter will be applied to simulated
data in the following chapter to test their relative strengths at
analysing data from CMB experiments. The best method(s) will then be
used to analyse the data and produce actual CMB maps 
from the experiments discussed in Chapter~\ref{chap4} in Chapter~\ref{chap7}.

\chapter{Testing the algorithms}
\label{chap6}

In this chapter the analysis techniques presented in the previous chapter
are tested by applying them to various CMB data sets. The best overall method
is then used in the following chapter to analyse data from the experiments 
presented in Chapter~\ref{chap3} and to produce maps of the sky at 
various frequencies.

\section[CLEAN vs. MEM]{Comparison between CLEAN and MEM}
\label{compmemcl}

Simulations were performed of the CLEAN algorithm 
and the MEM algorithm using the 5~GHz
Jodrell Bank interferometer as an example. The simulations using CLEAN
were performed by Giovanna Giardino (see Giardino 1995).
Data sets were
produced from the Green Bank catalogue by convolving them with the
interferometer beam. Gaussian noise was then added at varying levels
to the scans to simulate the atmosphere and instrumental noise. Eleven
declinations were simulated and ten `observations' of each declination
were made. The six noise levels chosen were 15, 25, 35, 45, 55 and
65~$\mu$K as these correspond to the range of noise levels expected in
the real data set. Only the central RA range ($161\deg - 240\deg$) was
analysed to reduce computing time. Figure~\ref{fig:simmemcl} shows the
input simulated data (dotted line) with the MEM reconstruction (solid
line) and the CLEAN reconstruction (dashed line). This particular simulation is
for Declination $30\deg$ and a noise level of 25~$\mu$K. It was chosen
at random from the full set of simulations. As can be
seen from the figure the MEM result appears to follow the data more closely
than the CLEAN result.

\begin{figure}
\caption {Comparison at one declination of the MEM reconstruction 
(solid line), the CLEAN reconstruction (dashed line) and the 
simulated data (dotted line).}
\label{fig:simmemcl}
\end{figure}

Figure~\ref{fig:diff} shows the mean of the difference between the
noise free simulated map and the reconstructed maps from MEM and
various CLEAN reconstructions. For the ideal case the mean would be
zero for the reconstruction (i.e. it reconstructed the exact simulated
data). As can be seen the CLEAN results deviate from zero by a few
micro Kelvin whereas the MEM result is centred around zero. CLEAN
also gets worse as the noise is increased because it cannot tell the
difference between a noise peak and a data peak whereas the MEM
process does not get any better or worse. 

\begin{figure}
\caption{Mean of difference between the noise free map and the 
reconstructed maps. The MEM reconstructed differences are shown by the
filled circles. CLEAN has been run for different values of the 
parameter $N_{iter}$ and $\gamma$. ${N_{iter}}=10000$ for all curves except
for the hollow squares, for which ${N_{iter}}=20000$. Stars refer to 
$\gamma=0.02$, hollow squares to $\gamma=0.06$, filled triangles to $\gamma=0.1$ 
and filled squares to $\gamma=0.25$ ($\gamma$ is the loop gain in each iteration 
of the CLEAN algorithm). The large error bars are due to the
number of realisations per noise level being limited to 10.}
\label{fig:diff}
\end{figure}

The errors in the final map have also been calculated (by use of a
Monte Carlo technique) and these are shown in Figure~\ref{fig:sigma}. 
As can be seen the different CLEAN processes all have
a similar error in their reconstructions but MEM has a much lower
error. This means that the MEM reconstruction will also be more
consistent between data sets with different noise realisations, which
is essential when analysing data that has many scans to be
simultaneously analysed (as in the case of Tenerife). The MEM error
line is also flatter than the CLEAN results (and also flatter than a
line with unit gradient) which means that not only does MEM perform
better than CLEAN on all noise scales but MEM actually does relatively better at
extracting signals when the noise level is higher. 

\begin{figure}
\caption{Standard deviation of the difference between the amplitude
of the noise free map and the maps reconstructed with CLEAN and MEM. 
The symbols are as in the previous figure.}
\label{fig:sigma}
\end{figure}

The final test of the algorithm is to check whether the fluctuations
reconstructed by the two methods are present in the original map. It
may be the case that the MEM method has reconstructed the correct
amplitude of the signal but has put all of the features in the
incorrect locations. To test this a correlation coefficient
method which correlates the levels of the fluctuations between the
input and output maps was used. Two maps will have a correlation coefficient of
one if they are identical. As can be seen in Figure~\ref{fig:corr} the
MEM reconstructions are very close to unity and so it is seen with
confidence that MEM is reconstructing the simulations very
accurately. The CLEAN results, however, shows that at low noise level
it closely follows the simulations but at high noise levels the correlation drops
dramatically. From this analysis it is concluded that MEM out--performs a
simple CLEAN routine on all reconstruction properties of the maps. 
The CLEAN technique is, therefore, not used in subsequent analysis.

\begin{figure}
\caption{The correlation coefficient between the noise free map and
the reconstructions from MEM and CLEAN. The symbols are the same 
as in the previous figures.}
\label{fig:corr}
\end{figure}

\section{The Planck Surveyor simulations}
\label{coban}

The MEM Galactic extraction algorithm was tested on simulated
observations from the Planck Surveyor satellite (see Hobson \et 1998a). 
To constrain any of the foregrounds in CMB data it is
necessary to have a large frequency coverage. The Planck Surveyor
satellite covers a range from 31~GHz to 847~GHz and has a high
angular resolution (see Chapter~\ref{chap3}) so that it will sample 
all of the foregrounds that were mentioned in Chapter~\ref{chap2}. 

\subsection{The simulations}
\label{cobsamsim}

The simulations described here include six different components as
input for the observations (see Bouchet, Gispert, Boulanger \& Puget
1997 and Gispert \& Bouchet 1997). The main component ignored in these
simulations are extra--galactic point sources. Very
little information is known about the distribution of point sources
at the observing frequencies of the Planck Surveyor. The usual method
for predicting point source contamination is to use observations at
IRAS frequencies ($> 1000$~GHz) or low frequency surveys ($< 10$~GHz) and
extrapolate to intermediate frequencies. This has obvious problems and
so predictions are unreliable. For small sky
coverage point source subtraction is performed by making
high--resolution, high--flux--sensitivity observations of the point
sources at frequencies close to the CMB experiment (see for example,
O'Sullivan \et 1995). For all sky observations, point source removal is
more complicated as it is difficult to make the required observations
of the point sources. For the Planck satellite it is anticipated that
the point sources can be subtracted to a level of 1~Jy at each
observing frequency, and it may be possible to subtract all sources
brighter than 100~mJy at intermediate frequencies where the CMB
emission peaks (De Zotti \et 1997). De Zotti \et (1997) find that the
number of pixels affected by point sources to be low and that the
level of fluctuations due to unsubtracted sources is also very low. 
Therefore, no modelling of
extra--galactic point sources will be made here. Recently, surveys
at 350~GHz and 660~GHz (Smail, Ivison \& Blain, 1997)
have confirmed previous estimates of the 
contribution made by point sources (De Zotti {\em et al.} 1997).
However, these surveys were in specially 
selected regions (gravitationally lensed objects) and so may be an 
over-estimate of the actual point source contribution. The MEM algorithm
described here has been applied to simulations with point sources 
(Hobson {\em et al.} 1998b) and it was seen that their inclusion alters the
conclusions reached here very little.

The six components used are the CMB, kinetic and thermal SZ,
dust, bremsstrahlung and synchrotron. A detailed discussion of the
simulations used is given by Bouchet \et (1997) and Gispert \& Bouchet
(1997). These have reasonably well
defined spectral characteristics and this information can be used,
together with the multifrequency observations, to distinguish between
them. Both MEM and Wiener filtering are used to attempt a reconstruction
of the components from simulated data taken with 14 months of
satellite observation. The simulations are constructed on $10\deg
\times 10\deg$ fields with $1.5\arcmin$ pixels.  
Two models of the CMB are used. The first used is a
standard CDM model with $H_\circ = 50 {\rm kms}^{-1} {\rm Mpc}^{-1}$
and $\Omega_b =0.05$. The second is a string
model produced by Francois Bouchet. The SZ component (thermal and
kinetic) was produced by Aghanim \et (1997) 
using a Press-Schechter formalism (Press \& Schechter 1974) 
which gives the number density of
clusters per unit redshift, solid angle and flux density interval. The
gas profiles of individual clusters were modelled as King 
$\beta$--model (King 1966), and their peculiar velocities were drawn at random
from an assumed Gaussian velocity distribution with a standard
deviation at $z=0$ of 400~km~s$^{-1}$. Galactic dust emission and
bremsstrahlung have been shown to have a component which is
correlated with 21cm emission from HI (Kogut \et 1996, Boulanger \et
1996). To model this correlation two IRAS 100~$\mu$m maps were used;
one for HI correlated emission and one for HI uncorrelated. The 
components are then related by 

\begin{equation}
HI_{corr} (\nu) = A \left[ 0.95 f_D (\nu) + 0.5 f_B (\nu) \right]
\label{eq:corr}
\end{equation}

\noindent
and

\begin{equation}
HI_{uncorr} (\nu) = B \left[ 0.05 f_D (\nu) + 0.5 f_B (\nu) \right],
\label{eq:uncorr}
\end{equation}

\noindent
where $f_D$ is the frequency dependence of the dust emission and $f_B$
is the frequency dependence of the bremsstrahlung.
Therefore, the dust component was modelled by addition of 
95\% of the correlated HI IRAS map and 5\% of the uncorrelated map 
and extrapolating to lower frequencies assuming a
black body temperature of 18~K and an emissivity of 2 for the dust. 
The bremsstrahlung (or
free--free) component was modelled by using 50\% of the HI correlated
IRAS map and 50\% of the uncorrelated IRAS map. 
The combined map was scaled so that the {\em rms}
amplitude of the bremsstrahlung 
at 53~GHz was 6.2$~\mu$K and a temperature spectral index
of $\beta=2.16$ was assumed. No spatial template is available at
sufficiently high resolution for the synchrotron maps. The simulations
were modelled by using the low frequency Haslam \et maps at
408~MHz, which have a resolution of $0.85\deg$ and adding small scale
structure that follows a $C_l \propto l^{-3}$ power spectrum. 
A temperature spectral index of $\beta=2.9$ was assumed. 

The components used were all converted to equivalent thermodynamic 
temperature, for comparison purposes, from flux using the equation 

\begin{equation}
\Delta I(\mu) = {{\Delta T x^4 e^x }\over{(e^x -1)^2}}
\label{eq:ttof}
\end{equation}

\noindent
where $x=hv/kT$ and $T=2.726K$. 
The flux is originally in units of ${\rm Wm}^{-2} {\rm Hz}^{-1}
{\rm sr}^{-1}$ and all programs use flux units rather than
temperature. The six input components (CMB, kinetic SZ, thermal SZ,
dust, bremsstrahlung and synchrotron) are shown in Figures~\ref{fig:cobinp} 
(for the CDM simulation) and \ref{fig:cobinp2} (for the
string simulation). Each component is plotted at 300~GHz and has been
convolved with a Gaussian of $4.5\arcmin$ FWHM, the highest resolution
of the Planck Surveyor. It is noted that the IRAS 100~$\mu$m maps used
as templates for the Galactic dust and free--free emission appear
quite non--Gaussian and the imposed correlation between these two
foregrounds is clearly seen. The azimuthally averaged power spectra of
the input maps (CDM realisation) 
are shown in Figure~\ref{fig:cobinppow}. It is easily
seen that the power in the maps is suppressed for $\ell >2000$. This is
due to the finite resolution. 

\begin{figure}[p!]
\caption{The six input maps used in the simulations for the data
taken by the Planck Surveyor. This is for the CDM model of the CMB. 
They are shown at 300~GHz in $\mu$K. a) CDM simulation, b) Kinetic SZ, c)
Thermal SZ, d) Dust emission e) Free-free emission and f) Synchrotron emission.}
\label{fig:cobinp}
\end{figure}
\begin{figure}[p!]
\caption{The six input maps used in the simulations for the data
taken by the Planck Surveyor. This is for the string model of the
CMB. They are shown at 300~GHz in $\mu$K. a) String simulation, b) Kinetic SZ, c)
Thermal SZ, d) Dust emission e) Free-free emission and f) Synchrotron emission.}
\label{fig:cobinp2}
\end{figure}

\begin{figure}
\caption{The azimuthally-averaged power spectra of the input maps
shown in Figure~\ref{fig:cobinp} at 300 GHz.}
\label{fig:cobinppow}
\end{figure}

The final design of the Planck Surveyor satellite is still to be
decided and significant improvements have recently been made to the
proposed sensitivities. Therefore, these recent improvements will be
used here to simulate observations (see Table~\ref{ta:cobras} for
full details on the frequency channels used).
Simulated observations were produced by integrating the emission
due to each physical component across each waveband, assuming the
transmission is uniform across the band. At each frequency the
beam was assumed to be Gaussian with the appropriate FWHM. 
Care is needed to include the effect of the waveband
integration in the MEM and Wiener algorithms but this can easily be done in the
conversion matrix described in Section~\ref{extract}. The
frequency channels now consist of a noise-free flux in ${\rm Wm}^{-2}$.

Isotropic Gaussian noise was added to each frequency
channel with the typical {\em rms} expected from 14
months of observations. Figure~\ref{fig:levelspl} shows the {\em rms}
fluctuations at each observing frequency due to each physical
component after convolution with the appropriate beam. The {\em rms}
noise per pixel at each frequency is also plotted. It is seen that
only the dust and CMB emissions are above the noise level (although
the thermal SZ is extremely non--Gaussian and has many peaks above the
noise level). The kinetic SZ has the same spectral characteristics as
the CMB, but the effect of the beam convolution at the different
frequencies on a point source distribution is more pronounced. The data was
created on maps with pixels that assumed a spatial
sampling of FWHM/2.4 at each frequency. That meant that for the
$10\deg \times 10\deg$ sky area there were $320\times 320$ 
pixels at the highest frequency and $44\times 44$ 
pixels at the lowest frequency. As the calculations were
performed in the Fourier plane it was necessary to repixelise the maps
onto a common resolution although it is noted that this is not
necessary for the MEM algorithm in real space. The final ten
frequency channels (now with pixel noise added) are shown in Figure~\ref{fig:data}. 

\begin{figure}
\caption{The rms thermodynamic temperature fluctuations at each
Planck Surveyor 
observing frequency due to each physical component, after convolution
with the appropriate beam and using a sampling rate of FWHM/2.4. The
rms noise per pixel at each frequency channel is also plotted.}
\label{fig:levelspl}
\end{figure}

\begin{figure}
\caption{The ten channels in the simulated data from the
Planck Surveyor. The noise and lower resolution is clearly seen in the
low frequency channels. The noise levels represent an improvement for the LFI
since Bersanelli \et 1996 was published. The units are in $\mu$K
equivalent integrated over the band.}
\label{fig:data}
\end{figure}

\subsection{Singular Value Decomposition results}
\label{svdres}

The simple linear inversion of Singular Value Decomposition (S.V.D.) was used
 on the two different data sets (one for the CDM realisation and one
 for the string realisation of the CMB) from the simulated Planck
Surveyor data. The final reconstructed map for the CDM and strings simulations
are shown in Figures~\ref{fig:svd} and \ref{fig:svds} and are
summarised in Table~\ref{ta:svds}. The grey scales
of the reconstructions and the input maps are the same. It was not
 possible to attempt a separate reconstruction of the CMB and kinetic
 SZ effect as these have the same frequency dependence. However, six
 plots are shown for easy comparison with the input maps. 
As can be seen from the figures and tables, the
S.V.D. performs quite well on both the 
CMB (CDM and strings) and the dust channels but fails
to reconstruct the other channels. As the
input maps used are known it is possible to calculate the residuals
for each reconstruction. This is defined as 

\begin{equation}
e_{rms} = \left[ {1\over N} \sum_{i=1}^N \left(T_{\rm rec}^i - T_{\rm
true}^i \right)^2 \right]^{1/2},
\label{eq:erms}
\end{equation}

\noindent
where $T_{\rm rec}^i$ is the reconstructed temperature of pixel $i$,
$T_{\rm true}^i$ is the input temperature and $N$ is the total number
of pixels. The values for the residuals are shown in
Table~\ref{ta:ermssvd}. The desired accuracy of the Planck Surveyor is
$5\mu$K (Bersanelli \et 1996) and it is seen that the SVD analysis of
the data falls far short of this sensitivity.

\begin{table}
\begin{center}
\begin{tabular}{|cc|r|r|}\hline \hline
Component & & Input & SVD reconstruction \\ \cline{3-4}
 & & \multicolumn{2}{c|}{$\Delta T$ in $\mu$K} \\ \hline \hline
CMB strings & Max. &  305.53 &  261.50 \\
            & Min. & -338.20 & -386.63 \\
            & Rms. &   69.43 &   77.81 \\ \hline
CMB CDM     & Max. &  272.45 &  315.30 \\
            & Min. & -277.82 & -319.15 \\
            & Rms. &   73.94 &   81.03 \\ \hline
Dust        & Max. &  647.88 &  626.26 \\
            & Min. & -316.68 & -354.97 \\ 
            & Rms. &  174.15 &  173.72 \\ \hline
T-SZ        & Max. &   75.08 &   88.97 \\
            & Min. &    0.00 &  -59.48 \\
            & Rms. &    4.82 &   15.40 \\ \hline
Free-free   & Max. &    2.04 &    4.96 \\
            & Min. &   -1.61 &   -9.42 \\
            & Rms. &    0.67 &    2.74 \\ \hline
Synchrotron & Max. &    0.23 &    1.70 \\
            & Min. &   -0.24 &   -0.94 \\
            & Rms. &    0.08 &    0.51 \\ \hline \hline
\end{tabular}
\end{center}
\caption{Results from the Singular Valued Decomposition analysis of
simulated data taken by the Planck Surveyor for a string CMB signal
and CDM CMB signal.}
\label{ta:svds}
\end{table}

\begin{table}
\begin{center}
\begin{tabular}{|l|c|c|} \hline
Component & Residual error ($\mu$K) & Gradient \\ \hline
CMB (CDM)        & 37 & $1.02 \pm 0.01$ \\
CDM (strings)    & 37 & $0.78\pm 0.01$ \\
Thermal SZ       & 19 & $0.57\pm 0.02$ \\
Dust             & 12 & $0.966\pm 0.002$ \\
Free-Free        & 2.5 & $1.45\pm 0.01$ \\
Synchrotron      & 0.5 & $2.28\pm 0.13$ \\ \hline
\end{tabular}
\end{center}
\caption{The rms of the residuals and the gradients of the best-fit
straight line through the origin for the comparison plots in the SVD
reconstructions shown in Figs~\ref{fig:svd} and \ref{fig:svds}.}
\label{ta:ermssvd}
\end{table}

\begin{figure}[p!]
\caption{The results from the S.V.D. analysis of simulated data taken
by the Planck Surveyor for a CDM CMB signal. The maps have been convolved
with a $4.5\arcmin$ Gaussian beam as this is the lowest resolution
that the experiment is sensitive to.}
\label{fig:svd}
\end{figure}

\begin{figure}[p!]
\caption{The results from the S.V.D. analysis of simulated data taken
by the Planck Surveyor for a string CMB signal.}
\label{fig:svds}
\end{figure}

The residuals are a rather crude method of quantifying the accuracy of
the reconstructions. A more useful technique is to look at the
amplitude of the reconstructed maps as compared to the input maps. 
Usually this plot consists of a collection of points but to make
things clearer three contour levels are plotted. The 68\%, 95\% and
99\% distribution of the reconstructed amplitudes are shown as a
function of the input amplitude. 
These plots can be used to give an estimate on the accuracy of the 
reconstructed maps. A perfect reconstruction would be a diagonal line
with unit gradient. Figures~\ref{fig:svdint} and \ref{fig:svdsint}
show the plots for the SVD analysis. The spread of points in this plot
do not respond to the respective residual values for each map as the
residual calculation also takes into account how far away from the
diagonal the points are whereas this plot only shows the deviation away
from the best fit line. The gradient of the best fit line is shown in
Table~\ref{ta:ermssvd}. From this table it is easily seen that the
dust and CDM realisation are reproduced quite well (the gradient is close 
to unity). However, it is seen
that the reconstruction of the CDM realisation of the CMB is more
accurate than that of the strings realisation. This is due to the extra
Gaussian features from the noise that are introduced during the SVD
analysis having a larger effect on the non-Gaussian string
reconstruction than on the Gaussian CDM reconstruction. It is clear
that other methods of analysis should be investigated to obtain better
results. 

\begin{figure}[p!]
\caption{A comparison between the input flux in each pixel and the
output flux in that pixel for the SVD reconstructed maps with full
correlation information for the CDM model of the CMB.
A perfect reconstruction would be a
diagonal line. The three contours enclose 68\%, 95\% and 99\% of the
pixels. 
If no contours are seen then no reconstruction was possible.}
\label{fig:svdint}
\end{figure}

\begin{figure}[p!]
\caption{A comparison between the input flux in each pixel and the
output flux in that pixel for the SVD reconstructed maps 
for the string model of the CMB.}
\label{fig:svdsint}
\end{figure}

\subsection{MEM and Wiener reconstructions}
\label{cobrasmem}

In the previous chapter it was seen that, in
the Fourier domain, it is possible to give the MEM and Wiener 
algorithms either the full
correlation matrix of the components, an estimate of the correlation
matrix, or no information on the spatial distribution (expect for a
starting guess on the total power in the map). 
Two different levels of information were tested. The first
gives the methods the full correlation matrix (including
cross-correlations) by using the input maps to construct the average
correlations. The second assumes that nothing about the spatial
distribution is known and the correlation matrix is assumed to be
flat. These two cases represent the two extremes that 
the real analysis of Planck Surveyor satellite data will take. They are
presented as useful constraints and the actual analysis performed will
be somewhere between the results presented here. For MEM, the
Bayesian $\alpha$ was found for each of the cases and the model was
updated between iterations until convergence was reached (less than
5\% change in any of the pixel flux levels). With no correlation information the
ICF was also updated between iterations. For Wiener no update was
attempted as classic Wiener has a tendency to suppress power and so
updating would cause the reconstruction to tend to zero. 

The results of the MEM and Wiener analyses can be seen in
Figures~\ref{fig:memfft}-\ref{fig:wiennoicfsint}. Each analysis is
grouped into four figures. The first and second show the
reconstructions of the MEM and Wiener algorithms. The third and fourth show the
accuracy of the reconstruction (similarly to the SVD analysis). 
Each of the reconstructions have been
convolved to the experimental resolution ($4.5\arcmin$). 
A comparison of these figures with the true input
components in Figures~\ref{fig:cobinp} and \ref{fig:cobinp2} 
clearly shows that the dominant input
components (i.e. the CMB and the dust emission) 
are faithfully reconstructed in all cases. Table~\ref{ta:resultwvm} 
shows the range and {\em rms} of the reconstructions. 
As seen MEM more closely follows the data than
Wiener in all cases and the difference is more marked in the
non-Gaussian components. Table~\ref{ta:ermsmem} shows the residuals
for each of the maps given by Equation~\ref{eq:erms}.

\begin{table}
\begin{center}
\begin{tabular}{|cc|rrrrr|} \hline \hline
Component & & \multicolumn{1}{c}{(a)} & \multicolumn{1}{c}{(b)} &
\multicolumn{1}{c}{(c)} & \multicolumn{1}{c}{(d)} & \multicolumn{1}{c|}{(e)} \\ \cline{3-7}
& & \multicolumn{5}{c|}{$\Delta T$ in $\mu$K} \\ \hline \hline
CMB strings & Max. &  305.53 &  298.34 &  285.16 &  280.56 &  255.98 \\
            & Min. & -338.20 & -328.64 & -327.33 & -330.10 & -306.04 \\
            & Rms. &   69.43 &   69.25 &   68.97 &   69.04 &   63.34
\\ \hline
CMB CDM     & Max. &  272.45 &  275.18 &  276.44 &  277.54 &  247.80 \\
            & Min. & -277.82 & -290.56 & -289.59 & -291.99 & -255.88 \\
            & Rms. &   73.94 &   73.88 &   73.70 &   73.78 &   67.13
\\ \hline
Dust        & Max. &  647.88 &  645.63 &  645.37 &  643.94 &  646.43 \\
            & Min. & -316.68 & -320.87 & -321.22 & -324.32 & -317.14 \\
            & Rms. &  174.15 &  174.15 &  174.14 &  174.15 &  174.02
\\ \hline
K-SZ        & Max. &    5.16 &    0.87 &    0.84 &  -      &  -      \\
            & Min. &  -12.23 &   -1.64 &   -1.35 &  -      &  -      \\
            & Rms. &    0.86 &    0.25 &    0.20 &  -      &  -
\\ \hline
T-SZ        & Max. &   75.08 &   53.16 &   39.52 &   48.22 &   17.26 \\
            & Min. &    0.00 &   -5.01 &    1.05 &   -3.56 &    1.06 \\
            & Rms. &    4.82 &    4.29 &    2.84 &    4.22 &    1.72
\\ \hline
Free-free   & Max. &    2.04 &    2.16 &    1.44 &    0.96 &    0.00 \\
            & Min. &   -1.61 &   -1.72 &   -0.73 &   -0.85 &   -0.00 \\
            & Rms. &    0.67 &    0.57 &    0.50 &    0.39 &    0.00
\\ \hline
Synchrotron & Max. &    0.23 &    0.24 &    0.14 &    0.02 &    0.00 \\
            & Min. &   -0.24 &   -0.25 &   -0.10 &   -0.02 &   -0.00 \\
            & Rms. &    0.08 &    0.09 &    0.04 &    0.08 &    0.00
\\ \hline \hline
\end{tabular}
\end{center}
\label{ta:resultwvm}
\caption{Results from the Planck simulations. (a) are the input
values, (b) are the reconstructed values from the full MEM with ICF
information, (c) are the reconstructed values from the full Wiener
filter with ICF information, (d) are the reconstructed values from the
full MEM with no ICF information and (e) are the reconstructed values
from the full Wiener filter with no ICF information.}
\end{table}

\begin{table}
\begin{center}
\begin{tabular}{|l|c|c|c|c|} \hline
Component & (a) ($\mu$K) & (b) ($\mu$K) & (c) ($\mu$K) & (d) ($\mu$K) \\ \hline
CMB (CDM)        & 5.9 & 6.0 & 6.1 & 7.5 \\
CMB (strings)    & 6.2 & 6.4 & 7.2 & 10.2 \\
Kinetic SZ       & 0.9 & 0.9 & - & -\\
Thermal SZ       & 3.9 & 4.1 & 4.4 & 4.6 \\
Dust             & 1.6 & 1.9 & 1.9 & 2.1 \\
Free-Free        & 0.3 & 0.4 & 0.4 & 0.5 \\
Synchrotron      & 0.1 & 0.1 & 0.1 & 0.1 \\ \hline
\end{tabular}
\end{center}
\caption{The rms of the residuals in the a) MEM with full ICF
information, b) Wiener with full ICF information, c) MEM with no ICF
information and d) Wiener with no ICF information 
reconstructions.}
\label{ta:ermsmem}
\end{table}

Perhaps most importantly, the
CMB has been reproduced extremely accurately (to within $6\mu$K). 
The residual errors on the MEM reconstruction of the CMB, 
as compared to the Wiener reconstruction, are slightly smaller.
The residual errors for the other components are similar for the MEM
and Wiener reconstructions, but are always slightly lower for the MEM
algorithm. The difference between the two methods is seen to be 
greatest for the components that are known to be non-Gaussian in nature (dust,
free-free, synchrotron and the thermal SZ effect). 
There is little
difference between the reconstructions of the kinetic SZ (see below for 
discussion on the SZ reconstructions). 
For the full ICF information case 
the free-free emission, which is highly
correlated with the dust, has been reconstructed reasonably well,
containing most of the main features present in the true input
map. It is noted that the Wiener reconstruction of the free-free
emission follows the dust
emission more closely than the MEM reconstruction which has also reconstructed
some free-free emission that is uncorrelated with the dust. For the case of
no prior ICF information it is seen that no significant reconstruction of the
free-free emission is made with the
Wiener analysis whereas a smooth version has been reconstructed
by MEM. The recovery of the synchrotron emission is less
impressive. With full information MEM and Wiener can reconstruct the synchrotron
to some degree but the significance of the features is not very high
(this is seen in the very broad correlations between input amplitude
and reconstructed amplitude in Figure~\ref{fig:memfftint}). 

Again, the residual errors on the MEM and Wiener reconstructions do
not show all of the information available about the
reconstructions. The extent of the deviation of the correlation plots
away from the diagonal also contains
information. Table~\ref{ta:memgrad} shows the gradients of the best
fit lines for the MEM and Wiener reconstructions with full and no
prior ICF information. From these best fit lines it is seen that the
CMB and dust emission are both reconstructed very well. However, it is
seen that MEM always outperforms the Wiener on the CMB reconstruction
when no prior ICF information was given. This is due to Wiener underestimating
the temperature of each pixel in the CMB channel. 
It is clearly seen that the thermal SZ is
consistently reconstructed with a lower amplitude than the true
amplitude. Even though the Wiener filter has been given the full prior
ICF information it is still outdone by MEM with no prior information
in the case of this highly non-Gaussian effect. The range of values 
reconstructed by Wiener is lower than that for MEM (the spread is smaller
around the best fit line), however, the residual errors for the Wiener are
always larger than those for MEM 
because of the smaller amplitudes reconstructed. 
The fit for MEM and Wiener can be improved by assuming the reconstruction is
at a lower resolution (which is likely to be the case as there is
little information on the SZ at the highest frequencies where the
resolution is highest). 
For example, the reconstruction of the thermal SZ effect has a much stronger
correlation with the input maps if it is assumed that the resolution
of the data is $10.6\arcmin$ (the resolution of the frequency channel
where the thermal SZ effect is most dominant in the Planck Surveyor
data). At this resolution the error is then $3\mu$K. 

\begin{table}
\begin{center}
\begin{tabular}{|l|c|c|c|c|} \hline
 & \multicolumn{2}{c|}{Full prior ICF} & \multicolumn{2}{c|}{No prior
ICF} \\ \cline{2-5}
Component        & MEM & Wiener & MEM & Wiener  \\ \hline
CMB (CDM)        & $1.00$ & $1.00$ & $1.00$ & $0.97$ \\
CDM (strings)    & $0.98$ & $0.97$ & $0.97$ & $0.89$ \\
Thermal SZ       & $0.55$ & $0.27$ & $0.50$ & $0.24$ \\
Kinetic SZ       & $0.07$ & $0.05$ & - & - \\
Dust             & $1.00$ & $1.00$ & $1.00$ & $1.00$ \\
Free-Free        & $0.48$ & $0.37$ & $0.60$ & $0.22$ \\
Synchrotron      & $0.62$ & $0.44$ & $0.47$ & $0.13$ \\ \hline
\end{tabular}
\end{center}
\caption{The gradients of the best-fit straight line through the
origin for the correlation plots of the reconstructions.}
\label{ta:memgrad}
\end{table}

\begin{figure}[p!]
\caption{The six reconstructed channels from the MEM analysis of the
Planck simulated data for a CDM model of the CMB 
using full correlation information.}
\label{fig:memfft}
\end{figure}

\begin{figure}[p!]
\caption{The six reconstructed channels from the Wiener analysis of the
Planck simulated data for a CDM model of the CMB 
using full correlation information.}
\label{fig:wien}
\end{figure}

\begin{figure}[p!]
\caption{A comparison between the input flux in each pixel and the
output flux in that pixel for the MEM reconstructed maps with full
correlation information for the CDM model of the CMB.}
\label{fig:memfftint}
\end{figure}

\begin{figure}[p!]
\caption{A comparison between the input flux in each pixel and the
output flux in that pixel for the Wiener reconstructed maps with full
correlation information for the CDM model of the CMB.}
\label{fig:wienint}
\end{figure}

\begin{figure}[p!]
\caption{The six reconstructed channels from the MEM analysis of the
Planck simulated data for a strings model of the CMB 
using full correlation information.}
\label{fig:memffts}
\end{figure}

\begin{figure}[p!]
\caption{The six reconstructed channels from the Wiener analysis of the
Planck simulated data for a strings model of the CMB 
using full correlation information.}
\label{fig:wiens}
\end{figure}

\begin{figure}[p!]
\caption{A comparison between the input flux in each pixel and the
output flux in that pixel for the MEM reconstructed maps with full
correlation information for the strings model of the CMB.}
\label{fig:memfftsint}
\end{figure}

\begin{figure}[p!]
\caption{A comparison between the input flux in each pixel and the
output flux in that pixel for the Wiener reconstructed maps with full
correlation information for the strings model of the CMB.}
\label{fig:wiensint}
\end{figure}

\begin{figure}[p!]
\caption{The six reconstructed channels from the MEM analysis of the
Planck simulated data for a CDM model of the CMB 
using no correlation information. The plot shows the kinetic SZ
(figure (b)) for ease of comparison even though no attempt was made to
reconstruct this effect.}
\label{fig:memnoicf}
\end{figure}

\begin{figure}[p!]
\caption{The six reconstructed channels from the Wiener analysis of the
Planck simulated data for a CDM model of the CMB 
using no correlation information.}
\label{fig:wiennoicf}
\end{figure}

\begin{figure}[p!]
\caption{A comparison between the input flux in each pixel and the
output flux in that pixel for the MEM reconstructed maps with no
correlation information for the CDM model of the CMB.}
\label{fig:memnoicfint}
\end{figure}

\begin{figure}[p!]
\caption{A comparison between the input flux in each pixel and the
output flux in that pixel for the Wiener reconstructed maps with no
correlation information for the CDM model of the CMB.}
\label{fig:wiennoicfint}
\end{figure}

\begin{figure}[p!]
\caption{The six reconstructed channels from the MEM analysis of the
Planck simulated data for a strings model of the CMB 
using no correlation information.}
\label{fig:memnoicfs}
\end{figure}

\begin{figure}[p!]
\caption{The six reconstructed channels from the Wiener analysis of the
Planck simulated data for a strings model of the CMB 
using no correlation information.}
\label{fig:wiennoicfs}
\end{figure}

\begin{figure}[p!]
\caption{A comparison between the input flux in each pixel and the
output flux in that pixel for the MEM reconstructed maps with no
correlation information for the strings model of the CMB.}
\label{fig:memnoicfsint}
\end{figure}

\begin{figure}[p!]
\caption{A comparison between the input flux in each pixel and the
output flux in that pixel for the Wiener reconstructed maps with no
correlation information for the strings model of the CMB.}
\label{fig:wiennoicfsint}
\end{figure}

Visually, the string realisation of the CMB 
appears to be reconstructed very accurately with all the non-Gaussian 
features still apparent in all cases. 
Comparing the residual errors from the string realisation with that of the 
CDM realisation, it is seen that even when MEM is used the non-Gaussian 
process is reconstructed slightly less accurately (a 10\% increase 
in the residual errors in both cases considered). 
However, when the Wiener filter is applied the 
difference is much more enhanced (a 10\% and 50\% increase in the residual errors for 
the case of full and no prior correlation information respectively).
In all cases the gradient of the MEM reconstruction is closer to unity than 
that of Wiener, but it is seen that the Gaussian realisation is more 
accurately reconstructed.
 
As a test of the power of MEM the frequency dependence of various
components was also checked. There was an obvious minimum in the value
of $\chi^2$ for the actual values of the dust temperature and
emissivity as there is a lot of information on the dust emission from
the higher frequency channels. The free--free and synchrotron spectral
indices were less tightly constrained as the $\chi^2$ depends mainly
on the CMB and dust emissions. It should therefore be possible to
constrain the spatial and frequency dependence of the dust emission
but more assumptions, or higher sensitivity, is required to constrain
the other Galactic components.

\subsection{SZ reconstruction}
\label{szrec}

The thermal SZ effect can be used to calculate the value of $H_\circ$ 
(Grainge {\em et al} 1993, Saunders 1997 etc.). 
However, for this to be possible the shape of the cluster
must be known. Any algorithm used to reconstruct the information about
the thermal SZ effect must closely follow the true shape of the underlying
cluster and not just reconstruct the {\em rms} or the
amplitude. It was seen in the reconstructed maps that the MEM
algorithm reconstructs the thermal SZ effect in more clusters than the
Wiener filter. With no information on the correlation function the
Wiener filter performs very badly in reconstructing the non-Gaussian
emission. Figure \ref{fig:szl} shows some of the typical thermal SZ
profiles reconstructed by the MEM and Wiener algorithms with full
and no correlation information. They are compared with the
input profiles convolved with a Gaussian of size $10.6\arcmin$, the
resolution of the 100 GHz channel (the frequency at which the fractional
contribution of the thermal SZ effect is largest). 
It is easily seen that the full
MEM analysis does reconstruct the cluster profiles closer to their
truer shape than the Wiener filter. Thus, as expected, the Gaussian
assumption of the Wiener filter leads to poor reconstructions of
highly non-Gaussian fields as compared with MEM. It is also seen that
even with no prior information on the correlations, the MEM algorithm
still reconstructs a reasonable approximation to the true profiles.

At first sight it appears that the MEM reconstructions contain some spurious
features as compared to the input profiles. This is seen most dramatically in
the top panel of Figure~\ref{fig:szl} for the full prior ICF case. The 
central cluster appears to have an extra feature on the right hand side. 
In fact, this phenomenon illustrates the care that should be taken when 
interpreting SZ profiles found this way, since the feature is actually present
in the input map, but has almost been smoothed out by the $10.6\arcmin$
convolution. The reason it is still present in the reconstruction is that the
effective resolution of the MEM and Wiener reconstructions can vary across
the map, depending on the level of the pixel noise and the other components. 
Therefore, in some regions a degree of super-resolution is possible whereas in
regions where the pixel noise, or emission from the other components, is large
the super-resolution does not occur. This was tested with different pixel
noise realisations and it was seen that the areas of 
super-resolution did indeed move across the map. 

\begin{figure}
\caption{Typical reconstructions of SZ profiles. The solid line is the
input cluster convolved to $10.6\arcmin$ (the resolution at which the
sensitivity to the SZ effect is a maximum). The dashed line is the
reconstruction from the MEM analysis using (a) full correlation
information and (b) no correlation information. 
The dotted line is the reconstruction from the Wiener
analysis using (a) full correlation information and (b) no correlation
information. As can be seen the Wiener
result is a much smoother reconstruction than the MEM result,
especially in the case with no correlation information when the Wiener
does not reconstruct any significant features but the MEM still
follows the true profiles very well.}
\label{fig:szl}
\end{figure}

The kinetic SZ can be used to put constraints on the interactions of
galaxy clusters through their velocity distributions. No reconstruction
was attempted when no information about the correlations was given, as
the frequency spectrum of the kinetic SZ is the same as that for the
CMB. Even in the case when full correlation information was given the
kinetic SZ emission is not recovered particularly well. The reconstruction as
compared to the input kinetic SZ emission is shown in
Figure~\ref{fig:ksz}. Comparing the spatial location of the recovered
kinetic SZ profiles to that of the thermal SZ it is seen that only the
large kinetic SZ effects that occur on pixels also associated with a
large thermal SZ are reconstructed. Therefore, it is seen that the
recovery of the kinetic SZ is highly dependent on the information
given to the MEM algorithm prior to the analysis and the data is not
sufficiently sensitive to allow the kinetic SZ to be extracted by
itself. This is easily seen as the largest kinetic SZ effect (in the
lower right quadrant of the figures) is not reconstructed as it is not
associated with a large thermal SZ effect.
For a better reconstruction extra information is
required, as well as the frequency and average spatial spectra. This
may come in the form of the positions of the clusters (through the
thermal SZ effect) and then using the information that the SZ effect
occurs mainly on angular scales below any contribution from the
CMB. However, this is very difficult to incorporate in an automatic
algorithm and so no further analysis was attempted. 

\begin{figure}
\caption{Comparison between the input simulation of the kinetic SZ
effect convolved to $23\arcmin$ (the resolution at which the
fractional contribution of the kinetic SZ is highest in the data) and
the MEM reconstruction of the effect.}
\label{fig:ksz}
\end{figure}

\subsection{Power spectrum reconstruction}
\label{powerspec}

Figures~\ref{fig:powermem} and \ref{fig:powerwien} show the
azimuthally averaged 
power spectra for the MEM and Wiener reconstructions with full prior 
correlation information
respectively. The 68\% confidence intervals obtained from the inversion
of the Hessian are
also shown. It is seen that the 68\% confidence intervals always encompass the
input power spectra. It should be noted that the confidence intervals shown are
for the reconstructed maps given that the data is representative of
the full sky and do not take into account sample or cosmic variance.
For the components with the most amount of information (namely the 
CMB and dust) the confidence intervals are correspondingly
smaller whereas for components with little information (as in the case
of synchrotron) the confidence intervals are much 
larger. It is seen that the MEM and Wiener confidence 
intervals are fairly similar but this is to be expected as the
Hessian calculation for both comes from a Gaussian assumption.

The reconstruction of the CMB is very good out to an $\ell$ of
about 1500 for both MEM and Wiener. Beyond $\ell \sim 1500$ Wiener
begins to visibly underestimate the true power but MEM is still
accurate to $\ell \sim 2000$. The dust emission is
reconstructed very well out to an $\ell$ of about 3000 for the MEM
reconstruction and 2000 for the Wiener reconstruction, the extra
information over the CMB coming from the increased resolution at the higher
frequencies. The white noise spectra of the thermal SZ and kinetic SZ is much more
closely followed by the MEM reconstruction although both algorithms
have lost some of the power from the free-free and synchrotron
reconstructions. 

\begin{figure}[p!]
\caption{The reconstructed power spectra (dark line) with errors
compared to the input power spectra (faint line) of the Planck
Surveyor CDM CMB simulation. The errors are calculated using the Gaussian
approximation for the peak of the probability distribution. 
This reconstruction was produced by MEM with full ICF
information.} 
\label{fig:powermem}
\end{figure}

\begin{figure}[p!]
\caption{The reconstructed power spectra (dark line) with errors
compared to the input power spectra (faint line) of the Planck
Surveyor CDM CMB simulation. The errors are calculated using the Gaussian
approximation for the full probability distribution. 
This reconstruction was produced by Wiener filtering with full ICF
information.} 
\label{fig:powerwien}
\end{figure}

For the case with no prior ICF information the difference between the MEM and
Wiener reconstruction is more easily seen. The power spectrum
reconstructions for this case are seen in
Figure~\ref{fig:powermemnoicf} and \ref{fig:powerwiennoicf}. Even in the CMB
reconstruction it is seen that the Wiener-produced power spectrum is
consistently below the true spectrum even at very low $\ell$. The MEM
reconstruction is accurate out to $\ell \sim 1500$ at which point it
drops rapidly to zero. This rapid drop, seen in most of the
reconstructions, is a result of continually updating the ICF in an
area for which there is little information in the data. For the
thermal SZ it is seen that the MEM reconstruction is reasonably
accurate out  to $\ell \sim 1000$ but the Wiener filter is only
reasonably accurate out to $\ell \sim 200$. No reconstruction of the
kinetic SZ was attempted. The dust power spectrum reconstruction is
accurate out to $\ell \sim 2000$ for MEM and $\ell \sim 3000$ for
Wiener. However, it is seen that the Wiener reconstruction has a spurious bump in the
power spectrum at $\ell \sim 2000 - 5000$ which
overestimates the power for the CMB and this is again seen for the
dust. Thus it is unclear whether the reconstruction from the 
Wiener filter beyond $\ell \sim 2000$ is accurate.
Out to $\ell \sim 70$ the MEM
reconstruction approximates the free-free true power spectrum but for
Wiener, and the two reconstructions of the synchrotron emission, the
power spectrum is always underestimated. 

\begin{figure}[p!]
\caption{The reconstructed power spectra (dark line) with errors
(dotted line)
compared to the input power spectra (faint line) of the Planck
Surveyor CDM CMB simulation. The errors are calculated using the Gaussian
approximation for the peak of the probability distribution. 
This reconstruction was produced by MEM with no ICF
information.} 
\label{fig:powermemnoicf}
\end{figure}

\begin{figure}[p!]
\caption{The reconstructed power spectra (dark line) with errors
(dotted line)
compared to the input power spectra (faint line) of the Planck
Surveyor CDM CMB simulation. The errors are calculated using the Gaussian
approximation for the full probability distribution. 
This reconstruction was produced by Wiener filtering with no ICF
information.} 
\label{fig:powerwiennoicf}
\end{figure}

\section{The MAP simulations}
\label{mapan}

Four years prior to the launch of the Planck Surveyor NASA is due to
launch its MAP satellite. This is a cheaper alternative to Planck,
with less frequency coverage and lower angular resolution. As a test
of the relative strength of the two experiments the same analysis that
was performed on the Planck Surveyor was performed on the MAP
satellite. The same simulations as 
the Planck analysis were used so that a direct comparison between the two
satellites would be possible. The input maps for MAP were the same as
those for the Planck, although the resolution of MAP is about 4 times
worse than that of Planck and so the results are expected to be
smoother (see Figure~\ref{fig:mapinps}). The resolution of the MAP 
satellite has improved considerably since these simulations were performed 
and the latest design has a best resolution of about 2 
times that of Planck (see Jones, Hobson, Lasenby \& Bouchet 1998 for 
simulations with the current design).
The true test of the sensitivity of MAP is again in the
flux reconstruction of features that it is sensitive to. A fit to the
six channels was again attempted. Only the CDM model of the CMB is
shown here as the string simulations show very similar results.

\begin{figure}[p!]
\caption{The six input maps convolved to the highest resolution of the
MAP satellite for comparison with the reconstructions. No
reconstruction of the kinetic SZ was attempted as the signal at this
resolution is negligible but it is still plotted to allow easy
comparison with the Planck results.}
\label{fig:mapinps}
\end{figure}

\subsection{MEM and Wiener results}
\label{mapmem}

The MEM and Wiener analyses of the MAP simulations were carried out
with the same levels of information as the Planck analysis.  
The results of the analyses can be seen in
Figures~\ref{fig:memfftmap}-\ref{fig:wiennoicfintmap}. Again, each analysis is
grouped into four figures. The first and second show the
reconstructions of the MEM algorithm and the Wiener filter
respectively. The third and fourth show the
accuracy of the reconstruction.

It is easily seen that there is no information in the MAP data
on any of the foregrounds that the Planck Surveyor is sensitive
to. Table~\ref{ta:resultmap} shows the {\em rms} of the reconstructions
from the simulations.
The lower frequency coverage of MAP results in no information
available on the dust emission (this is seen when no correlation
information is given and the reconstruction is just set to zero by
both MEM and Wiener). Even though the low frequency channels should
contain information on the free-free and synchrotron emissions none is
recovered. This is due to the lower resolution and sensitivity that
MAP has. MAP was never designed to extract information on the SZ
effect and does not contain channels near the critical 217~GHz
frequency. Hence, no information on either of the SZ effects 
was reconstructed. However,
the CMB is reconstructed fairly well. There is a very strong
correlation for both the MEM and Wiener filter. Little difference is
seen between these two reconstructions as the resolution is not large
enough to pick them out and everything appears Gaussian (this was true
in both the CDM and string simulations; the level of Gaussianity in
the string simulation at this resolution is tested in
Chapter~\ref{chap8}). The 68\% confidence intervals on the CMB
reconstruction are $19\mu$K and $29\mu$K for the reconstructions with
full correlation information and no correlation information
respectively. The gradient of the best-fit lines for the correlation
plots are 0.97 and 0.91 for
the full prior ICF information and no prior ICF information
respectively. It is seen that the
reconstructions with full prior correlation information are more accurate than
those with no assumed correlations by $10\mu$K. 
Therefore, a much better reconstruction
of the CMB can be achieved with added information on its spatial
distribution in contrast to the Planck Surveyor simulations where
there was enough information in the data to reconstruct the CMB at a
high degree of accuracy 
(the full and no prior correlation information reconstructions having a 68\%
confidence interval of $6\mu$K and $7\mu$K respectively). 
The Planck Surveyor appears to be three times
more sensitive to the CMB than MAP (although the sensitivity and
resolution of both satellites is not finalised). 

\begin{figure}[p!]
\caption{The six reconstructed maps from the MEM analysis of the
MAP simulated data for a CDM model of the CMB 
using full correlation information.}
\label{fig:memfftmap}
\end{figure}

\begin{figure}[p!]
\caption{The six reconstructed maps from the Wiener analysis of the
MAP simulated data for a CDM model of the CMB 
using full correlation information.}
\label{fig:wienmap}
\end{figure}

\begin{figure}[p!]
\caption{A comparison between the input flux in each pixel and the
output flux in that pixel for the MEM reconstructed maps with full
correlation information for the CDM model of the CMB. The contours are
as before.}
\label{fig:memfftintmap}
\end{figure}

\begin{figure}[p!]
\caption{A comparison between the input flux in each pixel and the
output flux in that pixel for the Wiener reconstructed maps with full
correlation information for the CDM model of the CMB.}
\label{fig:wienintmap}
\end{figure}

%
%
%

\begin{figure}[p!]
\caption{The six reconstructed maps from the MEM analysis of the
MAP simulated data for a CDM model of the CMB 
using no correlation information.}
\label{fig:memnoicfmap}
\end{figure}

\begin{figure}[p!]
\caption{The six reconstructed maps from the Wiener analysis of the
MAP simulated data for a CDM model of the CMB 
using no correlation information.}
\label{fig:wiennoicfmap}
\end{figure}

\begin{figure}[p!]
\caption{A comparison between the input flux in each pixel and the
output flux in that pixel for the MEM reconstructed maps with no
correlation information for the CDM model of the CMB.}
\label{fig:memnoicfintmap}
\end{figure}

\begin{figure}[p!]
\caption{A comparison between the input flux in each pixel and the
output flux in that pixel for the Wiener reconstructed maps with no
correlation information for the CDM model of the CMB.}
\label{fig:wiennoicfintmap}
\end{figure}

%
%
%

\begin{table}
\begin{center}
\begin{tabular}{|cc|rrrrr|} \hline
Component & & \multicolumn{1}{c}{(a)} & \multicolumn{1}{c}{(b)} &
\multicolumn{1}{c}{(c)} & \multicolumn{1}{c}{(d)} & \multicolumn{1}{c|}{(e)} \\ \cline{3-7}
& & \multicolumn{5}{c|}{$\Delta T$ in $\mu$K} \\ \hline
CMB CDM     & Max. &  175.37 &  151.39 &  189.37 &  148.59 &  266.65 \\
            & Min. & -171.30 & -178.33 & -222.52 & -165.08 & -318.52 \\
            & Rms. &   55.14 &   54.63 &   61.79 &   51.11 &   80.38 \\
Dust        & Max. &  532.67 &  177.51 &  186.61 &    3.59 &   36.75 \\
            & Min. & -273.58 & -216.32 & -152.90 &   -3.59 &  -36.31 \\
            & Rms. &  163.87 &   74.27 &   63.03 &    1.46 &   10.20 \\
T-SZ        & Max. &   22.76 &   23.28 &   19.46 &    7.08 &   14.46 \\
            & Min. &    0.00 &   -1.73 &    2.47 &    3.68 &  -11.60 \\
            & Rms. &    2.04 &    5.82 &    4.08 &    0.78 &    3.58 \\
Free-free   & Max. &    1.54 &    0.72 &    0.74 &    0.79 &    0.84 \\
            & Min. &   -1.42 &   -0.54 &   -0.55 &   -0.65 &   -0.66 \\
            & Rms. &    0.63 &    0.24 &    0.24 &    0.27 &    0.22 \\
Synchrotron & Max. &    0.21 &    0.10 &    0.09 &    0.06 &    0.09 \\
            & Min. &   -0.22 &   -0.07 &   -0.07 &   -0.05 &   -0.07 \\
            & Rms. &    0.08 &    0.03 &    0.03 &    0.02 &    0.02 \\ \hline
\end{tabular}
\end{center}
\caption{Results from the MAP simulations. (a) are the input
values, (b) are the reconstructed values from the full MEM with ICF
information, (c) are the reconstructed values from the full Wiener
filter with ICF information, (d) are the reconstructed values from the
full MEM with no ICF information and (e) are the reconstructed values
from the full Wiener filter with no ICF information.}
\label{ta:resultmap}
\end{table}

\section{MEM and Wiener: the conclusions}
\label{trashwien}

The simulations of the Planck Surveyor and MAP satellite data were
analysed by both MEM and Wiener filtering. It was found that there
were no differences between the two in the case of a fully Gaussian
data set (see the MAP analysis where all non-Gaussian effects were
negligible). However, when there was a non-Gaussian effect present,
whether in a foreground or in the map to be reconstructed, there was a
significant improvement when MEM was used as opposed to Wiener. 
The difference between the reconstructions using full and no prior 
information was less marked for the MEM analysis than for the Wiener analysis.
This is a very useful property because 
a smaller range of possible reconstructions means that the analysis
is less sensitive to the prior information (which may be incorrect in the real
analysis). It has
been shown that the Wiener approach can be considered as the quadratic
approximation to MEM and so even in the Gaussian case Wiener will be
no better than MEM. Therefore, MEM will be used to analyse the
Tenerife and Jodrell data in the next Chapter. The following section
will test the power of MEM to analyse the Tenerife data set.

\section[Tenerife simulations]{Testing the MEM algorithm on the Tenerife data set}
\label{sim}

Before applying the MEM algorithm to the real data, simulations were carried
out to test its performance. Two-dimensional sky maps were simulated using a
standard cold, dark matter model (${H_o}=50\kms$, $\Omega_b=0.1$)
with an {\em rms} signal of $22\mu$K per pixel
(normalised to COBE second year data, ${Q_{rms-PS}}=20.3\mu$K, see 
Tegmark \& Bunn 1995). Observations from the sky maps were
then simulated by convolving them with the Tenerife $8\deg$ FWHM
beam. Before noise was
added the positive/negative algorithm was tested by analysing the data 
and then changing the sign of the data and reanalysing
again. In both cases the same, but inverted, reconstruction was found for the MEM
output and so it is concluded that this 
method of two positive channels introduces
no biases towards being positive or negative. Various 
noise levels were then added to 
the scans before reconstruction with MEM. 
The two noise levels considered here
are $100\mu$K and $25\mu$K on the data scans,
which represent the two extrema of the 
data that are expected from the various Tenerife configurations
($100\mu$K for the 10 GHz, FWHM=$8\dg$ data and $25\mu$K for the 15 and
33 GHz, FWHM=$5\dg$ data). 

Figure~\ref{fig:simscan} shows the convolution of one of the simulations 
with the Tenerife beam and the result obtained from MEM with the two 
noise levels. The plots are averaged over 30 simulations and the bounds
are the $68\%$ confidence limits (simulation to simulation variation). As seen,
MEM recovers the underlying sky simulation to good accuracy for both 
noise levels, with the $25\mu$K result the better of the two as 
expected. Figure~\ref{fig:3plot} 
shows the reconstructed intrinsic sky from two of the simulations after 60
Newton--Raphson iterations of the MEM algorithm  
as compared with the real sky simulations 
convolved in an $8\deg$ Gaussian beam. Various common features are seen in 
the three scans like the maxima at ~RA $150\dg$, minima at ~RA $170\dg$
and the partial ring feature between RA $200\dg$ and $260\dg$ with
central minima at ~RA $230\dg$, ~Dec. $+35\dg$.
All features larger than the {\em rms} are reconstructed in both
the $25\mu$K and $100\mu$K noise simulations. However, there is a some
freedom in the algorithm to move these features within a beam
width. This can cause spurious features to appear at the edge of the
map when the guard region (about $5\dg$) around the map contains a
peak (this can be seen in the map as a decaying tail away from the edge).
For example, the feature at RA $230\dg$, Dec $50\dg$ has been
moved down by a few degrees out of the guard region
in the $100\mu$K noise simulation so it
appears more prominently on the ring feature.

\begin{figure}
\caption{The  solid line shows the sky simulation convolved
with the Tenerife
$8\dg$ experiment. The bold dotted line in the top figure 
shows the MEM reconstructed sky after
reconvolution with the Tenerife beam, averaged over 
simulations of the  MEM output from a simulated experiment with $25\mu$K
Gaussian noise added to each scan. Also shown are the $68\%$ confidence limits 
(simulation to simulation variation; dotted lines) on this
reconvolution. The bold dotted line in the bottom figure 
shows the reconvolution averaged over 
simulations with $100\mu$K Gaussian noise added to each scan. 
The $68\%$ confidence
bounds (dotted lines) are also shown for this scan.}
\label{fig:simscan}
\end{figure}

\begin{figure}
\caption{The top figure is the simulated sky convolved with an $8\deg$
Gaussian beam. The middle and bottom figures are the reconstructed skies from
the $25\mu$K and $100\mu$K noise simulations (see text) respectively.
They are all convolved with a final Gaussian of the same size.}
\label{fig:3plot}
\end{figure}

There is a tendency for 
the MEM algorithm to produce super-resolution (Narayan \& Nityananda 1986)
of the features in the sky 
so that even though the experiment may not be sensitive to small angular
scales the final reconstruction appears to have these features in
it. Even though this effect is only minor, care
must be taken not to interpret these features as actual sky features but 
instead the maps should be convolved back down with a Gaussian to the 
size of the features that are detectable by the experiment in
consideration. This has been done with the 
two lower plots in Figure~\ref{fig:3plot},
so that a direct comparison between all three is 
possible. By comparison of these plots it is seen that the 
reconstructed sky obtained from the MEM algorithm does give a good 
description of the actual sky. 

As an indicator of the error on the final sky reconstruction from the
MEM, a histogram of the fractional difference between the input and output map
temperatures is plotted in Figure~\ref{fig:histo}. If the initial
temperature at pixel $(i,j)$ is given by ${\rm T}_{input}$ and the
temperature at the same pixel in the output reconstructed map (after
convolution with a Gaussian beam to avoid superresolution) is given
by ${\rm T}_{recon}$ then the value of

\begin{equation}
{{{\rm T}_{recon}}-{{\rm T}_{input}}}\over{{\rm T}_{input}}
\label{eq:trecon}
\end{equation}

\noindent
is put into discrete bins and summed over all $(i,j)$. The final
histogram is the number of pixels within each bin. The output map
has been averaged over pixels within the beam FWHM as features have
freedom to move by this amount. A graph centred on -1 would mean that the output
signal is near zero and the amplitude is too small while a graph centred on 0
would mean the reconstruction is very accurate. As can be seen both
graphs (Figure~\ref{fig:histo} (a) and (c) for the $25\mu$K and
$100\mu$K noise simulations respectively) can be well
approximated by a Gaussian centred on a value just below zero. This
means that the MEM has a tendency to `damp' the data which is
expected and this `damping' increases with the level of noise ($\sim
10\%$ `damping' for the $25\mu$K simulation and $\sim 20\%$ for the $100\mu$K
simulation). From the integrated plots (Figure~\ref{fig:histo} (b) and
(d)) MEM can be expected to reconstruct all features with better than $50\%$
accuracy a half of the time for the $25\mu$K noise simulation and
a third of the time for the $100\mu$K noise simulation. 

\begin{figure}
\caption{ Histograms of the errors in the reconstruction of the
simulated sky maps. (a) shows the 
${{\rm T}_{recon}-{{\rm T}_{input}}}\over {{\rm T}_{input}}$ for the
$25\mu$K noise simulation and (c) shows the integrated 
$\left| {{\rm T}_{recon}-{{\rm T}_{input}}}\over {{\rm T}_{input}}
\right| $ 
for (a). (b) and (d) are the corresponding plots for the
$100\mu$K noise simulation. }
\label{fig:histo}
\end{figure}

\section{Discussion}
\label{memconc}

As seen from simulations performed in Section~\ref{coban}, the
positive/negative MEM algorithm performs very well recovering the amplitude,
position and morphology of structures 
in both the reconvolved scans and the two-dimensional deconvolved sky
map. No bias, other than the `damping' enforcement,
is introduced into the results from the methods described here and so  
this is the best of the methods tried to use when making maps using microwave 
background data, as the bare minimum of prior knowledge of the sky is
required. Even with the lowest signal to noise ratio (the $100\mu$K
noise simulation which corresponds to our worse case in 
the Tenerife experiments) all of the main features on the sky were 
reconstructed. Using this method it was possible to put constraints on the
galactic contamination for other experiments at higher frequencies, which is
essential when trying to determine the level of CMB fluctuations present.  

It is clear that this approach works well and provides a useful
technique for extracting the optimum CMB maps from both current and
future multi-frequency experiments. This will become of ever
increasing importance as the quality of CMB experiments improves.

\chapter[The sky maps]{Foregrounds and CMB features in the sky maps}
\label{chap7}

In this chapter I will use the maximum entropy method
to produce maps of the sky at various frequencies from the data
taken by the Jodrell Bank and Tenerife experiments. 
The possible origin of some of the features on the maps will be discussed. 

\section{The Jodrell Bank 5GHz interferometer}
\label{intro}

\subsection{Wide--spacing data}
\label{widespac}

The wide--spacing interferometer at Jodrell Bank has a baseline of
1.79m. The transfer function of the beam (i.e. the Fourier
transform of the beam) is shown in Figure~\ref{fig:5ghzpow}.
It can be seen that in the RA direction it is sensitive to $\sim
1.5\deg$ scales whereas in the declination direction it is sensitive
to $\sim 8\deg$ scales. At this frequency (5~GHz) the largest signals
present will be those from extra--galactic point sources on the
smaller angular scales and galactic synchrotron emission on the larger
angular scales. This data will therefore be used as a prediction for point
source contribution in the results from the other experiments 
and as a possible constraint
on galactic emission. Simulations performed for the
level of the CMB fluctuations that are expected in the data from a CDM dominated
Universe with $H_\circ = 50 {\rm km s}^{-1} {\rm Mpc}^{-1}$ and
$\Omega_b =0.03$ give $3.0\pm 1.1\mu$K (the error is the standard deviation
over 300 simulations). This is far below the noise
and so it is ignored in the analysis presented here. 

\begin{figure}
\caption{The cosine beam power sensitivity as a function of inverse
degrees. The sine beam has the same pattern but contains different
phase information.}
\label{fig:5ghzpow}
\end{figure}

The MEM algorithm was used to extract the best information from the
data using both sine and cosine channels as constraints as described
in the previous chapter. The full deconvolved map was not used as there is no 
simple consistent method to analyse the different scale dependencies
in the two directions (the map appears to be stripped in the RA
direction due to the larger resolution)
and thus compare the result with other data from
experiments such as Tenerife. The map presented here is the result
of combining the cosine and sine channel output from MEM to obtain the
amplitude and is therefore still convolved in a beam with $8 \deg$
FWHM. The analysis was restricted to the high Galactic latitude data
as there is too much confusion in the Galactic plane to allow any
constraints on the CMB. The Galactic plane
crossing is also an order of magnitude larger than the fluctuations
seen in the rest of the scan and so the MEM algorithm has a tendency
to fit this region better at the expense of the `interesting' regions.

To check that MEM was reconstructing the data correctly the results
were visually compared to the CLEAN results. This comparison is shown
in Figure~\ref{fig:memcldata} and it is seen that the MEM result
follows the noisy data more closely than the CLEAN result, as was
found in the simulations. 

\begin{figure}
\caption{Comparison between the MEM reconstruction (solid line) and the {\em Clean} reconstruction (dashed line) of the data (dotted line). The noise RMS in the scans are a) $25\mu$K, b) $18\mu$K and c) $36\mu$K.}
\label{fig:memcldata}
\end{figure}

Figures~\ref{fig:cosrecws} and \ref{fig:sinrecws} show the MEM
reconstructions of the data compared to the raw stacked data for the
cosine and sine channels respectively. It is easily
seen that the MEM reconvolution does follow the raw data very well in
each declination.
Figure~\ref{fig:recon} shows the MEM reconstructed 2--D sky map for
the high Galactic latitude region (RA $130\deg$ to $260\deg$). The
error on this map is calculated by use of Monte-Carlo simulations and
can be read directly from Figure~\ref{fig:sigma}. As the average error
on the input data is 37~$\mu$K the error on the 2--D sky map is
10~$\mu$K. The
point sources used as a check for the calibration correspond to the
three largest peaks in this
plot; 3C345 at RA $250\deg$, Dec. $39\deg$ with a flux of 6~Jy ($\sim
400\mu$K); 4C39 at RA
$141\deg$, Dec. $39\deg$ with a flux of 9~Jy ($\sim 550\mu$K); 3C286
at RA $200\deg$, Dec. $30\deg$ with a flux of 7~Jy ($\sim 450\mu$K). 
The amplitude of the data does not correspond to the exact predictions
as the sources are
variable to $\sim 30$\% but all agree to within this factor. Other
sources can be used to check the calibration and are clearly seen in
the data (e.g. 3C295 at RA $210\deg$,
Dec. $52\deg$).

\begin{figure}
\caption{MEM reconvolution (green line) of the cosine channel of the 
Jodrell Bank 5~GHz wide
spacing interferometer compared to the raw data
(black line with error bars showing the one sigma deviation across scans).}
\label{fig:cosrecws}
\end{figure}

\begin{figure}
\caption{MEM reconvolution (green line) of the sine channel of the 
Jodrell Bank 5~GHz wide
spacing interferometer compared to the raw data
(black line with error bars showing the one sigma deviation across scans).}
\label{fig:sinrecws}
\end{figure}

\begin{figure}
\caption{MEM reconstructed 2--D sky map of the wide--spacing Jodrell
Bank interferometer data at high latitude.}
\label{fig:recon}
\end{figure}

Figure~\ref{fig:GBmem} is a comparison between the MEM output of the
data and the predicted point source contribution by using the low
frequency GB catalogues. The 1.4~GHz and 4.9~GHz catalogues were
extrapolated by performing a pixel by pixel fit for the spectral
index. There is very good agreement between the MEM output and the
prediction as expected since the interferometer is
very sensitive to point sources. The difference between the GB
prediction and the data is shown in Figure~\ref{fig:GBmemdiff}. Notice
that the maximum amplitude of the difference
is 280~$\mu$K and the two largest peaks occur at the positions of
the two variable sources, 3C345 and 4C39. 
This difference may therefore be due to the variability of the sources.

\begin{figure}
\caption{Comparison between the GB catalogue (contour) and the MEM
reconstruction (grey-scale). A different grey scale is used on all
comparison plots (compared to the above plot) to enhance features in
the MEM reconstruction for easier comparison. The contour levels are
set at 10 equal intervals between the 0.0mK and 0.2mK.
The green contours are above 0.1mK and the red contours are below
0.1mK.
A larger region is also
plotted to allow easier comparison of features at the edges of the
maps.}
\label{fig:GBmem}
\end{figure}

\begin{figure}
\caption{Difference between the GB catalogue prediction and the MEM reconstruction. Notice that the main differences occur at the positions of the two main, 
variable point sources, 3C345 and 4C39.}
\label{fig:GBmemdiff}
\end{figure}

The other dominant source of radiation that could contribute at this
frequency and angular scale is that arising from synchrotron
sources. For example, the discrepancy between the GB
prediction and the data, for the region close to 3C345, could also be
due to Galactic emission as well as the variability of the source.
This is a region where Galactic emission has been detected
previously. 
To see the extent to which synchrotron contaminates the maps, 
the low frequency surveys of Haslam {\em et al}, 1982,
and Reich \& Reich, 1988, were extrapolated up to 5~GHz. This is done
to obtain a crude estimate on the Galactic spectral index as it has
already been noted that the artefacts in these surveys make
extrapolation difficult. 
Figure~\ref{fig:408ws} shows the extrapolated 408~MHz survey map
compared to the MEM output and 
Figure~\ref{fig:1420ws} shows the extrapolated 1420~MHz
survey map. The two extrapolations were done by
assuming a synchrotron power law with spectral index of -2.75. 
As the two surveys are already convolved in a beam with $0.85\deg$
FWHM (the 1420~MHz survey was convolved to the same resolution as the
408~MHz survey), care must be taken when carrying out the extrapolation. The
interferometer beam will reduce the amplitude of the prediction by a
factor that depends on the fringe visibility and this can be
calculated from the Fourier transforms. The Fourier transform of the
cosine channel can be shown to be 

\begin{equation}
\hat{R(u,v)}={1\over 2} \sigma^2 \left[ {\rm exp} \left( - {\sigma^2
\over 2} ((u-u_\circ )^2 +v^2) \right) +{\rm exp} \left( - {\sigma^2
\over 2} ((u+u_\circ )^2 +v^2) \right) \right] 
\label{eq:ftofjod}
\end{equation}

\noindent
where $u_\circ=2\pi b/\lambda$ is the fringe spacing for the
interferometer with baseline $b$, operating at wavelength $\lambda$,
and $\sigma$ is the dispersion of the
interferometer. The Fourier transform of the sine
channel is exactly the same except that the difference of the two
exponentials is required. Multiplying Equation~\ref{eq:ftofjod} with
the Fourier transform of a Gaussian (so that a convolution is applied
in real space) corresponding to the FWHM of the 
surveys ($0.85\deg$) and calculating the maximum visibility of the
source (i.e. when the telescope is pointing directly at the source)
it is found that

\begin{equation}
V={\sigma^2 \over a^2 + \sigma^2} S_\circ {\rm exp} \left[ - \left(
{u_\circ^2 \over 2} \right) \left( {\sigma^2 a^2 \over a^2 + \sigma^2
} \right) \right]
\label{eq:ftvisi}
\end{equation}

\noindent
where $a$ is the dispersion of the Galactic survey used ($0.85\deg$
FWHM corresponds to $0.33\deg$ dispersion)
and $S_\circ$ is the actual flux
of the source. With $\lambda=6$cm, $b=1.79$~m,
$\sigma=3.4\deg$ and $a=0.33\deg$, as in the case of the wide spacing
data, it is found that $V=0.91
S_\circ$. Therefore, after the convolution of the 1420 MHz survey
with the interferometer beam has been performed it is necessary to
multiply by a factor of $1/0.91=1.1$ and all results quoted here have
taken this into account.
 
\begin{figure}
\caption{Comparison between the 1420~MHz survey map extrapolated with a uniform 
spectral index and the MEM reconstruction. The grey-scale shows the 
MEM and the contours show the 1420~MHz survey.}
\label{fig:1420ws}
\end{figure}

\begin{figure}
\caption{Comparison between the 408~MHz survey map extrapolated with a uniform 
spectral index and the MEM reconstruction. The grey-scale shows the 
MEM and the contours show the 408~MHz survey. The contour levels are
the same as those in Figure~\ref{fig:GBmem}.}
\label{fig:408ws}
\end{figure}

The lack of obvious correlations between the results and those of
the low frequency surveys (except where the point sources are seen in
both surveys) may be due to errors in the surveys. As
previously discussed, the baselevels of the low frequency surveys are
uncertain to about 10\% and as the area considered is in the region of the
survey where the intensity is at a minimum this is where the baselevel is 
expected to have maximum effect on the extrapolations. However, if it
is assumed that the
extrapolation is correct then the absence of correlation must result
from a steepening of the spectral index of the synchrotron from the
low frequency surveys and so the galactic emission is not as
predicted. The frequency dependence and the spatial variance in the
steepening of the spectrum is unknown without any intermediate
frequency surveys. Column 2 of Table~\ref{ta:synch} 
summarises the {\em rms} of the data from this
experiment in the high Galactic latitude region (RA: $130\deg$ to
$260\deg$). The error on the data is calculated by using the
combination of errors for the cosine, $\sigma_c$, and sine,
$\sigma_s$, channels.

\begin{equation}
\sigma_{amp}^2 = \left( {{{\delta A}\over{\delta y_c}} \sigma_c}
\right)^2 + \left( {{{\delta A}\over{\delta y_s}} \sigma_s} \right)^2
\label{eq:errordat}
\end{equation}

\noindent 
where $y_s$ and $y_c$ are the sine and cosine channel responses respectively and
$A=\sqrt{y_s^2 +y_c^2}$. This gives 

\begin{equation}
\sigma_{amp}=\sqrt{{y_c^2 \sigma_c^2 + y_s^2 \sigma_s^2}\over{y_c^2 +y_s^2}}.
\label{eq:finerr}
\end{equation}

\noindent
These errors are shown in column 3 of Table~\ref{ta:synch}. Column 4
shows the {\em rms} of the MEM reconstruction of the data. The
remaining 3 columns are the predictions obtained by convolving each
survey with the interferometer beam. The Galactic survey results are
quoted at their observing frequencies.

\begin{table}
\begin{center}
\begin{tabular}{|c|c|c|c|c|c|c|} \hline
Dec. & Data & Error & MEM & GB & Galactic & Galactic  \\
 & & & recons. & prediction & (1420~MHz) & (408~MHz) \\ \hline
$30\deg$ & 0.141 & 0.028 & 0.133 & 0.097 & 2.48 & 138.0 \\
$32\deg$ & 0.169 & 0.054 & 0.123 & 0.078 & 2.05 & 109.9 \\
$35\deg$ & 0.152 & 0.023 & 0.154 & 0.093 & 1.63 & 101.4 \\
$37\deg$ & 0.223 & 0.033 & 0.199 & 0.156 & 1.44 & 100.3 \\
$40\deg$ & 0.188 & 0.020 & 0.189 & 0.181 & 1.35 & 105.3 \\
$42\deg$ & 0.168 & 0.039 & 0.127 & 0.127 & 1.39 & 113.3 \\
$45\deg$ & 0.096 & 0.021 & 0.085 & 0.067 & 1.62 & 115.3 \\
$47\deg$ & 0.145 & 0.067 & 0.087 & 0.068 & 1.76 & 124.0 \\
$50\deg$ & 0.092 & 0.030 & 0.094 & 0.075 & 2.08 & 258.7 \\
$52\deg$ & 0.137 & 0.032 & 0.108 & 0.076 & 2.27 & 194.1 \\
$55\deg$ & 0.123 & 0.036 & 0.116 & 0.055 & 2.03 & 112.3 \\ \hline
Total    & 0.153 & 0.037 & 0.140 & 0.110 & 1.95 & 145.6 \\ \hline
\end{tabular}
\end{center}
\caption{Summary of the results found from the 5~GHz, wide spacing
interferometer at
the high Galactic latitude region (RA: $130\deg$ to $260\deg$). All
values are in mK. The predictions were found by convolving the survey
maps with the interferometer beam (see text).}
\label{ta:synch}
\end{table}

\subsection{Narrow--spacing data}
\label{jodns}

After 1994 the baseline of the interferometer was changed to
0.702~m. This meant that the experiment was now more sensitive to the
large scale Galactic fluctuations. Together, the two data sets 
can be used to put a very tight constraint
on the Galactic emission as well as a point source prediction for
other experiments. Simulations performed showed that the level
of CMB fluctuations that are expected in the data from a CDM dominated
Universe with $H_\circ = 50 {\rm km s}^{-1} {\rm Mpc}^{-1}$ and
$\Omega_b =0.03$ are $5.2\pm 1.6\mu$K (the error is the standard deviation over 300
simulations). This is still below the noise and is
ignored here. 

The same procedure was used as in the wide--spacing data
analysis. Again, the comparison between the MEM reconvolved data and
the raw data show very good agreement and this can be seen in
Figures~\ref{fig:cosrecns} and ~\ref{fig:sinrecns}. 
Figure~\ref{fig:nsmem} shows the MEM reconstructed 2--D sky
map for the high Galactic latitude region (RA $130\deg$ to
$260\deg$). Using Figure~\ref{fig:sigma} the error on the 2--D sky map
is 8~$\mu$K. The point sources seen with the wide--spacing
interferometer are again clearly visible in the narrow--spacing
data. Figure~\ref{fig:nsGB} shows the comparison between the point
source prediction and the MEM reconstruction. 
However, there are now larger sources that are not due to point
source contributions (e.g. the region about RA $170\deg$,
Dec. $35\deg$). These are most likely produced by the extra
sensitivity to large scale structure that the narrow--spacing data
has and are therefore most likely to be Galactic in origin. The
dilution of the beam given by Equation~\ref{eq:ftvisi} becomes $V=0.56
S$. Therefore, after the convolution of the two low frequency surveys
for comparison, it is necessary to multiply by a factor of $1/0.56 =
1.8$ and all results quoted here have taken this into account. 

\begin{figure}
\caption{MEM reconvolution (green line) of the cosine channel of the 
Jodrell Bank 5~GHz narrow
spacing interferometer compared to the raw data
(black line with error bars showing the one sigma deviation across scans).}
\label{fig:cosrecns}
\end{figure}

\begin{figure}
\caption{MEM reconvolution (green line) of the sine channel of the 
Jodrell Bank 5~GHz narrow
spacing interferometer compared to the raw data
(black line with error bars showing the one sigma deviation across scans).}
\label{fig:sinrecns}
\end{figure}

\begin{figure}
\caption{MEM reconstructed 2--D sky map of the narrow--spacing Jodrell
Bank interferometer data at high latitude.}
\label{fig:nsmem}
\end{figure}

\begin{figure}
\caption{Comparison between the GB catalogue (contour) and the MEM reconstruction (grey-scale).}
\label{fig:nsGB}
\end{figure}

Figure~\ref{fig:1420ns} and Figure~\ref{fig:408ns} show the two low
frequency surveys extrapolated with a synchrotron power law compared
to the narrow--spacing MEM reconstruction. It is seen that there is
little correlation between the 1420~MHz survey and the MEM
reconstruction (ignoring the point source contributions that are seen
in both maps) but there are some common features in the 408~MHz and
the MEM reconstruction (although these are saturated in the
contours). For example, there is a possible Galactic feature at RA
$170\deg$, Dec. $35\deg$ which is common to both the 408~MHz and MEM
reconstruction but it appears at a higher level in the 408~MHz survey
(the contour levels are saturated) which may be an indication that it
is a steepened synchrotron source. The 1420~MHz survey is considered to be of
poorer quality in the low level Galactic emission (high Galactic
latitude) than the 408~MHz survey as it suffers from more striping
effects. This could account for the discrepancy between the two
predictions. Table~\ref{ta:nstab} summarises the results for this
experiment in the high Galactic latitude region (RA $130\deg$ to
$260\deg$). 

\begin{figure}
\caption{Comparison between the 1420~MHz survey map extrapolated with a uniform 
spectral index and the MEM reconstruction. The grey-scale shows the 
MEM and the contours show the 1420~MHz survey.}
\label{fig:1420ns}
\end{figure}

\begin{figure}
\caption{Comparison between the 408~MHz survey map extrapolated with a uniform 
spectral index and the MEM reconstruction. The grey-scale shows the 
MEM and the contours show the 408~MHz survey.}
\label{fig:408ns}
\end{figure}

\begin{table}
\begin{center}
\begin{tabular}{|c|c|c|c|c|c|c|} \hline
Dec. & Data & Error & MEM & GB & Galactic & Galactic  \\
 & & & recons. & prediction & (1420~MHz) & (408~MHz) \\ \hline
$30\deg$ & 0.164 & 0.031 & 0.149 & 0.116 & 3.14 & 130.1 \\
$32\deg$ & 0.229 & 0.032 & 0.196 & 0.116 & 3.01 & 189.9 \\
$35\deg$ & 0.211 & 0.032 & 0.206 & 0.106 & 2.36 & 190.7 \\
$37\deg$ & 0.224 & 0.034 & 0.186 & 0.141 & 1.91 & 179.9 \\
$40\deg$ & 0.150 & 0.018 & 0.162 & 0.164 & 1.81 & 171.8 \\
$42\deg$ & 0.173 & 0.030 & 0.121 & 0.124 & 1.86 & 162.5 \\
$45\deg$ & 0.114 & 0.022 & 0.110 & 0.074 & 1.93 & 161.2 \\
$47\deg$ & 0.147 & 0.027 & 0.134 & 0.077 & 2.24 & 189.6 \\
$50\deg$ & 0.177 & 0.032 & 0.161 & 0.099 & 2.79 & 207.6 \\
$52\deg$ & 0.176 & 0.042 & 0.168 & 0.112 & 3.04 & 192.2 \\
$55\deg$ & 0.158 & 0.029 & 0.147 & 0.090 & 2.57 & 166.2 \\ \hline
Total    & 0.178 & 0.030 & 0.169 & 0.119 & 2.59 & 190.8 \\ \hline
\end{tabular}
\end{center}
\caption{Summary of the results found from the 5~GHz, narrow spacing
interferometer at
the high Galactic latitude region (RA: $130\deg$ to $260\deg$). All
values are in mK. The predictions were found by convolving the survey
maps with the interferometer beam.}
\label{ta:nstab}
\end{table}

\subsection{Joint analysis}
\label{jodjoint}

It is possible to combine the two data sets to extract the most likely
common underlying sky for both the narrow and wide spacings. This is a
very good test of the consistency of the experiment. The joint MEM analysis
described in the Chapter~\ref{chap5} can be used to combine the narrow
and wide--spacing and extract the most likely underlying sky common to
the two experiments. Table~\ref{ta:combi} summarises the results from
the combination of the two data sets. By comparing this to Table~\ref{ta:synch} 
and \ref{ta:nstab} it can be seen that the results from
the joint analysis are almost identical to those from the individual
analyses. This means that the narrow and wide--spacing data sets are
consistent and can be used to put constraints on Galactic
emission. Any discrepancies between the analysis are a result of the
variability in the point sources. 

\begin{table}
\begin{center}
\begin{tabular}{|c|c|c|} \hline
Dec. & WS MEM result & NS MEM result \\ \hline
$30\deg$ & 0.134 & 0.150 \\
$32\deg$ & 0.122 & 0.193 \\
$35\deg$ & 0.155 & 0.204 \\
$37\deg$ & 0.198 & 0.185 \\
$40\deg$ & 0.189 & 0.162 \\
$42\deg$ & 0.127 & 0.122 \\
$45\deg$ & 0.085 & 0.111 \\
$47\deg$ & 0.087 & 0.134 \\
$50\deg$ & 0.094 & 0.161 \\
$52\deg$ & 0.108 & 0.168 \\
$55\deg$ & 0.116 & 0.147 \\ \hline
Total     & 0.140 & 0.168 \\ \hline
\end{tabular}
\end{center}
\caption{Summary of the results found from the joint MEM analysis of
 the two 5 GHz interferometer data sets at
the high Galactic latitude region (RA: $130\deg$ to $260\deg$). All
values are in mK.}
\label{ta:combi}
\end{table}

To predict the level of Galactic fluctuations at higher frequencies
the spectral index dependence of the foregrounds is required. To
calculate this it is possible to compare the results from the joint
analysis to the predictions from the lower frequency surveys. Firstly
the point sources must be subtracted from the map reconstruction to
leave the Galactic contribution. The prediction from the Green Bank
catalogue was subtracted from the narrow and wide--spacing data to
leave the residual signal. Due to the variability of the sources this
residual will be an upper limit on the Galactic contribution and only
by having continuous source monitoring would the results be
better. We will use the residual signal for the
narrow--spacing data (where the Galactic signal is expected to be higher and so a
smaller error will be obtained). The {\em rms} of this signal 
is $73\pm 23\mu$K (where the error on the MEM reconstruction comes
from comparison with simulations and a 30\% variability in the major
sources is assumed). When comparing this to the signal of $2.59\pm
0.26$~mK (error from a 10\% error in the survey) at 1420~MHz and
$190.8\pm 19$~mK at 408~MHz it is found that the average spectral index
from 1420~MHz to 5~GHz is $2.8\pm 0.4$ and from 408~MHz to 5~GHz is
$3.1 \pm 0.4$. These results are in agreement with previous
predictions of the Galactic spectral index at a range of angular
scales. Bersanelli \et (1996) found
that the spectral index between 1420~MHz and 5~GHz was $2.9\pm 0.3$ on
a $2\deg$ angular scale and Platania \et (1997) found that the
spectral index between 1420~MHz and a range of frequency
data between 1~GHz and 10~GHz gave a spectral index of $2.8\pm 0.2$ on
an $18\deg$ angular scale. 
This is an indication of a steepening synchrotron
spectral index. To take the analysis one step further it is possible
to compare the surveys pixel by pixel to obtain a map of the spectral
index dependencies. 

Figure~\ref{fig:408nscomp} shows the spectral index variation as
predicted by comparison between the 408~MHz survey and the
narrow--spacing data over the
high Galactic region. The {\em rms} spectral index is $3.3\pm 0.5$
where the error is now the {\em rms} over the map. This is consistent
with the above result. Figure~\ref{fig:1420nscomp} shows the result
for the 1420~MHz survey prediction. The {\em rms} spectral index is
$3.1\pm 0.9$. The regions where there is a large deviation away from
the {\em rms} spectral index (which results in the large variance) 
are usually associated with variable point sources
(e.g. 3C345) that have not been fully removed from the 5~GHz
interferometer or Galactic survey data. All the results so far are
consistent with a steepened synchrotron source being the dominant
contributor to the data. It is also possible that the shallower index
between 1420~MHz and 5~GHz than between 408~MHz and 5~GHz could be due
to the increasing importance of free--free emission, although further
frequency measurements are necessary to confirm this. 
Therefore, the low frequency surveys should not be used
(especially the 1420~MHz survey) as a prediction for Galactic
contribution without taking into account the synchrotron
steepening. The new 5~GHz data presented here represents an
intermediate step in the frequency coverage between the low frequency
Galactic surveys and the higher frequency CMB experiments. It is
therefore a very useful check on Galactic models and can be used to
make estimates of Galactic contamination in CMB experiments. 

\begin{figure}
\caption{The variation in the derived spectral index using the
narrow--spacing data and the 408~MHz survey. The contour level is at
the {\em rms} of the map.}
\label{fig:408nscomp}
\end{figure}

\begin{figure}
\caption{The variation in the derived spectral index using the
narrow--spacing data and the 1420~MHz survey. The contour level is at
the {\em rms} of the map.}
\label{fig:1420nscomp}
\end{figure}

It should be noted that the level of Galactic emission predicted here
may be too large as the artefacts in the
surveys may enhance the fluctuations. Also, the 1420~MHz and 408~MHz
surveys contain contributions from the point sources (for example
3C295) and so the level of Galactic fluctuations predicted will be too
large. If it is assumed that 50\% of the 408~MHz survey {\em rms} is
due to these effects (i.e. $135\pm 20$~mK at 408~MHz for the narrow
spacing data)
then the spectral index constraint is $\beta=3.0\pm 0.4$ which still 
corresponds to a steeping synchrotron spectrum. This does not take
into account the residual effect from the point sources in the 5 GHz
prediction and so should be taken as a lower limit on the spectral
index. This does not alter the
conclusion that the low frequency surveys should not be used by
themselves as a
prediction for galactic emission as these artefacts, or the presence of
point sources, only make the prediction worse. 

If it is assumed that the signal remaining after subtracting the Green
Bank point source prediction from the data is due to Galactic emission
alone then it is possible to make a prediction for the level of
contamination in the Tenerife data. As the interferometer is only
sensitive to a certain range of angular scales it is necessary to
assume something about the spatial variation of the synchrotron
source. A spectrum of $l^{-3}$ is used spatially and the steepened
synchrotron spectrum is assumed to hold until the higher Tenerife
frequencies. Therefore, considering the $5\deg$ FWHM Tenerife
experiments, it is expected that the 10~GHz 
data will be contaminated by 30~$\mu$K of synchrotron emission,
the 15~GHz data will be contaminated by 10~$\mu$K of synchrotron
emission and the 33~GHz data will be contaminated by 1~$\mu$K of
synchrotron emission. However, these figures are very approximate as
the interferometer is sensitive to two different scales in the RA and
Dec. direction and this has not been taken into account and will
reduce the contamination level. Also the
increasing importance of the free--free emission has been ignored
which will increase the Galactic contamination level. 

\section{The Tenerife experiments}
\label{tenres}

The MEM programs were applied to all Tenerife data described in
Chapter~\ref{chap4}. This is an ongoing process and the results shown
here are not final but represent the current iteration of the
analysis. 

\subsection{Reconstructing the sky at 10.4~GHz with $8\deg$ FWHM}
\label{recon}

To apply the MEM deconvolution process described to
the data from the 10.4~GHz, $8\deg$ FWHM, Tenerife experiment 
it is necessary to select parameters that not only
achieve convergence of the iterative scheme, but also make the fullest
use of the data.  The amplitude of the fluctuations that are of
interest is at least two orders of magnitude smaller than the
magnitude of the signal produced during the major passage through the
Galactic plane region ($\sim 45$~mK at $\sim$ Dec. $+40\deg$). Clearly, any
baseline fitting and reconstruction will be dominated by this feature
at the expense of introducing spurious features into the regions
which are of interest. For this reason the data
(Table~\ref{ta:nogal}) corresponding to the principal Galactic plane
crossing are not used in the reconstruction.

\begin{table}
\begin{center}
\begin{tabular}{|c|c|} \hline
Declination & RA range excised (degrees)\\ \hline
$+46.6^{\circ}$ & 275-340\\
$+42.6^{\circ}$ & 275-340\\
$+39.4^{\circ}$ & 275-340\\
$+37.2^{\circ}$ & 275-340\\
$+27.2^{\circ}$ & 280-320\\
$+17.5^{\circ}$ & 265-310\\
$+07.3^{\circ}$ & 255-310\\
$+01.1^{\circ}$ & 255-310\\
$-02.4^{\circ}$ & 260-310\\
$-17.3^{\circ}$ & 255-300\\
\hline
\end{tabular}
\end{center}
\caption{The Galactic plane regions excised at each declination.}
\label{ta:nogal}
\end{table}

In contrast, the anti-centre crossing ($\sim$ RA $60^{\circ}$ at
$\sim$ Dec. $+40\deg$ ) corresponding to 
scanning through the Galactic plane, but looking
out of the Galaxy, is at an acceptable level 
($\lta 5$~mK) and is a useful check on the performance and consistency
of the observations.
With the parameters set as in Chapter~\ref{chap5}, Table~\ref{ta:params}, 
$\chi^2$ demonstrates a rapid convergence. For example, the
change in $\chi^2$ after 120 iterations of MEM is
$\Delta \chi^2/\chi^2 \simeq -9\times 10^{-4}$ 
while the change in $\chi_{base}^2$ is
$\Delta \chi_{base}^2/\chi_{base}^2 \simeq -2\times 10^{-4}$. 
 
The fitted baselines are subtracted from the raw data set to
provide data free from baseline effects, allowing the scans for a given
declination to be stacked together to provide a single high sensitivity
scan. 
Figure~\ref{fig:stack} shows the stacked results for the $8\deg$
experiment at each declination
compared with the reconvolution of the MEM result with the beam. 
The weak Galactic crossing is clearly visible at RA=$50\deg-100\deg$. 
At lower declinations this crossing shows a complex structure 
with peak amplitudes $\sim$ a few mK. Only positions on
the sky with more than ten independent measurements have been
plotted. The data with better sensitivities are those at Dec.=$+39.4\deg$ 
and $+1.1\deg$.

\begin{figure}
\caption{The stacked scans at each declination displayed as a function 
of right ascension. Again the plots are the second difference binned 
into $4\dg$ bins and the $68\%$ confidence limits. The main Galactic
plane crossing has been excluded, and only positions on the sky in
which we have more than $\sim$10 independent measurements have been
plotted. Also shown (solid line) is the reconvolved result from MEM overlayed 
onto each declination scan.}
\label{fig:stack}
\end{figure}

The sky is not fully sampled with this data set (as seen in 
Figure~\ref{fig:stack}) but
the MEM uses the continuity constraints on the data to 
reconstruct a two-dimensional sky model. 
In Figure~\ref{fig:8degpn}, the sky reconstruction is shown. Although
a rectangular projection has been used for display, the underlying
computations use the full spherical geometry for the beams (as
described in Chapter~\ref{chap3}).
The anti-centre crossings of the Galactic plane are clearly visible on the
right hand side of the image, while one should recall that
the principal Galactic crossing has been excised from the data.
It is clearly seen that there is apparent continuity of structure 
between adjacent independent 
data scans which are separated by less than the $8\dg$ beam
width (see the higher declination strips in the plot where the data are
more fully sampled). Where the data are not fully sampled (the lower
declinations) the MEM has reverted to zero as expected and this is
seen as `striping' along declinations in the reconstructed map. 

\begin{figure}
\caption{MEM reconstruction of the sky at 10.4 GHz, as seen by the Tenerife
$8.4\dg $ FWHM experiment.}
\label{fig:8degpn}
\end{figure}

\subsection{Non-cosmological foreground contributions}
\label{noncos}

\subsubsection{Point sources}
\label{points}

The contribution of discrete sources to this
data set has been estimated using the
K\"{u}hr \et (1981) catalogue, the VLA calibrator list and the Green 
Bank sky surveys (Condon \& Broderick 1986); sources $\lta 1$ Jy at 10.4~GHz
were not included in the analysis. The response of the 
instrument to these point sources has been modelled by
converting their fluxes into antenna temperature (1 Jy is equivalent to
12~$\mu$K for the experiment), convolving these with the triple beam of
the instrument and sampling as for the real data  (see the details
in Guti\'errez \et 1995). 
The two main radio sources at high Galactic latitude,
expected in the Tenerife scans are 
3C273 (RA=$186.6\deg$, Dec.=$+02\deg19^{\prime}43^{\prime\prime}$) 
with a flux density at 10~GHz 
of $\sim 45$~Jy; this object should contribute with a peak amplitude
$\Delta T\sim 500$~$\mu$K in the triple beam to the data at Dec.=$+1.1\deg$, and 
3C84 (RA=$49.1\deg$, Dec.=$+41\deg19^{\prime}52^{\prime\prime}$) with a
flux density at 10~GHz of $\sim 51$~Jy.
Figure~\ref{fig:compar} presents a comparison between the MEM result reconvolved
in the Tenerife triple beam, the data and
the predicted contribution of the radio source 3C273.
A diffuse Galactic contribution near the
position of this point source accounts for the differences in
amplitude and shape of the radio source prediction and the data (see below).
The radio sources 3C273 and 3C84 have also been detected in the deconvolved map
of the sky shown in Figure~\ref{fig:8degpn}. For example, 3C273 is
clearly seen in the reconstructed map. 
Also clearly detected are 3C345 
(RA=$250.3\deg$, Dec.=$+39\deg54^{\prime}11^{\prime\prime}$)and 4C39
(RA=$141.0\deg$, Dec.=$+39\deg15^{\prime}23^{\prime\prime}$)
in both the reconvolved scans and the deconvolved map. Many other
features are seen in the deconvolved map but these may be swamped by the 
Galactic emission so it cannot be said with confidence that any 
originate from point sources. For example, features at 
Dec.$\sim+40\deg$, RA$\sim 180\deg$,
Dec.$\sim+17.5\deg$, RA$\sim 240\deg$ and Dec.$\sim+1.1\deg$, RA$\sim 220\deg$ do 
not correspond to any known radio sources (see Figure~\ref{fig:stack}). The additional
contribution by unresolved radio sources has been estimated to be
$\Delta T/T\sim 10^{-5}$ at 10.4~GHz (Franceschini \et 1989) in a single 
beam. This will be less in the Tenerife switched beam and is not
considered in the analysis presented here.

\begin{figure}
\caption{Comparison between the MEM reconstructed sky convolved in the 
Tenerife beam (solid line), the predicted point source contribution at
Dec.=$1.1\dg$ (dashed line) and the Tenerife data (dotted line with one 
sigma error bars shown). The source observed is 3C273 
(RA$={12^h}{26^m}{33^s}$, Dec.$=+{02\dg}{19'}{43''}$).}
\label{fig:compar}
\end{figure}

\subsubsection{Diffuse Galactic contamination}
\label{diffuse}

The contribution of the diffuse Galactic emission in the data can be
estimated in principle using the available maps at frequencies below
1.5~GHz. The 408~MHz (Haslam \et 1982) and 1420~MHz (Reich
\& Reich 1988) surveys were used; unfortunately  the usefulness of these maps is
limited because a significant part of the high Galactic latitude
structure evident in them is due to systematic effects as already
discussed (also see Davies, Watson
\& Guti\'errez 1996).  Only in regions (such as crossings of the
Galactic plane) where the signal dominates clearly over the
systematic uncertainties, is it possible to estimate the expected signals
at higher frequencies. With this in mind, these two maps were converted 
to a common resolution ($1\deg \times 1\deg$ in right ascension and
declination respectively) and convolved in the triple beam
response.

This contribution at 408 and 1420~MHz can be compared with the
data at 10.4~GHz to determine the spectral index of the Galactic
emission in the region where these signals are high enough to dominate
over the systematic effects in the low frequency surveys.
A power law spectra ($T \propto \nu^{-\beta}$) 
for the signal with an index independent of the
frequency, but varying spatially was assumed.  The signals in
the Galactic anti-centre are weaker than those for the Galactic plane
crossing and are mixed up with several extended
structures, but even in this case it is possible to draw some conclusions about
the spectral index in this region. It was found that  
$\beta=3.0\pm 0.2$ between 408/1420~MHz and $\beta =
2.1\pm0.4$ between 1420/10400~MHz which indicates that free-free
emission dominates over synchrotron at frequencies $\gta 1420$~MHz in
the Galactic plane. Taking this together with the results from the
5~GHz interferometer it is seen that synchrotron dominates for
frequencies up to 5~GHz and then free-free will dominate.  
One of the stronger structures in the region away from the galactic plane 
is at RA$\sim 180\deg - 200\deg$, Dec$\sim 0\deg$ and
therefore the main contribution should be to the data at Dec=$1.1\deg$.
This structure at 408 MHz, assuming a slightly steepened synchrotron spectral
index of $\beta =2.8$, gives a predicted peak amplitude at 10.4~GHz of
$\sim 500$~$\mu$K; it is believed that this is responsible for the
distortion between the measurements at Dec=$1.1\deg$ and the predictions
for the radio source 3C 273. 

\subsection{The Dec $35\deg$ 10 and 15~GHz Tenerife data.}
\label{bunncomp}

A first direct comparison of the Tenerife and COBE DMR data at
Dec.$=+40\deg$, which also included the 33~GHz data, 
was made by Lineweaver \et (1995) who demonstrated a clear
correlation between the data-sets and showed the presence of common
individual features. Bunn, Hoffman \& Silk (1995) applied a Wiener filter
to the two-year COBE DMR data assuming a CDM model. They obtained a
weighted addition of the results at the two more sensitive frequencies
(53 and 90~GHz) in the COBE DMR data, and used the results of this
filtering to compute the prediction for the Tenerife experiment over
the region $35\deg \le$~Dec.$\le 45\deg$. At high Galactic latitude
the most significant features predicted for the Tenerife data are two
hot spots with peak amplitudes $\sim 50-100$ $\mu$K around
Dec.=$+35\deg$ at RA$\sim 220\deg$ and $\sim250\deg$. 
A comparison
between the reconvolved results of the data from the Tenerife
15~GHz, $5\deg$ FWHM experiment, using Maximum Entropy and the COBE data
has been made, 
and this prediction is plotted in Figure~\ref{fig:10_15_dec35}. The solid
line shows the reconvolved results at 15~GHz after subtraction of the
known point source contribution. The two most intense structures in
these data agree in amplitude and position with the predictions from 53
and 90~GHz (dashed line), with only a slight shift in position for the
feature at RA=$250\deg$. A possible uncertainty by a factor as large as
2 in the contribution of the point-source 1611+34 would change only
slightly the shape and amplitude of this second feature. As a test the
multifrequency MEM was applied to the 10~GHz and 15~GHz Tenerife data
at this declination. The program is currently in development and so it
was not possible to apply it to the full two dimensional data set in
the time allowed. However, the application of multi-MEM to this
declination can be used as a test of the power that it will have in
analysing the full two dimensional data set. 
Figure~\ref{fig:bumps} clearly shows the two features predicted by
Bunn, Hoffman \& Silk (1995) at Dec $+35\deg$
as reconstructed by the multifrequency MEM algorithm using
the 10~GHz and 15~GHz data simultaneously.

\begin{figure}
\caption{Comparison between the 15 GHz data (solid line)
and
the COBE prediction by Bunn {\em et al} 1995 (dashed line).}
\label{fig:10_15_dec35}
\end{figure}

\begin{figure}
\caption{A multifrequency MEM reconstruction of the two fluctuations
(white spots on the bottom colour contour plot),  
which Bunn {\em et al} (1995) predicted should be seen by the Tenerife
experiments using COBE results, from the 10~GHz and 15~GHz channels. 0
and 50 on the y--axis correspond to $65\deg$ and $15\deg$ in
declination respectively. 0 and 50 on the x--axis correspond to
$210\deg$ and $260\deg$ in right ascension respectively. The vertical
axis is in arbitrary units. Only the Dec $+35\deg$ data was used to
constrain this reconstruction so the fluctuations fall to zero away
from this declination.}
\label{fig:bumps}
\end{figure}

\subsection{The full $5\deg$ FWHM data set}
\label{tenfull}

The MEM deconvolution was applied to the full data set at each
frequency. This represents a large portion of the sky at the two lower
frequencies and so it is possible to produce sky maps covering a large
area. Figures~\ref{fig:10memtharr} and \ref{fig:15memtharr} show the
MEM reconvolved data compared to the raw stacked data at each
declination for the 10~GHz and 15~GHz Tenerife experiments
respectively. At both frequencies all of the declinations were
analysed simultaneously utilising the continuity across the sky. As
can be seen the MEM result falls within the one sigma confidence
limits at each declination. There are some discrepancies between
declinations where, because the data were taken at different times,
the variability of the sources leads to a different flux
contribution. This can be seen clearly at 
RA $250\deg$ where the variable source (the predicted source
contribution is shown as the red
line) has become smaller in amplitude between the data acquisition of
Dec. $37.5\deg$ and that of Dec. $40\deg$. The only way to allow for
this is to make simultaneous observations of all sources and subtract
their flux from the raw data. This is work in progress (see
Figure~\ref{fig:3c345}).

\begin{figure}
\caption{Comparison of the MEM reconvolved data (green line) and the
raw stacked data (black line) for each of the declinations at
10~GHz. Also shown (red line) is the expected point source
contribution from an extrapolation of the 1.4~GHz and 5~GHz Green Bank
point source catalogue.}
\label{fig:10memtharr}
\end{figure}

\begin{figure}
\caption{Comparison of the MEM reconvolved data (green line) and the
raw stacked data (black line) for each of the declinations at
15~GHz. Also shown (red line) is the expected point source
contribution from an extrapolation of the 1.4~GHz and 5~GHz Green Bank
point source catalogue.}
\label{fig:15memtharr}
\end{figure}

As only one declination is currently available at 33~GHz no map
reconstruction was possible and so the MEM algorithm was only used to
subtract the long term baseline variations. The result from this
subtraction was shown in Figure~\ref{fig:tharr33}. However, the two
dimensional map reconstructions at 10~GHz and 15~GHz, which are fully
sampled in both declination and right ascension in this region, are
shown in Figures~\ref{fig:10map} and \ref{fig:15map}. The switched
beam pattern has been removed from the data to produce these map
reconstructions but they are still convolved in a $5\deg$ beam. Only the central
region away from the Galactic plane crossings is shown as this is the
area where it may be possible to identify CMB features. 

\begin{figure}
\caption{The reconstruction of the sky at 10~GHz using MEM on the
Tenerife $5\deg$ Tenerife experiment data. The pixels are $1\deg
\times 1\deg$.}
\label{fig:10map}
\end{figure}

\begin{figure}
\caption{The reconstruction of the sky at 15~GHz using MEM on the
Tenerife $5\deg$ Tenerife experiment data. The pixels are $1\deg
\times 1\deg$.}
\label{fig:15map}
\end{figure}

\subsubsection{Point source contribution}
\label{pointten}

The main radio sources in this region of the sky are 3C345 (RA
$250\deg$, Dec. $39\deg$, $\sim 8$~Jy at 10~GHz and 15~GHz), 4C39 (RA
$141\deg$, Dec. $39\deg$, $\sim 9$~Jy at 10~GHz and 15~GHz) and
3C286 (RA $200\deg$, Dec. $30\deg$, 4.5~Jy at 10~GHz and 3.5~Jy at
15~GHz). The two larger sources are clearly visible with peak
amplitudes of $\sim 400\mu$K at 10~GHz. 4C39 is seen with a peak
amplitude of $300\mu$K and 3C345 with an amplitude of $250\mu$K at
15~GHz. The expected amplitude (using
Equation~\ref{eq:fluxtoT}) is $300\mu$K for 3C345 and $350\mu$K for
4C39 at both frequencies. The discrepancy between the observed flux
and the expected flux (taken from the Kuhr catalogue) is easily
accounted for when it is noted that both sources are $\sim 50\%$
variable. 3C286 is less well defined at 10~GHz as it occurs at the
very edge of the observed region but there is still evidence for a
source at the expected position with a peak amplitude of $\sim
100\mu$K (the expected peak is $170\mu$K). In the 15~GHz reconstructed
map 3C286 is more clearly defined and can be seen with a peak
amplitude of $140\mu$K (the expected peak is $130\mu$K). All other
sources in the region have peak amplitudes at least five times smaller
than the three discussed here and are therefore well below the noise. 

\subsubsection{Galactic source contribution}
\label{galten}

As the comparison between the low frequency Galactic maps and the
5~GHz interferometer maps was so poor it was decided that an
extrapolation up to 10~GHz would be impossible. Therefore, a
comparison of the 5~GHz narrow spacing map and the 10~GHz Tenerife map
was made by eye to check for any possible common features. One feature
which is clearly detected in both the 5~GHz (at 0.3mK in the
interferometer beam) and 10~GHz (at 0.2mK in the $5\deg$ Gaussian beam) 
maps lies just above Dec. $35\deg$ at RA
$170\deg$ (this feature is also detected in the 1420~MHz and 408~MHz
maps as well). This feature was assumed to be Galactic in origin in the
5~GHz map and also seems to be Galactic in origin in the 10~GHz map
(as it has vanished in the 15~GHz map). Taking into account the
different beam sizes (FWHM of the Jodrell interferometer is $8\deg$
and that of the 10~GHz Tenerife experiment is $5\deg$) it is possible
to calculate an approximate spectral index for the feature. The peak
amplitude at 5~GHz was $\sim 280\mu$K and at 10~GHz was $\sim 140\mu$K
and so the spectral index is $\beta = 2.3\pm 0.5$. This would indicate
a Galactic origin and it is more likely that the feature is free-free
emission. This agrees with the findings of the $8\deg$ FWHM experiment
that the majority of Galactic emission between 5~GHz and 10~GHz was
free-free emission in origin.

The 10~GHz map is generally expected to contain more Galactic features
(indeed the {\em rms} values of the 10~GHz data are larger than the
15~GHz data which would indicate additional emission processes are
contributing to the data)
but it is very difficult to assign each feature to being Galactic or
cosmological in origin without a simultaneous analysis of the 10~GHz
and 5~GHz (and possibly lower frequency data). This is now work in
progress with the new multi-MEM procedure but due to the size and
complexity of the problem it was not possible to produce any results
in time for the publication of this thesis.

\subsubsection{CMB features in the map}
\label{cmbten}

It is very difficult to decide whether a particular feature is
cosmological in origin or whether it originates from one of the
foregrounds considered here. Without the completion of the full
analysis now in progress it is only possible to speculate on the
origin of the features detected. By comparing the various frequencies from
the Tenerife experiments or by comparison with other experiments it is
possible to make a good `guess' at whether a particular feature is CMB
or not. It was seen that the 15~GHz Tenerife data set is expected to have a
maximum Galactic contribution of 10~$\mu$K which is well below the
noise and so the 15~GHz reconstruction is predominantly 
CMB. The 15~GHz data can be used by itself, as a first approximation, to put
constraints on CMB features. 
One example of a possible CMB feature is visible at RA
$180\deg$ and Dec. $40\deg$. This feature is detected at the same
amplitude in the 10~GHz, 15~GHz and 33~GHz data sets and so is a clear
candidate for being CMB in origin (this was first reported in Hancock
\et 1994). Another example are the features at Dec. $35\deg$ between
RA $210\deg$ and $250\deg$ at 15~GHz which appear as two positive features
separated by a large negative feature. These also appear in the COBE
53~GHz and 90~GHz maps and were used for the prediction by Bunn \et
(1995) which was shown in Figure~\ref{fig:10_15_dec35}. With such a
large frequency coverage indicating that the features do have the
correct spectral dependence to be CMB in origin these are probably the
best candidates in literature today for CMB features.


\vspace{2cm}

The following chapter introduces some of the techniques used to
analyse the maps produced here in an attempt to automatically
characterise the features without the need for a comparison by eye.

\chapter{Analysing the sky maps}
\label{chap8}

In the preceding chapter the data were processed by MEM to produce a
two dimensional partial sky map of the CMB fluctuations. In this
chapter I will explore some of the main procedures that are in use to
analyse CMB maps to get the maximum amount of information from
them.

\section{The power spectrum}
\label{powers}

A simple way of comparing maps is to look at their power spectra. By
Fourier transforming the temperature fluctuation distribution 
underlying theories can be tested
by comparing the predicted spectrum with the observed. However, there
are problems in implementing this. The main problem is the spherical
nature of the sky. A simple Fourier transform is not possible unless
the sky area is small enough so that the spherical nature of the sky
can be ignored. To
overcome this, a high resolution experiment can be used to survey a
small area of sky. The power spectra from $10\deg \times 10\deg$
patches of simulated Planck Surveyor data were plotted in
Chapter~\ref{chap6}.  As the sky area covered decreases sample
variance will quickly dominate the 
errors. Sample variance was not plotted in the power spectra of
Chapter~\ref{chap6} as the purpose of these plots was not to compare
the result with theory but to compare the reconstructed power spectra
with the true, input power spectra. 
Otherwise, it is necessary to use spherical harmonics to
transform the map into $\ell$ space (see Chapter \ref{chap2}). 
This has problems caused by the nature
of data acquisition. Without full sky coverage (which is never possible
because of the Galactic plane contamination of the data, which needs to
be excluded when testing the CMB parameters) there will
always be artefacts present due to the absence of data. Window
functions (for example, the cosine bell) 
can be used to reduce their effects but they cannot be
completely eliminated. The window functions also have the effect of
reducing the number of data points that are used in the analysis (the
ones at the edge of the map are weighted down) and so have the effect
of increasing the errors on the final parameter estimation. 
So, instead of trying to predict what the
theory looks like using the data, it is better to use the theory to try
and predict what the data should look like, as all artefacts can then
be easily incorporated, and then compare this prediction with the real
data. This is done using the likelihood function
(see Chapter \ref{chap4} for likelihood results). 

The power spectrum is a useful test for the cosmological parameters in 
a given theory. However, it is 
fairly straight-forward to construct a map with Gaussian fluctuations and one
with non-Gaussian fluctuations (based on the string model for example) 
that have the same power spectrum. Therefore, it is necessary to use
further tests for non-Gaussianity and the remainder of this chapter 
attempts to summarise some of these tests.

\section{Genus and Topology}
\label{genus}

When presented with a map of any description the eye automatically
searches for shapes within that image. It would therefore be logical
to construct an algorithm that will do this but in a statistical
manner. By using the topology of an object it is possible
to group together shapes with similar mathematical
properties. There are many ways of defining the
topological parameters of an object. The shapes in a two dimensional
map can be characterised by their area, circumference or
curvature. The mean curvature of a map is also known as the Genus. 

\subsection{What is Genus?}
\label{what}
 
In three dimensions (see Gott {\em et al} 1986) an object 
will have a genus of +1 if 
it is similar to a torus and a genus of
-1 if it is similar to a sphere. In two dimensions (see Gott {\em et al} 1990)
an object will have a genus of +1 if it is similar 
to a coin and a genus of -1 if it is similar to a ring. The genus of a map
is simply the sum of the genus of each of the shapes in that map. If
a two dimensional map is a perfect sponge shape, so that there
are an equal number of coin and ring shapes, then the total average
genus is zero and we have a perfect Gaussian field. In one
dimension the genus is simply taken as the number of up crossings
above a certain threshold (see Coles \& Barrow 1987).  

Genus was first applied in cosmology to large scale structure
surveys. Many such surveys (e.g. de Lapparent, Geller \& Huchra 1986,
Schectman {\em et al} 1992, Jones {\em et al} 1994) are presently
being analysed in this manner and the results from these will be of
great interest in their own right. However, topology offers a unique
way for comparing the fluctuations present in the large scale
structure with those in the microwave background. This comparison
between today's anisotropies and their precursors will lead to
information on the evolution of the universe through gravitational
interactions. I have developed algorithms to apply the genus statistic
to pixellised CMB maps and the results from the application of these 
algorithms will be presented here. Firstly, it is useful to derive
the expected form of the genus for the case of a purely Gaussian process.

In two dimensions the genus, G, of the surface 
is given by 
\begin{equation}
G={\rm No.\; isolated\; high\; density\; regions}-{\rm No.\; isolated\; low\;
density\; regions}
\label{eq:gendef}
\end{equation}

\noindent
In the case of a CMB map this corresponds to setting a
threshold temperature and calculating the number of fluctuations above
that threshold minus those below the threshold. 
The genus can also be defined in terms of the curvature of the
contours that enclose the shapes.

Consider a contour, C, enclosing an excursion region (defined as the
region in a map above, or below, the threshold) counterclockwise. The
curvature of the contour, along its length $s$, is defined as 

\begin{equation}
\kappa(s)={1\over R}
\label{eq:kappa}
\end{equation}

\noindent
where $R$ is the radius of a circle which, when placed so that its
perimeter lies on s, has the same curvature as the contour. The radius
of curvature, $R$, is defined as being positive if the circle is on
the same side of the contour as the enclosed region and negative if
the circle is on the other side. 
The total curvature is the integral along $s$,

\begin{equation}
K=\int_C \kappa ds
\label{eq:kaptot}
\end{equation}
 
\noindent
The genus is then defined as

\begin{equation}
G={1\over 2\pi} \int_C \kappa ds.
\label{eq:gen2d}
\end{equation}

\noindent
For example, consider a contour enclosing a simple, circular, 
high temperature region (like a coin). The radius of curvature around
the contour will always be equal to the radius of the coin and, as the
circle is on the same side of the contour as the enclosed region, it
will be positive. Therefore,
$\kappa$ is a constant ($1/R$) and the integral around $s$ is equal to the
perimeter of the coin ($2\pi R$). The genus is therefore equal to 1. 
For a circular contour surrounded by a high temperature region
(like the inside of a ring), the radius of curvature is still equal to
the radius of the ring but now it is defined as being negative (the
enclosed region is on the outside of the circle). Therefore, the genus
is now equal to -1. There will
also be contours that cross the edge of the map being analysed and in
this case the genus will be fractional. 

The genus of an object is usually quoted as a function of the
threshold level set in computing the excursion region.  
There are many different methods to derive the expected functional
form for the genus of a two dimensional map. Adler (1981) derived the
form of the genus for general geometrical problems and Bardeen {\em et
al} (1986) and Doroshkevich (1970) use the Euler--Poincare statistic
to derive the frequency of high density peaks in Gaussian fields. All
CMB maps are produced on a pixellised grid and so it is more
advantageous to follow the derivation set out in Hamilton {\em et al}
(1986; hereafter HGW86) which applies 
the genus approach to three dimensional smoothed
large scale structure surveys.

HGW86 use tessellated polyhedra to analyse large scale structure
data. By smoothing the density function of the large scale structure
they are able to calculate the mean density in octahedra. This gives
the objects in their maps regular, repeating shapes that make it
easier to calculate the genus. In two dimensions, as in CMB maps, the
simplest form of tessellation to use are square pixels (although this
does lead to an ambiguity in assigning genus; see below). Each 
octahedron (or pixel in two dimensions) is then either above or below
the threshold level set. The surface of the excursion region is then
the surface of the octahedra. As stated in HGW86, 
the curvature of any polyhedra is only
non--zero at its vertices and so it is relatively easy to find the
total curvature by summing up the vertex contributions. As the number
of polyhedra approaches infinity (so that they are infinitely small)
then the genus calculated in this way approaches the true genus of the
objects being analysed. 

Consider a map, in two dimensions, made up of square pixels of size
$d\times d$. Each pixel has four vertices which touch four other
pixels. The number of vertices per unit volume, $N_{vol}$, is therefore 

\begin{equation}
N_{vol}={N_{vert}\over{Area \times N_{pixels}}} = {1\over d^2}.
\label{eq:genden}
\end{equation}

\noindent
where $N_{pixels}$ is the number of pixels at each vertex.
Now consider the possible configurations about the vertices. Figure
\ref{fig:gen2d} shows the sixteen possibilities around each
vertex. The genus of the vertex in each case is easily calculated. For (a)
the genus is zero in both 4 low density and 4 high density cases. For
(b) the genus is +1/4 for the 3 low density and 1 high density case,
and -1/4 for the 3 high density and 1 low density complimentary
case. For (c) the genus is zero. For (d) the genus is slightly
ambiguous. If we consider the two high density pixels to be connected
and separating the two low density pixels then the genus is -2/4 but
if the two low density regions are connected and they separate the two
high density regions then the genus is +2/4. These two possibilities 
are shown in Figure
\ref{fig:higvslow}. To account for this ambiguity the genus is
assigned randomly as $\pm 2/4$ for this case. 
For a CMB map the density of a pixel is just the
average temperature within that pixel and from now on will be referred
to as such.

\begin{figure}
\caption{The 16 different configurations possible around a vertex in a
pixellised map. The shaded region corresponds to pixels that are above
the threshold level and are, therefore, within a high density
excursion region.}
\label{fig:gen2d}
\end{figure}

\begin{figure}
\caption{Case (d) in Figure \ref{fig:gen2d} has two possible
configurations around the vertex leading to two different genus.}
\label{fig:higvslow}
\end{figure}

The genus defined above can be shown to be correct in the simplest
cases: consider a single high temperature pixel in a sea of low temperature
pixels. Each vertex will then correspond to case (b) in Figure
\ref{fig:gen2d} and contribute +1/4 to the total genus. Summing over
each of the vertices gives $4\times +1/4 = +1$ which is the expected
result from Equation \ref{eq:gendef}. The total expected theoretical
genus for any pixellised map
can be calculated by summing the genus contribution for each vertex
multiplied by the probability that the vertex has that configuration. 

Assume that the temperature distribution in the CMB map is Gaussian and define
fluctuations from the mean as 

\begin{equation}
\delta={{T-\bar{T}}\over{\bar{T}}}
\label{eq:densgen}
\end{equation}

\noindent
where $\bar{T}$ is the average temperature in the full sky map. In the
case of CMB maps $\bar{T}=2.73$K, the temperature of the blackbody
spectra, but this is already subtracted from the maps in most
cases. The probability of each vertex configuration can now be
calculated. Label the four pixels around a vertex in a clockwise
direction as 1, 2, 3 and 4 and define the correlation function as 

\begin{equation}
\xi_{ij}=< \delta_i \delta_j >
\label{eq:esp}
\end{equation}

\noindent
where $i$ and $j$ are over the four pixels. Now define that
probability function 

\begin{equation}
f(\delta_1,\delta_2,\delta_3,\delta_4 )=  {1\over{4\pi^2 ({\rm det}[{\bf
\xi}])^{1\over2}}} {\rm exp} \left( -{1\over2} \sum_{ij}
\xi_{ij}^{-1} \delta_i \delta_j \right)
\label{eq:fprob}
\end{equation}

\noindent
where ${\rm det} [{\bf \xi}]$ is the determinant of the covariance
matrix $\xi_{ij}$. It is
easily seen that, from symmetry the matrix ${\bf \xi}$ can be
written as 

\begin{equation}
\left( \begin{array}{cccc}
\xi(0) & \xi_{12} & \xi_{13} & \xi_{12} \\
\xi_{12} & \xi(0) & \xi_{12} & \xi_{13} \\
\xi_{13} & \xi_{12} & \xi(0) & \xi_{12} \\
\xi_{12} & \xi_{13} & \xi_{12} & \xi(0) 
\end{array}
\right) .
\label{eq:epmatrix}
\end{equation}

\noindent
Note that $\xi_{ii}=\xi (0)$ which in turn is the square of the pixel
temperature {\em rms} from $\xi_{ii}=<\delta_i^2>$. 
In terms of the continuous correlation function over the map
$\xi_{12}=\xi (d)$ and $\xi_{13}=\xi (d\sqrt{2})$. As the pixels
become smaller and smaller a Taylor expansion can be made for $\xi
(r)$

\begin{equation}
\xi(r)= \xi(0) + {r^2 \over 2!} \xi^{(2)} + {r^4 \over 4!} \xi^{(4)}
+ \; .\: .\: .
\label{eq:taylxi}
\end{equation}

\noindent
where 

\begin{equation}
\xi^{(n)}=\left| {d^n \xi(r)\over dr^n} \right|_{r=0}.
\label{eq:xiderivs}
\end{equation}

\noindent
It is now possible to derive the functional form of the genus.

If the probability that pixel 1 is above the threshold temperature,
$\delta_c$, and pixels 2, 3 and 4 are below $\delta_c$ is equal to $p_1$ then 

\begin{equation}
p_1=\int_{-\infty}^{\delta_c} \int_{-\infty}^{\delta_c} \int_{-\infty}^{\delta_c}
\int_{\delta_c}^{\infty} f(\delta_1,\delta_2,\delta_3,\delta_4)
d\delta_1 d\delta_2 d\delta_3 d\delta_4 .
\label{eq:p1prob}
\end{equation}

\noindent
Similarly, the probability that pixels 1 and 2 are above $\delta_c$
and pixel 3 and 4 are below $\delta_c$ is given by

\begin{equation}
p_{12}=\int_{-\infty}^{\delta_c} \int_{-\infty}^{\delta_c}
\int_{\delta_c}^{\infty} \int_{\delta_c}^{\infty}
f(\delta_1,\delta_2,\delta_3,\delta_4) d\delta_1 d\delta_2 d\delta_3
d\delta_4 . 
\label{eq:p12prob}
\end{equation}

\noindent
The remaining probabilities follow in an analogous way. By symmetry
$p_1 =p_2 =p_3 =p_4$, $p_{12}=p_{14}=p_{23}=p_{34}=p_{13}=p_{24}$ and
$p_{123}= p_{234}=p_{341}=p_{124}$. Combining this with Equation
\ref{eq:genden} the genus per unit area is given by 

\begin{equation}
G={1\over d^2} \sum_i g_i p_i
\label{eq:gentotal}
\end{equation}

\noindent
where $i$ runs over the vertices and $g_i$ is the genus of the pixel
with configuration $i$ and probability $p_i$. Using the symmetry of
the probabilities the genus is

\noindent
\begin{displaymath}
G={1\over d^2} 
( 0\times p_{none}+ \left({+{1\over4}}\right)\times 4p_1
+0\times p_{12} +{1\over2}\times{\left({+2\over4}\right)} \times
2p_{12} + 
\end{displaymath}

\begin{equation}
{1\over2}\times{\left({-2\over4}\right)} \times 2p_{12}
+\left(-{1\over4}\right) \times 4p_{123}+0\times p_{1234} ) =
{{(p_1 - p_{123})}\over d^2}
\label{eq:totprob}
\end{equation}

\noindent
and so only $p_1$ and $p_{123}$ need to be calculated. The algebra
required in Equation \ref{eq:p1prob} is very long and in the past has
been relegated to computer programs like MACSYMA (see Melott {\em et
al} 1986, hereafter MCHGW). MCHGW perform the analysis for hexagon
shaped pixels which eliminates the ambiguity in case (d), Figure
\ref{fig:gen2d}. However, hexagonal pixels are not used in general CMB
experiments (although see Tegmark 1996 for a recent hexagonal pixel
projection of the COBE data) 
and so the analysis here is restricted to square pixels.
MCHGW quote the result for square pixels
to fourth order in pixel size after the Taylor
expansion of all $\xi$ terms have been performed. They find, to second
order in pixel size,

\begin{equation}
G={1\over (2\pi)^{3\over2}} \left(-{\xi^{(2)} \over \xi(0)} \right)
\nu e^{-{\nu^2 \over 2}} \left( 1+ {d^2 \over 24} \left(-{\xi^{(2)}
\over \xi(0)} \right) [3-\nu^2 - {\xi^{(4)} \xi(0) \over (\xi^{(2)}
)^2} ] + \; .\: .\: .\:. \right)
\label{eq:melotgen}
\end{equation}

\noindent
where $\nu$ is given by

\begin{equation}
\nu={\delta_c \over \xi(0)^{1\over2}}.
\label{eq:nugen}
\end{equation}

\noindent
and is the number of standard deviations of the temperature {\em rms} 
that $\delta_c$ is from the mean of the map. As the pixel size
approaches zero the genus approaches the result for a non--pixellised
map which was shown by Adler (1981, p115) to be 

\begin{equation}
G\propto \nu\>exp\>(-{{\nu^2}\over2}).
\label{eq:Gadler}
\end{equation}

\noindent
The term in $d^2$ in Equation \ref{eq:melotgen} can, therefore, 
be thought of as an
error on the genus calculated from a pixellised map due to the
pixellisation. 

For random Gaussian fluctuations the area fraction covered by
fluctuations above a certain threshold, $\delta_c$, is given by 

\begin{equation}
f=\int^\infty_\nu {1\over{\sqrt{2\pi}}} e^{-\nu^2/2} d\nu = {1\over 2}
{\rm erfc} \left({\nu\over\sqrt{2}}\right),
\label{eq:fracgen}
\end{equation}

\noindent
where erfc($x$) is the complementary error function. This is an easier
definition to use when implementing the algorithm in the case of a
pixellised map, as it is trivial to set the $f$ highest temperature
pixels as the excursion region. It is this definition which is used in
the analysis to construct the contours, but all graphs will be plotted
as a function of $\nu$.

The expected form for a random Gaussian temperature map has been
calculated. The genus from the data, calculated on pixellised maps using
the definitions shown in Figure \ref{fig:gen2d}, 
can be compared to this and if it
differs significantly from the expected curve then the underlying
field is non--Gaussian. 

\subsection{Simulations}
\label{simgens}

To test the genus algorithm and its power at distinguishing between
the origin of fluctuations, simulated data was used. This also tests
the relative merit of each of the experiments for distinguishing
between Gaussian and non--Gaussian origins for the temperature
fluctuations. Simulations for the Planck Surveyor experiment were used. 
The genus of the
full sky will closely follow that of the theoretical curve 
(to within cosmic variance) if it is Gaussian distributed. However,
most experiments do not have sufficient sky coverage to allow for
this. Therefore, the simulations performed here are for a smaller
patch of the sky. The simulations were made at 300~GHz. 

Figure \ref{fig:cdmgen} shows 9 regions analysed using genus for a
Cold Dark Matter simulation 
compared to the theoretical curve for a Gaussian process. 
The regions used for these plots are
$50\times 50$ pixels (a total of $1.25\dg \times 1.25\dg$). Figure
\ref{fig:stringsgen} shows 9 regions analysed for a string
simulation. There is no significant difference to the eye between
these plots. Figure \ref{fig:szgen} shows the results for the SZ
analysis. The SZ genus appears to be
shifted to the left of the theoretical genus. This shift implies that
there are more excursion regions above the {\em rms} of the map than
below it (the area under the curve in the positive $v$ region is larger than
that in the negative $v$ region). The SZ effect is made up of point
source features and so there are indeed more excursion regions above
the {\em rms}. 

\begin{figure}
\caption{Genus of CDM simulations (error bars are the {\em rms} over the
simulations) compared with the theoretical expected genus for a
Gaussian distributed function (dashed line).}
\label{fig:cdmgen}
\end{figure}

\begin{figure}
\caption{Genus of string simulations (error bars are {\em rms} over the
simulations) compared with the theoretical expected genus for a
Gaussian distributed function (dashed line).}
\label{fig:stringsgen}
\end{figure}

\begin{figure}
\caption{Genus of SZ simulations (error bars are the {\em rms} over the
simulations) compared with the theoretical expected genus for a
Gaussian distributed function (dashed line).}
\label{fig:szgen}
\end{figure}

To test the significance of the genus analysis a $\chi^2$ fit to the
recovered genus was performed. The $\chi^2$ level was minimised with
respect to the amplitude of the theoretical genus. Table
\ref{ta:genus} shows the results for each of the maps and whether that
map is assigned to be Gaussian or not. As can be seen from the table each
channel is assigned correctly to Gaussian or non-Gaussian for the full
range of $\nu$. However, when the central range of $\nu$ is
considered, the SZ effect is the only one to be assigned as
non-Gaussian. This is to be expected as the non-Gaussian features of
the string simulations are on small angular scales
(line discontinuities) and these will
only show up in the genus at the extremities of the map where there are
not enough fluctuations for the average effect to appear Gaussian (by the central
limit theorem). Therefore, it is possible to say that for a highly
non-Gaussian process (like the SZ effect) the genus algorithm can
easily distinguish the non-Gaussian effects, whereas for a process in
which the central limit theorem dominates (a large area of CMB
anisotropies produced by strings) genus can only distinguish the
non-Gaussian effects in the peaks of the distribution.

\begin{table}
\begin{center}
\begin{tabular}{|c|ccc|c|} \hline
Channel & & $\chi^2$ & & Gaussian? \\ 
& $-1 < \nu < 1$ & $-2 < \nu < 2$ & $-3 < \nu < 3$ & \\ \hline
CDM & $1.0\pm 0.5$ & $1.1\pm 0.6$ & $3.3\pm 2.6$ & Yes \\
Strings & $1.3\pm 0.8$ & $1.2\pm 0.5$ & $30\pm 16$ & No \\
SZ & $18 \pm 4$ & $60\pm 7$ & $673\pm 89$ & No \\
\hline
\end{tabular}
\end{center}
\caption{$\chi^2$ obtained from the calculated genus for each of the
simulations compared to the predicted genus for a Gaussian distributed
function.}
\label{ta:genus}
\end{table}

As a further use of the genus algorithm, simulations of the CMB maps
produced by strings were used to compare the
proposed satellite experiments. Table~\ref{ta:genusofks} shows the
minimised $\chi^2$ for the genus from 50 string simulations produced by
Pedro Ferreira at different resolutions. One of the simulations is
shown in Figure~\ref{fig:pedres} at the four different resolutions
considered here. The Planck Surveyor is
expected to have a maximum resolution of 4.5 arc minutes and the MAP
satellite will have a maximum resolution of 17 arc
minutes\footnote{Since the simulations presented here were performed
the resolution of the MAP satellite has improved to 12 arc minutes}. From the
table it can be seen that with a beam of $4.5^\prime$ the non-Gaussian
nature of the strings is clearly seen at 15 times the expected
$\chi^2$ for a Gaussian process, whereas at
$17^\prime$ the non-Gaussian nature is still seen but at a much
reduced level. Again, it is seen that without a very high resolution
experiment ($\sim 1^\prime$) the non-Gaussian nature of the string
simulation is only seen in the extrema of the temperature
distribution. It should be noted that no noise was added to these
simulations so these represent the best scenario for any
experiment with these resolutions. 

\begin{table}
\begin{center}
\begin{tabular}{|c|ccc|c|} \hline
Beam FWHM & & $\chi^2$ & & Gaussian? \\ 
& $-1 < \nu < 1$ & $-2 < \nu < 2$ & $-3 < \nu < 3$ & \\ \hline
1.0 & $4.5\pm 1.6$ & $3.3\pm 1.1$ & $69\pm 13$ & No \\
4.5 & $0.9\pm 0.5$ & $0.9\pm 0.4$ & $15\pm 5$ & No \\
10.0 & $0.9\pm 0.5$ & $0.9\pm 0.4$ & $4.8\pm 1.7$ & No \\
17.0 & $1.0\pm 0.5$ & $0.9\pm 0.4$ & $3.1\pm 0.8$ & No \\
\hline
\end{tabular}
\end{center}
\caption{$\chi^2$ obtained from the calculated genus over 50 string
simulations provided by Pedro Ferreira. They are convolved with
different beams and are over a $2\deg \times 2\deg$ patch of the sky.}
\label{ta:genusofks}
\end{table}

\begin{figure}
\caption{One of the string simulations produced by Pedro Ferreira at
a) $1.0^\prime$, b) $4.5^\prime$, c) $10^\prime$ and d) $17^\prime$
resolution.}
\label{fig:pedres}
\end{figure}

It is also possible to use the genus algorithm to check for
differences between non-Gaussian theories. Three possible sources of
non-Gaussian effects are monopoles, strings and textures. Simulations
provided by Neil Turok were used to test the power of the genus
algorithm for distinguishing between these three theories of
non-Gaussian anisotropies. Figures~\ref{fig:genmons} to
\ref{fig:gentexs} show the genus for nine of the simulations of 
the monopoles, strings and textures. By eye there does not seem to be
a great deal of difference between the three defect models. However,
with the theoretical Gaussian model fitted (by minimising the $\chi^2$
as before), the difference is more
obvious. Table~\ref{ta:genusofnt} shows the results from the $\chi^2$
calculation for each model. As can be seen each model requires the
extremities of the temperature distribution to be distinguished from a
Gaussian process. It is also possible to see that the defect process
that deviates most from Gaussian is string theory. All three are
easily discernible from Gaussian at a very large significance. 
Table~\ref{ta:likelgen} shows the likelihood values for each of 
the non-Gaussian models used here. One test simulation of each process was 
compared to each input model to see if the genus statistic could correctly 
identify the model. For these particular cases, 
it is seen that the underlying input model for each test
simulation is correctly identified although there is a possibility that
the texture and monopole maps may be mistaken for each other.
The addition of noise will reduce the differences
slightly but it has been shown that MEM can reconstruct the CMB to a
very high degree of accuracy so it is not expected to effect the
results significantly. 

\begin{figure}
\caption{Genus from 9 of the 30 simulations of the CMB from a monopole
theory. The errors are the 68\% confidence limits over those 30 simulations.}
\label{fig:genmons}
\end{figure}

\begin{figure}
\caption{Genus from 9 of the 30 simulations of the CMB from a string
theory. The errors are the 68\% confidence limits over those 30 simulations.}
\label{fig:genstring}
\end{figure}

\begin{figure}
\caption{Genus from 9 of the 30 simulations of the CMB from a texture
theory. The errors are the 68\% confidence limits over those 30 simulations.}
\label{fig:gentexs}
\end{figure}

\begin{table}
\begin{center}
\begin{tabular}{|c|ccc|c|} \hline
Model & & $\chi^2$ & & Gaussian? \\ 
& $-1 < \nu < 1$ & $-2 < \nu < 2$ & $-3 < \nu < 3$ & \\ \hline
Monopoles & $1.0\pm 0.5$ & $1.5\pm 0.6$ & $274\pm 31$ & No \\
Strings & $1.0\pm 0.5$ & $0.9\pm 0.3$ & $1558\pm 106$ & No \\
Textures & $0.8\pm 0.4$ & $1.1\pm 0.5$ & $645\pm 85$ & No \\
\hline
\end{tabular}
\end{center}
\caption{$\chi^2$ obtained from the calculated genus over 30 
simulations provided by Neil Turok of the CMB expected from 
different defect models. They are convolved with
a $4.5^\prime$ beam to simulate the best results possible from the
Planck Surveyor.}
\label{ta:genusofnt}
\end{table}

\begin{table}
\begin{center}
\begin{tabular}{|cc|ccc|} \hline
&Input model & Monopole & Strings & Textures \\
Test model & & & & \\ \hline
Monopoles & & $0.16$ & $2.3\times 10^{-11}$ & $0.11$ \\
Strings & &   $2.0\times 10^{-15}$ & $0.44$ & $8.1\times 10^{-10}$ \\
Textures & &  $0.42$ & $6.5\times 10^{-8}$ & $0.63$ \\ \hline
\end{tabular}
\end{center}
\caption{The likelihood results for the three different non-Gaussian 
simulations. A peak value of 1 is obtained if the test model has the 
exact genus of the input model. The genus of the input model was found by
averaging over 32 test models of each defect process. It is seen that the 
strings model is easily distinguished from the other two defect processes (see
text).}
\label{ta:likelgen}
\end{table}

\subsection{The Tenerife data}
\label{tengen}

So far the genus algorithm has been applied to simulated data from
future experiments. Now the genus algorithm will be used on
simulated data from an existing experiment, the Tenerife switched-beam
experiment. Figure \ref{fig:gensims} shows the normalised genus
averaged over 30 maps taken from a Gaussian realisation of the CMB for
the Tenerife experiment.
This is seen to have the expected form of Equation \ref{eq:Gadler}. 

\begin{figure}
\caption{Average genus from 30 2 dimensional simulations of the CMB with the
experimental configuration of the Tenerife experiment over a $10\deg \times
100\deg$ area of the sky. A standard deviation (st dev) of 0 corresponds to
50\% of the area being high density and 50\% low and a standard deviation of
3 corresponds to 98\% of
the area being low density and 2\% high. The genus has been normalised
to one at its maximum.}
\label{fig:gensims}
\end{figure}

Simulated Tenerife
observations of one of the Gaussian realisations 
were performed for a $10\deg \times 100\deg$ area of the sky. 
The genus of the input map used is shown in Figure
\ref{fig:geninp}. The data was then analysed with the genus
algorithm and the average over 30 noise realisations is shown in Figure
\ref{fig:genout}. As can be seen from this plot, even though the
error bars are large, the point of intersection with the y--axis is
well recovered. This point corresponds to half the pixels being classed
as high temperature and half as low temperature. It is intrinsically
dependent on the smaller fluctuations in the data as well as the large
ones and MEM is seen to be performing very well on all amplitudes
in this reconstruction. In this realisation
it is seen that the two regions (high and low temperature) are not
completely equivalent as expected in a Gaussian case but as this is
just one realisation from an ensemble this is to be expected. Therefore, when
using the genus to analyse maps care must be taken to include both the 
errors from the noise realisations (Figure~\ref{fig:genout}) and the errors
from the sample variance (Figure~\ref{fig:gensims}) before any conclusion 
about the non-Gaussian nature of the underlying process is reached.  

\begin{figure}
\caption{Genus of one of the 2 dimensional simulations used as the
input for the next figure.}
\label{fig:geninp}
\end{figure}

\begin{figure}
\caption{Genus of the output map after 30 simulations with the
Tenerife configuration with different noise realisations. v=0 corresponds to
50\% of the area being high density and 50\% low, v=3 corresponds to 98\% of
the area being low density and 2\% high.}
\label{fig:genout}
\end{figure}

The genus algorithm can also be used to provide
extra proof that observations have detected real astronomical
fluctuations and are not noise dominated. The amplitude of a genus
curve is proportional to the amount of structure present within the
map so even though the noise map has the same form as the CMB map (for
Gaussian CMB) the amplitude will be different. By
simulating noise maps Colley, Gott \& Park (1996) and Smoot \et (1994) show that 
the COBE maps have more structure in them than expected from pure
noise at a level of over four standard deviations. They also show that
there is no significant deviation away from Gaussian fluctuations,
although this only rules out highly non--Gaussian processes, as
most expected non--Gaussian fluctuations will approach Gaussian on
this angular scale due to the central limit theorem. The genus
algorithm will now be applied to the Tenerife data set in a similar
way. 

\subsubsection{The real data}
\label{realgenten}

As the 10~GHz data set is expected to be contaminated
by Galactic emission only the 15~GHz data set will be used
here. The map shown in Figure~\ref{fig:15map} was used to test the
power of the genus algorithm. The region between RA $131\deg$ and
$260\deg$ was analysed to test for Gaussianity. Figure~\ref{fig:gen15}
shows the result for this analysis. A preliminary attempt at
subtracting the effect of the two point sources in this region was
made prior to the genus analysis and it can be seen that the
non-Gaussian behaviour expected due to these sources does not show in
this figure. The average $\chi^2$ for the difference between the
theoretical curve and the genus from the 15~GHz data is 0.38. 
It is seen that the MEM
reconstruction of the Tenerife 15~GHz data is completely consistent
with a Gaussian origin (less than one sigma deviation away from the
theoretical Gaussian curve). However, this does not mean that it is
inconsistent with a non-Gaussian origin. As in the COBE analysis only
highly non-Gaussian processes can be ruled out as most expected
non-Gaussian fluctuations will approach Gaussian at the scales that
Tenerife is sensitive to. 

\begin{figure}
\caption{Genus of the 15~GHz Tenerife MEM reconstructed map. The 68\%
confidence limits shown are calculated from the Monte-Carlo
simulations (results in Figure~\ref{fig:genout}). }
\label{fig:gen15}
\end{figure}

\subsection{Extending genus: the Minkowski functionals}
\label{minkowksi}

Recent advances (Schmalzing \& Buchert 1997, Kerscher \et 1997,
Mecke, Buchert \& Wagner 1994, and references therein) 
in the analysis of Large Scale Structure data sets
using integral geometry have led to an interest in this area of
statistics in the CMB community
(see Winitski \& Kosowsky 1998 and Schmalzing \& Gorski 1998). Large
Scale Structure and CMB data sets both need the following requirements
for a functional that can describe them: the functional must be
independent of the orientation or position in space (motion
invariance), must be additive (so that the functional of the 
combination of two data sets is the addition of the two separate
functionals minus their intersection) and must have conditional
continuity (the functional of a pixellised data set must approach the true
functional of the underlying process as the pixels are made
smaller). These three requirements taken together were shown by
Hadwiger (1957) to lead to only $d+1$ functionals in a $d$-dimensional space
that would completely describe the data set. These are the Minkowski
functionals. For a CMB data set (in 2-dimensional space) there are three
Minkowski functionals; surface area, boundary length and the Euler
characteristic (or genus). 

As has already been shown, the genus of the CMB maps holds a great
deal of information and so it is expected that the inclusion of the
two other Minkowski functionals will allow further discrimination
between theories. Tests on the three-dimensional Minkowksi
functions applied to Large Scale Structure data sets
(Jones, Hawthorn \& Kaiser 1998) have shown that the
addition of the other Minkowksi functionals does indeed increase the
power to discern between underlying theories. Other groups have
already began to test the Minkowski functionals on CMB data sets 
(Schmalzing \& Gorski 1998 apply them to the COBE data set). This is work in
progress. 

\section{Correlation functions}
\label{peakstat}

Instead of looking at the morphology of the temperature distribution
using Minkowski functionals (which include genus) it is possible to
use statistical techniques to measure the distribution in space of
pixel fluxes. This is done using correlation functions. It has been
shown that any correlation function can be expressed in terms of the
Minkowksi functionals (for example, see Mecke, Buchert \& Wagner
1994). However, the correlation functions do contain useful properties
and so they will be discussed here. I will summarise the two, three
and four point correlation functions and apply the two and four point
functions to various data sets.

\subsection{Two point correlation function}
\label{twop}

The two point correlation function is a measure of the average product
of temperature fluctuations in two directions. For two pixels in a CMB
map, $i$ and $j$, separated by an angle $\beta$, the two
point correlation function is given by

\begin{equation}
V(\beta) = \left< {\left({\Delta T_i \over T_\circ} \right) } {\left({ \Delta
T_j \over T_\circ} \right)} \right>
\label{eq:twop}
\end{equation}

\noindent
and it is easily seen that when $\beta=0$, so that $i=j$, $V(0)$ is
the variance of the data. The variance is very easily calculated and
is used as a check at each of the data reduction stages. It is also
fitted for in the likelihood function. 

On the raw data 
the two point correlation function can be used as a test of the origin
of the emission detected in a CMB experiment. By applying the weighted
two point correlation function to the Tenerife data it is possible to
test whether the data is consistent with noise or whether there is
some underlying signal present. The weighted two point correlation
function is given by

\begin{equation}
C(\theta) = {{\sum_{i,j} \Delta T_i \Delta T_j w_i
w_j}\over{\sum_{i,j} w_i w_j}}
\label{eq:autocorr}
\end{equation}

\noindent
where $w_i$ and $\Delta T_i$ are the weight ($1/\sigma_i^2$) and
double-differenced temperature of the Tenerife data set. In this form
the two point correlation function is also known as the
auto-correlation function.

Figure~\ref{fig:autocorr} 
presents the auto-correlation of the 15~GHz Tenerife data in the region
at RA=$161\deg-250\deg$. The error-bars were determined by
Monte-Carlo techniques. The errors on each data point were estimated by
assuming a random Gaussian process with the appropriate {\em rms} given
by the Tenerife data set (i.e. the noise). The data point was then displaced
by this amount and a new data set was constructed for which the
auto-correlation function was found. This was done over 1000 noise
realisations and the errors show the 68\% confidence limits over these
realisations. These techniques were also used to obtain the
confidence bands in the case of pure uncorrelated noise (long-dashed
lines) and the expected correlation (short-dashed line) in the case of
a Harrison-Zel'dovich spectrum for the primordial fluctuations with an
amplitude corresponding to the signal of maximum likelihood (see
Chapter~\ref{chap4}). 
Clearly this model gives an adequate description of the
observed correlation whilst the results are incompatible with pure
uncorrelated noise. The cross-correlation between the data at 10 GHz
and 15~GHz is inconclusive as it is dominated by the noisy character of
the 10~GHz data.

\begin{figure}
\caption{The auto--correlation of the 15~GHz data in the region at
RA=$161\deg - 250\deg$. The solid line is the best fit to the data,
the short dashed line shows the expected region of correlation for a
Harrison-Zel'dovich spectrum CMB model and the long dashed line shows
the expected region of correlation for the case of pure noise. The
errors are calculated using a Monte-Carlo technique.
This figure was produced by Carlos Gutierrez.}
\label{fig:autocorr}
\end{figure}

A variation to the two point correlation function
that is commonly used (see, for
example, Kogut \et 1995 and Kogut \et 1996) is the extrema correlation
function.
Instead of applying the two point correlation function to the full
data set, the peaks (and troughs) of the data set are found. A peak is
defined as any pixel `hotter' than the neighbouring pixels and a
trough is any pixel `colder' than the neighbouring pixels. The
correlation function between these peaks and troughs is then found. It
can be separated into three different analyses; a) peak-peak
auto-correlation (and trough-trough auto-correlation), 
b) peak-trough cross-correlation and c) extrema
cross-correlation (the correlation between all extrema regardless of
whether they are a peak or trough). 
Kogut \et (1995) use the extrema two point correlation function
to analyse the COBE 53~GHz map and use the likelihood function to
predict whether the peaks are from a Gaussian or non-Gaussian
source. They find that the COBE result is most likely to have
originated from a Gaussian distribution of fluctuations although the
significance of their analysis is very difficult to compute. 

The theoretical predictions for
the two point and three point correlation functions can be found in
Bond \& Efstathiou (1987) and Falk, Rangarajan \& Srednicki (1993)
show the predictions for the full two point correlation function and
the collapsed three point correlation function (see below) 
for inflationary cosmologies. 
The main problem with any of the correlation function analysis
techniques is that they do not take into account any noise or
foregrounds present. Therefore, it is necessary to perform the
analysis on the MEM processed map or simulations of the noise and foregrounds
must be performed to evaluate their effect. Kogut \et (1995) use Monte
Carlo simulations of the noise added onto different models for the CMB
(they do not include foregrounds) to evaluate the significance of
their result. 

\subsection{Three point correlation function}
\label{threep}

The three point correlation function is similar to the two point
correlation function except that it takes the product between three
pixels. The angular separation between pixels is now $\alpha$ (between
$i$ and $j$), $\beta$ (between $i$ and $k$) and $\gamma$ (between $j$
and $k$) which gives

\begin{equation}
S(\alpha,\beta,\gamma) = \left< {\left({\Delta T_i \over T_\circ}
\right) } {\left( {\Delta T_j \over T_\circ} \right)} {\left({\Delta T_k
\over T_\circ} \right)}  \right>
\label{eq:threep}
\end{equation}

\noindent
and when $\alpha=\beta=\gamma=0$, so that $i=j=k$, $S$ is defined as
the skewness of the data. The skewness of the data is slightly more
sensitive to non-Gaussian features than the two point correlation
function. 

The collapsed three point correlation function ($\beta=\alpha$ and 
$\gamma=0$) was used by Gangui \& Mollerach (1996) to analyse the COBE
results. They found that defects could not be ruled out using the COBE
data but a higher resolution experiment could distinguish between
Gaussian fluctuations and those arising from textures. Falk \et
(1993) also show that the collapsed three point correlation function
is not sensitive enough to be detected by COBE for generic models
(those without specially chosen parameters to make the three point
correlation function artificially large). 

\subsection{Four point correlation function}
\label{fourp}

The four point correlation function is very rarely used in
full. Instead it is used in the collapsed form when the separation
between pixels is zero. This is defined as the kurtosis (see, for
example, Gaztanaga, Fosalba \& Elizalde 1997) and is equal
to

\begin{equation}
K={{\left< {\left( {\Delta T \over T_\circ} \right)^4 } \right> - 3
\left< {\left( {\Delta T \over T_\circ} \right)^2 } \right>^2
}\over{\left< {\left( {\Delta T \over T_\circ} \right)^2 } \right>^2 }}
\label{eq:kurtosis}
\end{equation}

\noindent
The kurtosis is a good discriminatory test between Gaussian and non-Gaussian
features but it can only be applied effectively in data with little or
no noise and minimal foregrounds. For the Gaussian case the kurtosis
should tend to zero. 

Gaztanaga \et (1997) use the kurtosis to analyse data
from the MAX, MSAM, Saskatoon, ARGO and Python CMB experiments (all
are sensitive to angular scales around the first Doppler peak). They
find that there is a very large kurtosis for each of the experiments
and a Gaussian origin of the fluctuations is ruled out at the one
sigma level. However, the analysis does not allow for any systematic
errors or foreground effects and these could alter the results
greatly. 

\subsubsection{Application to simulated data}
\label{kurtapp}

The kurtosis was applied to the Planck simulation maps. Table
\ref{ta:kurt} shows the kurtosis values for the analysis. The 
kurtosis of the reconstructions is compared to that of the input 
maps convolved with the highest resolution of the experiment. As can be
seen the kurtosis of the SZ effect is very high as expected for a
strongly non-Gaussian feature. After the simulated observations and
analysis was performed it can be seen that the MEM result reconstructs
the kurtosis very well. This implies that MEM is reconstructing the
non-Gaussianity closely. The Wiener results are less impressive. In
each non-Gaussian process the Wiener reconstruction is worse than the
MEM reconstruction and most markedly for the SZ effect where the input
map had a kurtosis of -1.56 and Wiener filtering recovered a kurtosis
of 3.31. The results for the MAP simulation show that
the kurtosis can also distinguish between the Gaussian and
non-Gaussian origin of the CMB fluctuations at this lower resolution
but cannot reconstruct the dust or SZ very well. This is due to the
lack of frequency coverage for the latter processes and not due to the
resolution of the experiment. Therefore, MEM is better at
reconstructing the non-Gaussian features than Wiener filtering (as was
expected from the results in Chapter~\ref{chap6}) and both MAP and
Planck should be able to distinguish between a Gaussian and
non-Gaussian process for the origin of the CMB. It should be
noted that the string simulations used here did not contain a
Gaussian background which is expected to be present due to the effects
prior to recombination and any possible reionisation that may have
occurred. 

\begin{table}
\centering
\begin{tabular}{|ccccc|} \hline
Experiment & CDM CMB & Strings CMB & SZ effect & Dust map \\ \hline
Planck input & 0.03 & 1.53 & -1.56 & 0.15 \\
MEM reconstruction & 0.04 & 1.49 & -1.36 & 0.15 \\
Wiener reconstruction & 0.05 & 1.49 & 3.31 & 0.16 \\ \hline
MAP input & -0.17 & 1.61 & -0.62 & 0.17 \\
MEM reconstruction & -0.02 & 2.11 & -1.31 & -0.92 \\
Wiener reconstruction & 0.10 & 1.59 & -1.67 & -0.95 \\ \hline
\end{tabular}
\caption{The kurtosis for the input simulations and reconstructions
from the analysis of the Planck and MAP simulations shown in Chapter
6. The MEM and Wiener reconstructions are for the case of full ICF
information. The input maps for the two experiments are convolved to
their highest resolution.}
\label{ta:kurt}
\end{table} 

\vspace{2cm}
The following Chapter will attempt to summarise the results
presented in this thesis and bring them together in a coherent
fashion.

\markboth {{\uppercase{\itshape{}}}}{{\uppercase{\itshape{}}
}}
.

\vspace{3in}
{\center
\noindent
{\em In the beginning there was only darkness, dust and water. The darkness
was thicker in some places than in others. In one place it was so thick 
that it made man. The man walked through the darkness. After a while,
he began to think. }

\vspace{1cm}
\hspace{2.0in} Creation myth from the Pima tribe in Arizona 
}

\chapter{Conclusions}
\label{conclusion}

In this final chapter I will attempt to bring together the various aspects of the work
discussed in this thesis. A brief review of the main results found and
their implications for cosmology will be undertaken in the first
section, while the future of CMB experiments is discussed in the final
section. 

\section{Discussion}
\label{discuss}

Observations from the 5 GHz interferometer at Jodrell Bank and the 10,
15 and 33 GHz switched beam experiments at Tenerife have been
presented and analysed. Simulations of observations from the proposed 
Planck surveyor and MAP satellites have also been performed.

The Jodrell Bank interferometer covers an area of the sky between
declinations $+30\deg$ and $+55\deg$ while the Tenerife experiments cover an
area between declinations $+30\deg$ and $+45\deg$. There are over 100
independent measurements for the average pixel in right
ascension at each of the declinations sampled in both experiments. The
noise per beam for each of the experiments are $\sim 20\mu$K for the 5
GHz data, $\sim 50\mu$K for the 10 GHz data, $\sim 20\mu$K for the 15
GHz data and $\sim 30\mu$K for the 33 GHz data. Taken together these
form very good constraints on the CMB fluctuations as well as the
Galactic foregrounds and point source
contribution. Figure~\ref{fig:conc1} shows the level of Galactic
foreground emission expected in a $5\deg$ FWHM CMB experiment at
frequencies between 408~MHz and 33~GHz. The points used in the
generation of this plot are the 408~MHz and 1420~MHz surveys and
predictions from the 5~GHz Jodrell Bank interferometer and the 10~GHz,
$8\deg$ FWHM Tenerife experiments. It is seen that at frequencies
below 5~GHz synchrotron emission dominates the foregrounds, whereas at
frequencies above 5~GHz free-free emission dominates. 

\begin{figure}
\caption{Level of Galactic foreground in a $5\deg$ FWHM experiment
predicted by the 408~MHz, 1420~MHz, 5~GHz and 10~GHz ($8\deg$ FWHM) surveys.}
\label{fig:conc1}
\end{figure}

The level of CMB fluctuation
found using the 15~GHz Tenerife data set is $Q_{RMS-PS}=22^{+5}_{-3}$ 
$\mu$K (68 \% confidence) at an $\ell$ of $18^{+9}_{-7}$ which is
consistent with findings from the COBE data. Combining the likelihood 
results from the COBE and Tenerife data sets more stringent constraints on the
level of CMB can be found. This analysis gave $Q_{RMS-PS} = 
19.9^{+3.5}_{-3.2}\mu$K for the level of the Sachs-Wolfe plateau.
It was also possible to put constraints on the 
spectral index which gave $n=1.1^{+0.2}_{-0.2}$ (at 68\%
confidence). For an $n=1$ spectrum it was found that 
$Q_{RMS-PS}=22.2^{+4.4}_{-4.2} \mu$K which can be compared to the result
from the Dec. $40\deg$ 33~GHz Tenerife data which gives 
$Q_{RMS-PS} = 22.7^{+8.3}_{-5.7} \mu$K for $n=1$. These likelihood 
results are all consistent with a CMB origin for the structure within
the data and span a frequency range of between 15~GHz and 90~GHz. 
Common features between the COBE and Tenerife data were
found leading to the conclusion that actual CMB features are observed
in the two experiments. Figure~\ref{fig:conc2} shows the most recent
results from various CMB experiments around the world (figure provided by
Graca Rocha). The Tenerife
result calculated here is plotted. The solid line is the predicted curve for
a Cold Dark Matter
Universe with $\Omega_\circ=1$, $H_\circ=45$~Mpc~km$^{-1}$s$^{-1}$ 
and $\Omega_b$=0.1. Taking the Tenerife data together with the results from
other experiments, sensitive to smaller angular scales, shows evidence
for a Doppler peak as expected for inflation.

\begin{figure}
\caption{Recent results from various CMB experiments. 
Shown is the predicted CMB level from the 15~GHz
Tenerife experiment derived here. The solid line is the
prediction (normalised to COBE) for standard CDM with $\Omega_\circ =
1.0$, $\Omega_b = 0.1$ and $H_\circ = 45$km\ s$^{-1}$Mpc$^{-1}$. The
Saskatoon points have a 14\% calibration error.}
\label{fig:conc2}
\end{figure}

A new technique for analysing data from CMB experiments was
presented. Positive/negative Maximum Entropy was used to extract the most information out
of each data set. With the Tenerife experiment the long term
atmospheric baseline variations were removed from the data scans as
well as the triple beam pattern produced by the switching of the
beam. Sky maps at $5\deg$ resolution at 10~GHz and
15~GHz, and $8\deg$ at 10~GHz, were produced. With the Jodrell Bank
experiment it was possible to analyse the two data sets (from the two
different baselines) independently to produce two sky maps at 5~GHz
and $8\deg$ resolution. This analysis was compared to the CLEAN
technique and it was shown that the MEM outperforms CLEAN in all areas
of map reconstruction. Using the new MEM technique it was
possible to analyse the two baseline data sets simultaneously to show
that they were consistent.

The MEM technique was also applied to simulated data from both
the Planck Surveyor and MAP satellites. Using multi-frequency
information it was shown that it should be possible to extract
information on the CMB to high accuracy (6~$\mu$K for map
reconstruction and out to $\ell\sim 2000$ for power spectrum
reconstruction) with, or without, any knowledge on the spatial
distribution of the foregrounds. The multi-MEM
technique was also compared to Single-Valued Decomposition and the
Wiener filter and it was found that multi-MEM always outperforms SVD
and if any of the foregrounds (or the CMB itself) is non-Gaussian in
structure then multi-MEM also outperforms the Wiener filter. 

The final chapter introduced some of the techniques that are used to
analyse the CMB sky maps once they have been produced. It was seen
that all the techniques reviewed have advantages. The most
promising test for non-Gaussianity appears to be Minkowski functionals as these incorporate all
of the other techniques together in just three functionals (for a two
dimensional map). Using the auto-correlation function it was
shown that there is an excess signal present in the 15~GHz Tenerife
sky map that is not due to noise alone. The Genus analysis of this map
showed that it was consistent with a Gaussian process.

\section{The future of CMB experiments}
\label{future}

Within the next ten years a new generation of CMB experiments will
be in operation. These include both space based (like the Planck Surveyor
and MAP), balloon based (like TopHat and Boomerang) 
as well as ground based (like the Very Small Array which is based on a
similar design to the CAT interferometer). It has been shown
here that the Planck surveyor will produce very 
accurate maps of the CMB fluctuations. It will also
allow very tight constraints on the level of Galactic foreground emissions 
which can be subtracted from other experiments. The recent ESA report
on the Planck Surveyor show that the sensitivity
of the satellite has improved since the simulations presented here 
were performed and so the accuracy will be even higher.

Ground--based measurements have already proven to provide tight
constraints on the level of the CMB anisotropies and so the VSA (a 15
element interferometer that will measure the CMB anisotropy at high
angular resolution) should also perform very well. In conjunction with
existing ground based telescopes 
constraints on the CMB anisotropies will increase considerably prior
to the launch of either satellite. 
The Tenerife experiment will
continue to take measurements at all three frequencies with the aim of
having a final two dimensional map with very low noise and the Jodrell
Bank interferometer is currently taking data for a new baseline.
With such a wealth
of data at many frequencies (5 - 900~GHz) 
conventional analysis techniques will have
to be refined. The maximum entropy algorithm described here can cope
with multiple frequencies, varying pixel size, multiple component
fitting and a very high level of noise. 
This `multi-MEM' is now in the process of being applied to
the Tenerife and Jodrell Bank data described in this thesis. The data
from the two experiments are also being combined with data from COBE and
lower frequency surveys (the 408 MHz, 1420 MHz and 2300 MHz surveys)
using the multi-MEM in an analogous way to the Planck
Surveyor analysis presented here, to put better constraints on the CMB
at the large angular scales covered by Tenerife and COBE ($\ell < 30$).

Another area of CMB research which is becoming increasingly of
interest is the comparison with large scale structure. All theories
that explain the shape of the power spectrum of CMB anisotropies also
make predictions for the evolution of these anisotropies into the
present large scale structure. By attempting to match the two power
spectra on the different scales more constraints can be put on the
exact form of the underlying matter. This is now being done 
by various groups although research into this area is still in its
very early stages. 

With new high quality data and the ability to
extract the CMB signal from the foreground contamination, very tight
constraints on the cosmological parameters should be achievable.

\markboth {{\uppercase{\itshape{}}}}{{\uppercase{\itshape{}}
}}
.

\vspace{3in}
{\center
\noindent
{\em Eureka?}

\vspace{1cm}
\hspace{2.7in} C. Lineweaver on the COBE discovery 
}

\backmatter

{}

\end{document}